\documentclass[final,1p,times]{elsarticle}
\usepackage{amssymb}
\usepackage{amsmath}
\usepackage{amsthm}
\usepackage{graphicx}
\usepackage{bm}
\usepackage[dvipsnames]{xcolor}
\usepackage{epsfig}
\usepackage{amsfonts}
\usepackage{bbold}
\usepackage{multirow} 
\usepackage{footnote}
\usepackage{blkarray}
\makesavenoteenv{tabular}
\makesavenoteenv{table}
\biboptions{sort&compress}

\newcommand{\beqa}{\begin{eqnarray}}
\newcommand{\beeq}{\begin{equation}}
\newcommand{\eeqa}{\end{eqnarray}}
\newcommand{\eeqe}{\end{equation}}
\newcommand{\beq}{\begin{equation}}
\newcommand{\eeq}{\end{equation}}

\newcommand{\ome}{\omega}

\newcommand{\coherenceSTM}{
Delgado_Rossier_prl_2012,
Gauyacq_Lorente_jp_2015,
Delgado_Hirjibehedin_sc_2014,
Delgado_Loth_epl_2015,
Delgado_Rossier_inprep}

\newcommand{\STMexperiments}{
Eigler_Schweize_nature_1990,
wiesendanger1990,
yazdani1997probing,
Madhavan_Chen_science_1998,
Heinrich_Gupta_science_2004,
Hirjibehedin_Lutz_Science_2006,
Hirjibehedin_Lin_Science_2007,
Meier_Zhou_Science_2008,
Otte_Ternes_natphys_2008,
Chen_Fu_prl_2008,
Ji_Hua_prl_2008,
Wiesendanger_revmod_2009,
Tsukahara_Noto_prl_2009,
Otte_Ternes_prl_2009,
Fu_Zhang_prl_2009,
Serrate_Ferriani_nature_2010,
Loth_Bergmann_natphys_2010,
Khajetoorians_Chilian_nature_2010,
Loth_Etzkorn_science_2010,
kahle2011quantum,
Khajetoorians_Lounis_prl_2011,
Khajetoorians_Wiebe_science_2011,
Khajetoorians_Wiebe_natphys_2012,
Loth_Baumann_science_2012,
khajetoorians2013current,
Khajetoorians_Schlenk_prl_2013,
Heinrich_Braun_natphys_2013,
Oberg_Calvo_natnano_2013,
Bryant_Spinelli_prl_2013,
Nadj_Perge_science_2014,
Rau_Baumann_science_2014,
Spinelli_Bryant_natmat_2014,
Spinelli_Gerrits_arXiv_2014,
Yoshida_Aizawa_natnano_2014,
Baumann_Paul_science_2015,
Burgess_Malavolti_natcom_2015,
Bryant_Toskovic_nanol_2015,
Choi_Robles_arXiv_2015,
Jacobson_Herden_natcom_2015,
Spinelli_Gerrits_natcom_2015,
Yan_Choi_nanolett_2015,
Yan_Choi_natnano_2015,
choi2016magnetic,Steinbrecher_Sonntag_natcom_2016,
yan2016non}

\begin{document}
\begin{frontmatter}

\title{Spin decoherence of magnetic atoms on surfaces
}

\author[cfm,dipc,iker]{F. Delgado\corref{cor1}}
\ead{fernando.delgadoa@ehu.eus}
\address[cfm]{Centro de F\'{i}sica de Materiales, Centro Mixto CSIC-UPV/EHU, Paseo Manuel de Lardizabal 5, E-20018 Donostia-San Sebasti\'an, Spain }
\address[dipc]{Donostia International Physics Center (DIPC), Paseo Manuel de Lardizabal 4, E-20018 Donostia-San Sebasti\'an, Spain }
\address[iker]{IKERBASQUE, Basque Foundation for Science, E-48013 Bilbao, Spain}

\author[inl,ali]{J. Fern\'andez-Rossier}
\address[inl]{QuantaLab, International Iberian Nanotechnology Laboratory (INL), Av. Mestre Jos\'e Veiga, 4715-330, Braga, Portugal}

\address[ali]{Permanent address: Departamento de F\'{\i}sica Aplicada, 
Universidad de Alicante.}

\begin{abstract}
We review the problem of spin decoherence of  magnetic atoms deposited on a surface.  Recent  breakthroughs in scanning tunnelling microscopy (STM) make it possible to probe the spin dynamics of individual atoms, either isolated or integrated in  nanoengineered  spin structures.   
Transport pump and probe techniques with  spin polarized  tips  permit measuring the spin relaxation time $T_1$, while novel demonstration of  electrically driven  STM single spin resonance has provided  a direct  measurement of  the spin coherence time $T_2$ of an individual magnetic adatom.  Here we address the problem of spin decoherence from the theoretical point of view. First we provide a short general overview of decoherence in open quantum systems and we  discuss with some detail ambiguities that arise in the case of degenerate spectra, relevant for magnetic atoms. Second, we  address the physical mechanisms that allows probing  the spin  coherence  of magnetic atoms on surfaces. Third, we discuss the main spin decoherence mechanisms at work on a surface, most notably, Kondo interaction, but also spin-phonon coupling and dephasing by Johnson noise.   Finally,  we briefly discuss the implications in the broader context of quantum technologies.

\end{abstract}

\begin{keyword}
  decoherence \sep relaxation \sep Kondo \sep adatoms \sep spin-phonon
\PACS 72.15.Qm \sep 75.10.Jm \sep 75.30.GW \sep 75.30.Hx \sep 75.78.-n \sep 76.20.+q
\end{keyword}

\end{frontmatter}

\tableofcontents

\section{Introduction}
%
%
%
%

Major technological revolutions have occurred when the humankind has been able to harness natural resources, such as fire, electricity or nuclear energy.   We are now in the verge of the
 so called second quantum revolution, that aims to harness two of the weirdest natural resources,  coherence and entanglement.   This is a tall order that calls for  a great dose of ingenuity, because   keeping quantum states in coherent superpositions that could be used towards  our advantage requires to defeat a rather powerful enemy,  the infamous decoherence.  Here, we review the phenomenon of  spin decoherence in the the context of magnetic atoms deposited on surfaces. 

\subsection{The relevance of decoherence}

The interaction of quantum spins with their environment  introduces relaxation and decoherence 
in the otherwise  fully coherent evolution  of ideal closed quantum systems~\cite{slichter2013principles}. Spin relaxation and decoherence  play a central role in many branches of physics.  In the case of nuclear spins, the time scales associated to energy relaxation and decoherence, $T_1$ and $T_2$ respectively, provide a very meaningful information of the environment that forms the basis of magnetic resonance imaging techniques~\cite{Pykett_Newhouse_rad_1982}.   The spin relaxation and decoherence time scales set the limit of sensitivity in several existing  magnetometry techniques, such as optically detected magnetic resonance (ODMR)~\cite{Cavenett_aphys_1981,jelezko2006,Balasubramanian_Chan_nature_2008} and spin-exchange relaxation-free (SERF) atomic magnetometry~\cite{Kominis_Kornack_nature_2003}, and are also one of the major constraints in the implementation of  
spin-based quantum computers, such as donors in silicon~\cite{Kane_Nature_1998}, electrons in quantum dots~\cite{Loss_DiVincenzo_pra_1998} and even molecular magnets~\cite{Leuenberger_Loss_nature_2001,Ardavan_Rival_prl_2007}. 

Decoherence plays a prominent role in the modern interpretation of
the  foundations of quantum mechanics, the quantum measurement problem~\cite{Leggett_Chakravarty_rmphys_1987,Zurek_revmodphys_2003,Suter_Alvarez_rmp_2016,adler2003decoherence,
sep-qm-decoherence} and the quantum to classical transition~\cite{Zurek_physics_today_91}. 
 In the more specific context of magnetism,    decoherence  accounts for the emergence of the  classical  behavior~\cite{Delgado_Loth_epl_2015}. For instance, in the quantum realm,  the ground state  of an  integer spin $S$ nanomagnet with uniaxial anisotropy  can display  quantum spin tunnelling~\cite{garg1993topologically,Gatteschi_Sessoli_book_2006},  so that the ground state is non-degenerate, and it is separated from the first excited state by an energy gap known as quantum spin tunnelling splitting $\Delta_{\rm QST}$.   The wave functions of both the ground state and the first excited states are then linear superposition states~\cite{Delgado_Loth_epl_2015}
\begin{equation}
|\phi\rangle\propto |C_1\rangle+e^{i\theta} |C_2\rangle  
\label{Qstate}
\end{equation}
where $|C_1\rangle$ and $|C_2\rangle$ describe states with a well defined and mutually orthogonal  magnetic moments, and $\theta$ is a (real) phase.   These  states have null  expectation value of the magnetic moment, which highlights how different they are from our experience in the classical realm. By contrast, a half-integer spin  uniaxial magnet has  two equivalent  ground states, whose quantum states would be $|C_1\rangle$ and $|C_2\rangle$ respectively.   As we discuss below,   the coupling to the environment favors  classical states and makes states like (\ref{Qstate}) fragile~\cite{Delgado_Loth_epl_2015}, unless 
$\Delta_{\rm QST}$ is larger than all the relevant energy scales in the problem, which only happens for  small $S$ at cryogenic temperatures.  

The emergence of states with a non-zero atomic magnetization is even more intriguing in the case of insulating  antiferromagnets~\cite{anderson1997concepts,anderson1984basic,Donker2016}, such as MnO,  that display the so called N\'eel states, with a finite staggered magnetization, demonstrated  in a seminal neutron diffraction experiment~\cite{shull1951neutron}.    A good starting point to describe  insulating antiferromagnets  is the  Heisenberg model, which commutes with the total spin operator $\hat S_{\rm TOT}^2$. As a result, the ground state is an eigenstate of  $\hat S_{\rm TOT}^2$. For antiferromagnets this would be a state with  $S_{\rm TOT}=0$, which has a null expectation value for every atom in the lattice~\cite{anderson1997concepts} and it is thus very different from the broken symmetry N\'eel states~\cite{katsnelson2001neel} that are actually observed.  Therefore, other minor interactions such as magnetic anisotropy, together with  the coupling to the environment,  must account for the emergence of symmetry breaking N\'eel states, as we discuss with some detail below.

\subsection{Magnetic adatoms}
\begin{figure}[t]
  \begin{center}
       \includegraphics[width=1.\linewidth,angle=0]{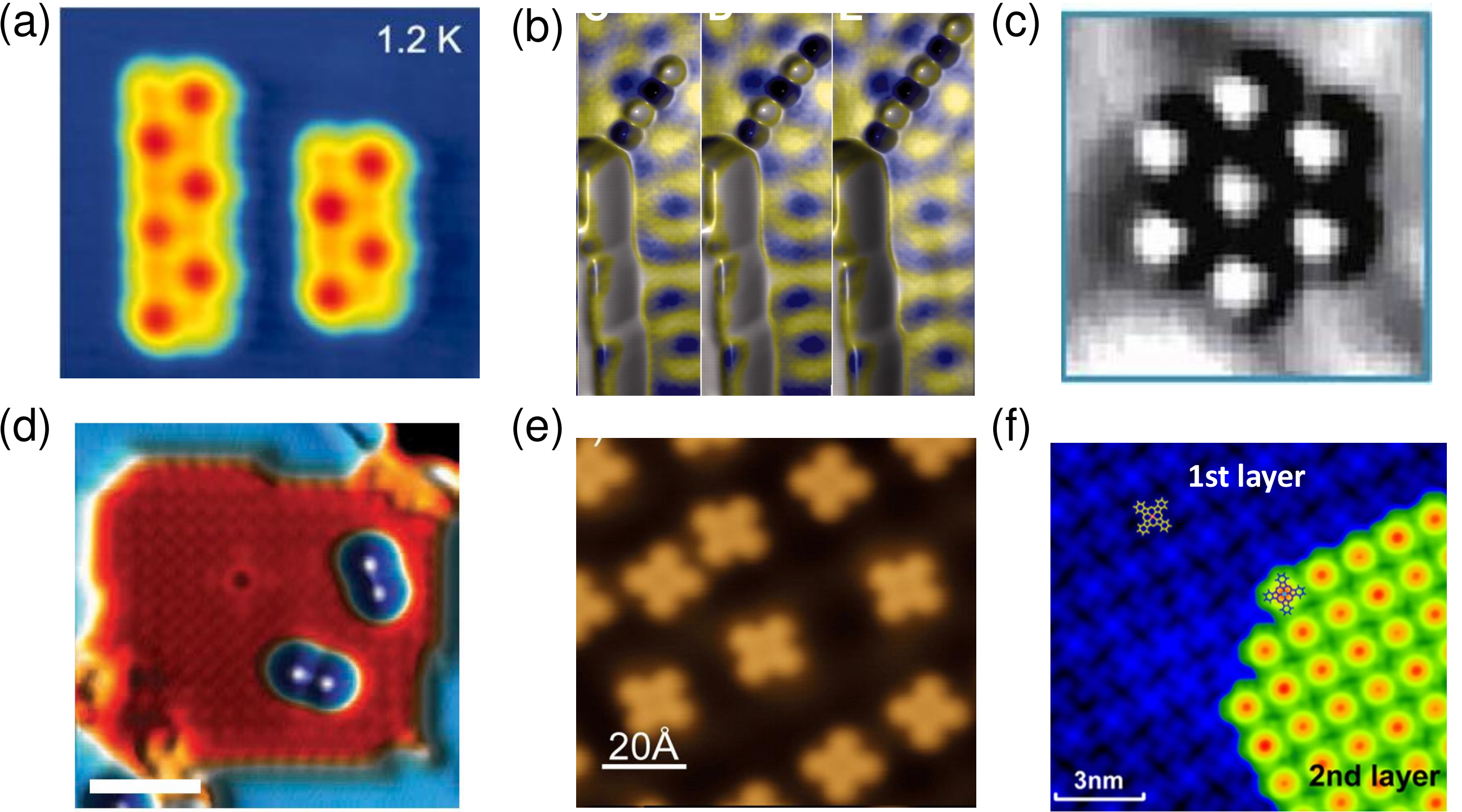}  
  \end{center}
  \caption{ STM topography images of small arrays of magnetic adatoms or molecules deposited on substrates. (a) From S. Loth {\em et al.}, Science  {\bf 335}, 196 (2012). Reprinted with permission from AAAS. ($2\times 6$) and ($2\times 4$) Fe arrays on a Cu$_2$N/Cu(100) substrate. (b) From A.A. Khajetoorians {\em et al.}, Science {\bf 332}, 1062 (2011). Reprinted with permission from AAAS. Spin-resolved topography image of the $dI/dV$ of 4, 5 and 6 Fe chains of on a Cu$(111)$ surface. (c)  From A.A. Khajetoorians {\em et al.}, Nature Physics {\bf 8}, 497 (2012), reprinted by permission of Macmillan Publishers Ltd: Nature Physics copyright (2012). Hexagonal array of seven antiferromagnetically coupled Fe atoms on Cu$(111)$. (d) Reprinted with permission from Bryant {\em et al.} {\bf 111}, 127203 (2013). Copyright 2013 by the American Physical Society.  Weakly coupled antiferromagnetic Fe dimer on Cu$_2$N/$Cu(100)$. (e) Reprinted with permission from B.W. Heinrich {\em et al.}, Nanoletters {\bf 13}, 4840 (2013). Copyright 2013 American Chemical Society. Fe porphyrin on Au$(111)$ (e) From Xi Chen {\em et al.}, Phys. Rev. Lett. {\bf 111}, 197208 (2008). Copyright 2008 by the American Physical Society. Stacking of two Co-Phthalocyanine layers on a Pb$(111)$ surface.
  }
\label{fig0}
\end{figure}

The extraordinary series of experimental breakthroughs~\cite{\STMexperiments} in the manipulation (see Fig.~\ref{fig0})   and probe of magnetic atoms, mostly using scanning tunnelling microscopes (STM) and in some instances X ray magnetic circular dichroism (XMCD), have allowed the study of the  crossover from quantum to classical regime~\cite{Loth_Baumann_science_2012} in nanoengineered spin structures as well  the exploration of coherent dynamics of individual magnetic atoms of surfaces  \cite{Baumann_Paul_science_2015}.
An early development that permitted probing magnetism with  STM   was
the so called spin polarized STM (SP-STM). Based on the same physical principles that tunnel magnetoresistance~\cite{wiesendanger1990},  SP-STM 
  yields the average magnetization of individual magnetic atoms on surfaces~\cite{Meier_Zhou_Science_2008,Wiesendanger_revmod_2009}. These experiments  could be analyzed in terms  non-quantized  magnetic moments,  typical of itinerant magnetic systems,  a picture in line with the results of density functional calculations for magnetic metals~\cite{Wortmann_Heinze_prl_2001}.

One of the first   experimental spectroscopic fingerprints of spin-related phenomena was 
  the observation of in-gap Yu-Shiba-Rusinov (YSR)  states~\cite{Yu_aps_1965,Shiba_ptp_1968,Rusinov_sjetp_1969} for Mn and Gd magnetic atoms on a superconducting Nb$(110)$ surface~\cite{yazdani1997probing}. Soon after this observation, a characteristic Kondo  dip at the Fermi energy in the $dI/dV$ for Cobalt on Au$(111)$ was observed~\cite{Madhavan_Chen_science_1998}, which implied the screening of the atomic magnetic moment and the formation of a correlated singlet state.  
  Both the Kondo peak and the YSR  states  are a  expected  consequences of the Kondo exchange interaction between the magnetic adatom and the conduction electrons of the substrate. These Kondo interactions are known to produce a finite spin lifetime for local spins~\cite{korringa1950nuclear,langreth1972theory,Delgado_Hirjibehedin_sc_2014}, and thereby a broadening in their spin spectral functions.  
  
  The development of single spin  inelastic electron tunnelling spectroscopy (IETS) gave a direct access to the  atomic spin excitations 
  of both individual magnetic atoms~\cite{Heinrich_Gupta_science_2004,Hirjibehedin_Lin_Science_2007} and atomic spin chains fabricated adding atoms one by one~\cite{Hirjibehedin_Lutz_Science_2006}. In most instances these  observations~\cite{Heinrich_Gupta_science_2004,
Hirjibehedin_Lutz_Science_2006,
Hirjibehedin_Lin_Science_2007,
Otte_Ternes_natphys_2008,Oberg_Calvo_natnano_2013,
Rau_Baumann_science_2014,Spinelli_Bryant_natmat_2014,
Baumann_Donati_prl_2015,Baumann_Paul_science_2015,Jacobson_Herden_natcom_2015}
were reported for  magnetic atoms  deposited on top of an atomically thin insulating  decoupling layer, such as Cu$_2$N/Cu(100)~\cite{Hirjibehedin_Lin_Science_2007,
Otte_Ternes_natphys_2008,Oberg_Calvo_natnano_2013,Choi_Gupta_jp_2014}, MgO/Ag(001)~\cite{Rau_Baumann_science_2014,Baumann_Donati_prl_2015,Baumann_Paul_science_2015},   or h-BN/Rh(111)~\cite{Jacobson_Herden_natcom_2015}, or on top of  doped semiconductors substrate~\cite{Khajetoorians_Chilian_nature_2010}, although they have also been observed in     Fe/Pt(111)~\cite{Khajetoorians_Schlenk_prl_2013,Khajetoorians_Steinbrecher_natcomm_2016}.  
The decoupling layer   decreases the strength of the Kondo coupling with the underlying surface,
preserving the localized atomic-like nature of the spin excitations in the magnetic adatoms. This localized nature is in line with the fact that,  in most cases~\cite{Hirjibehedin_Lin_Science_2007,
Otte_Ternes_natphys_2008,Otte_Ternes_prl_2009,Oberg_Calvo_natnano_2013,
Rau_Baumann_science_2014,Baumann_Donati_prl_2015,Baumann_Paul_science_2015,Jacobson_Herden_natcom_2015}, the experiments could be accurately modelled using quantized spin model Hamiltonians~\cite{Rossier_prl_2009}, such as single spin models, very often used in the context of transition metal ions in insulators~\cite{Abragam_Bleaney_book_1970} and molecular magnets~\cite{Gatteschi_Sessoli_book_2006},  and the Heisenberg model commonly used to study magnetic insulators~\cite{Manousakis_rmp_1991}, which describes spin exchange interactions between localized moments.

Broadening of the inelastic spin transitions,  beyond the thermal  5.4$k_BT$  factor~\cite{Jaklevic_Lambe_prl_1966},  has been measured for Fe on top of Cu(111)~\cite{Khajetoorians_Lounis_prl_2011},  in line with the predictions of theory for spin relaxation due to Kondo coupling~\cite{Delgado_Rossier_prb_2010} and also with more sophisticated theoretical treatments~\cite{dos2015relativistic,Jacob_Rossier_arXiv_2015}.  This broadening  is 
the spectral counterpart of a finite spin relaxation time $T_1$ in the time domain. 
The development of  electrical pump-probe technique with STM~\cite{Loth_Etzkorn_science_2010}
has made it possible to measure the rather fast  (ns) spin relaxation time $T_1$ of individual atoms~\cite{Loth_Etzkorn_science_2010,Baumann_Paul_science_2015}
and other atomically  engineered structures, such as antiferromagnetic chains and ladders~\cite{Loth_Baumann_science_2012,Yan_Choi_natnano_2015} or ferromagnetic chains~\cite{Spinelli_Bryant_natmat_2014}  and clusters~\cite{khajetoorians2013current}.

%
%
%

Compared to single dopants systems~\cite{koenraad2011single,rossier2013single},  such as nitrogen-vacancy (NV) centers  in diamond or single donor in silicon~\cite{zwanenburg2013silicon},    the study of spin coherence of magnetic adatoms is in its infancy. 
At the time of writing this review, there is only one experimental paper~\cite{Baumann_Paul_science_2015} that reports the measurement of the decoherence time $T_2$ of an individual Fe atom on MgO/Ag(001) inferred from a continuous wave resonance experiment, rather than 
%
the more sophisticated spin echo techniques  used to probe the $T_2$ of
 individual $P$ dopants~\cite{Pla_Tan_nature_2012,Muhonen_Dehollain_natnano_2014} or shallow NV centers~\cite{Rosskopf_Dussaux_prl_2014,Myers_Dartiailh_prl_2014,Romach_Muller_prl_2015}. 
   On the theory side, only a few papers address the problem of decoherence in this specific context~\cite{\coherenceSTM}.  
Yet,  the potential of magnetic adatoms to explore quantum spin dynamics is enormous for 
several reasons.   First, STM  makes feasible to fabricate  atom-by-atom  magnetic nanostructures~\cite{Hirjibehedin_Lutz_Science_2006,Khajetoorians_Wiebe_science_2011,
Loth_Baumann_science_2012,Spinelli_Bryant_natmat_2014,Choi_Gupta_jp_2014,Choi_Robles_arXiv_2015}, controlling at will the number of atoms, the spin $S$ of the atoms, and  their interatomic distance.  In turn, this  permits researchers to  determine  both the  strength and the sign of the exchange interaction between magnetic adatoms~\cite{Bryant_Spinelli_prl_2013,Spinelli_Gerrits_natcom_2015}.   The vast space for combinations of spin $S$,  exchange interactions, magnetic anisotropies and number of atoms results in very different types of magnetic behaviors, including both systems with broken-symmetry N\'eel states~\cite{Loth_Baumann_science_2012} or quantum disordered spin chains~\cite{Hirjibehedin_Lutz_Science_2006}.  Strong spin-correlation can lead, in some cases,
to the emergence of non-trivial $S=1/2$ objects at the edges~\cite{Delgado_Batista_prl_2012} as in chains of antiferromagnetically coupled $S=1$ atoms, objects that might be more robust with respect to decoherence than conventional $S=1/2$ spins.

Another  important resource to explore
 quantum coherent phenomena in  magnetic atoms on  surfaces is the very large potential to  engineer 
 their coupling to the substrate by means of the adequate choice of materials, including both substrate and coating layer.  Most of the experiments mentioned in this review are done using 4 types of metal (Cu, Ag, Pt, Rh), in some instances with a  coating layer, [Cu$_2$N/Cu(100), MgO/Ag(001) or h-BN/Rh(111)]. 
 Using different materials will bring many opportunities. Take the example of superconductors. 
 The standard BCS~\cite{bardeen1957theory} theory of superconductor predicts that spin relaxation lifetimes of a localized spin in a superconductor, such as nuclei,  can be significantly enhanced~\cite{hebel1959nuclear}.  This has been already experimentally demonstrated by means of STM spectroscopy  for  electronic spins in magnetic molecules on top of superconducting lead~\cite{Heinrich_Braun_natphys_2013}.   The discovery of zero-energy edge states  in ferromagnetic chains deposited on superconductors~\cite{Nadj_Perge_science_2014}, which might be a physical realization of the Majorana modes,  
  provides additional motivation to place magnetic nanostructures on top of electronically interesting substrates. In this regard, the exploration of adatom spin dynamics in the case of substrates with peculiar transport properties, such as  topological insulators, graphene,  or the edge states of Quantum Hall and Quantum Spin Hall systems  has barely been explored.

Molecules are another extremely powerful  resource to build  on surfaces nanostructures with non-trivial magnetic properties.   IETS has been used to probe, with atomic resolution, the spin excitations of individual magnetic molecules, such as Fe phthalocyanine (FePC)~\cite{Tsukahara_Noto_prl_2009}, with $S=1$,  Cobalt PC, with $S=1/2$~\cite{Chen_Fu_prl_2008},  and molecular magnets, such as Mn$_{12}$~\cite{kahle2011quantum}  and Fe$_4$~\cite{Burgess_Malavolti_natcom_2015}.   In the case of CoPC,  the self-assembly of molecular stacks also allowed 
exploring the spin properties of vertically spin chains, a physical realization of the Hubbard model~\cite{Chen_Fu_prl_2008,Delgado_Rossier_prb_2011}.
Transport experiments  both on open-shell and single-molecule magnets have also been reported using mechanical break junction techniques that, in some instances,  permits one to add a gate and  study different charge states of the molecule~\cite{park2002,vincent2012,thiele2014,Gaudenzi_Burzuri_nanolett_2016,
Frisenda_Gaudenzi_nanolett_2015}.   Molecular magnets with build-in spin chains~\cite{Ardavan_Rival_prl_2007}  have been shown to posses very long $T_1$ and $T_2$ times in diluted phases. Therefore, it would be very interesting to probe  them individually, using STM, and see how their coupling to a conducting electrode changes $T_1$ and $T_2$.

Whereas atomically engineered magnetic nanostructures are definitely  interesting by their own,  they can also be   used to probe the spin dynamics of nearby structures~\cite{yan2016non}. This concept can be downscaled to the atomic limit:   probing the spin of a single magnetic atom can provide information of a chemically different neighboring atom~\cite{Otte_Ternes_prl_2009,Steinbrecher_Sonntag_natcom_2016,natterer2016} as well as the dynamics of magnetic nanostructure~\cite{choi2016magnetic}. In addition, the combination of different magnetic atoms in the same structure, such a spin chain,  may result in unexpected new properties that show the enormous potential of spin doping in correlated systems~\cite{Choi_Robles_arXiv_2015}. Limits for upscale remain to be explored:  non-magnetic atomically engineered structure where more than eight thousand atoms were nanostructured has been presented recently~\cite{Kalff_Rebergen_arXiv_2016}.

With this background, the main goal of this review  paper is to serve as a guide for future exploration of  spin coherence at the atomic scale. 
 To do so, we shall
try to promote cross fertilization between 
the traditional STM/atomic scale magnetism community on one side, with other fields  where coherent single/few spin physics has been successfully explored  using different instrumental techniques,  most notably,  optically detected magnetometry using NV centers, and silicon qubits.  The rest of this review is divided in four main blocks. 
The first one is devoted  to provide a general theory background that yields a proper definition of  coherence/decoherence and how to compute it.  Particular attention will be given to two level systems (TLS), given that in many instances one will deal with systems with either a doubly degenerate ground state, such as half-integer spin magnetic atoms, or systems where the two lowest energy states are well separated from the higher energy excited states, such as  spin chains with strong uniaxial anisotropy~\cite{Loth_Baumann_science_2012,Spinelli_Bryant_natmat_2014,Delgado_Loth_epl_2015}.
Then, in the second block we illustrate the analysis of decoherence due to the main source in magnetic adatoms, the Kondo exchange coupling with the substrate electrons.
 In the third block we discuss other decoherence mechanism for magnetic adatoms, including spin-phonon coupling and Zeeman coupling to the random environmental magnetic field created by  Johnson-Nyquist  noise and shot noise. 
 In the fourth block, we briefly review the state of the art experimental status to measure spin dynamics with STM and we comment on the main challenges to observe Rabi oscillations at the single-atom spin level.
 Finally, we finish with a discussion and main results of our work.

\begin{savenotes}
\begin{table*}[t]
\caption{Typical orders of magnitude of the relaxation times $T_1$ and decoherence times $T_2$ in a variety of spin systems at $T\approx 1$ K in different environments.}
\centering
\begin{tabular}{l|c|c|c|c|c|}
\hline
System&     Shallow    Donors in Si & NV-Centers  & QD's & Magnetic adatom\\
\hline \hline
$T_1$(s) &  $10^{-2}- 10^ {3}$ ~\cite{Tyryshkin_Tojo_naturemat_2012,Morley_Warner_natmat_2010,George_Witzel_prl_2010}\footnote{ The $T_1$ grows exponentially with $1/T$. $T_2$ is bounded by $T_1$ at high temperatures but then saturates at low temperatures at a value dependent on the dopant concentration. The values in the table corresponds to $T\approx 1.8$ K.}
   & $1-10$  \cite{Takahashi_Hanson_prl_2008}  & 
   $ 10^{-4}$~\cite{Fujisawa_Austing_nature_2002,Kroutvar_Ducommun_nature_2004}  & $10^{-13}-10^{4}$ \cite{Khajetoorians_Lounis_prl_2011,Donati_Rusponi_science_2016,natterer2016}
   \footnote{It has been claimed that Ho adatoms on a Pt$(111)$ surface  leads to relaxation times exceeding the second time scale~\cite{Miyamachi_Schuh_nature_2013}. However, this claim is in clear contrast with XMCD measurements revealing no evidence of magnetic stability and a different ground state configuration for Ho/Pt$(111)$~\cite{Donati_Singha_prl_2014}, which violates the theoretical criterion proposed by Miyamachi et al.~\cite{Miyamachi_Schuh_nature_2013}. Moreover, more recent SP-STM and IETS-STM measurements have found no evidences of magnetic moment of Ho on this substrate~\cite{Steinbrecher_Sonntag_natcom_2016}. } \\
$T_2$(s) & $10^ {-4}-0.6$ \cite{Tyryshkin_Tojo_naturemat_2012,Morley_Warner_natmat_2010,George_Witzel_prl_2010}
  &  $10^{-3}- 10^ {-4}$~\cite{Takahashi_Hanson_prl_2008}  & $10^{-7}-10^{-5}$   &  $2.10^{-7}$ \cite{Rau_Baumann_science_2014} \\
\hline
\end{tabular}
\label{tablaT}
\end{table*}
\end{savenotes}

\section{Decoherence, a general overview\label{Chap2}}

\subsection{Quantum dissipative dynamics in open quantum systems: decoherence and relaxation\label{openQS}}
%
%
In principle, the study of the dynamics of a quantum state can be tackled in deceptively  simple terms, 
solving the time dependent
Schr\"odinger equation
\beq
i\hbar \frac{\partial |\psi(t)\rangle}{\partial t}=H|\psi(t)\rangle
\eeq
that describes the unitary evolution of the  state $|\psi(t)\rangle$. This  gives us the most complete 
information about the quantum system. In particular, the expectation value of any observable $\hat A$ can be then computed as
$\langle \hat A(t)\rangle = \langle \psi(t)|\hat A|\psi(t)\rangle$.    However, strictly speaking,
$H$ should be the Hamiltonian of the entire universe,  in which case the Schr\"odinger equation can not be solved, and even if it could, it would give us a bunch of unusable information.

\begin{figure}
  \begin{center}
       \includegraphics[width=0.7\linewidth,angle=0]{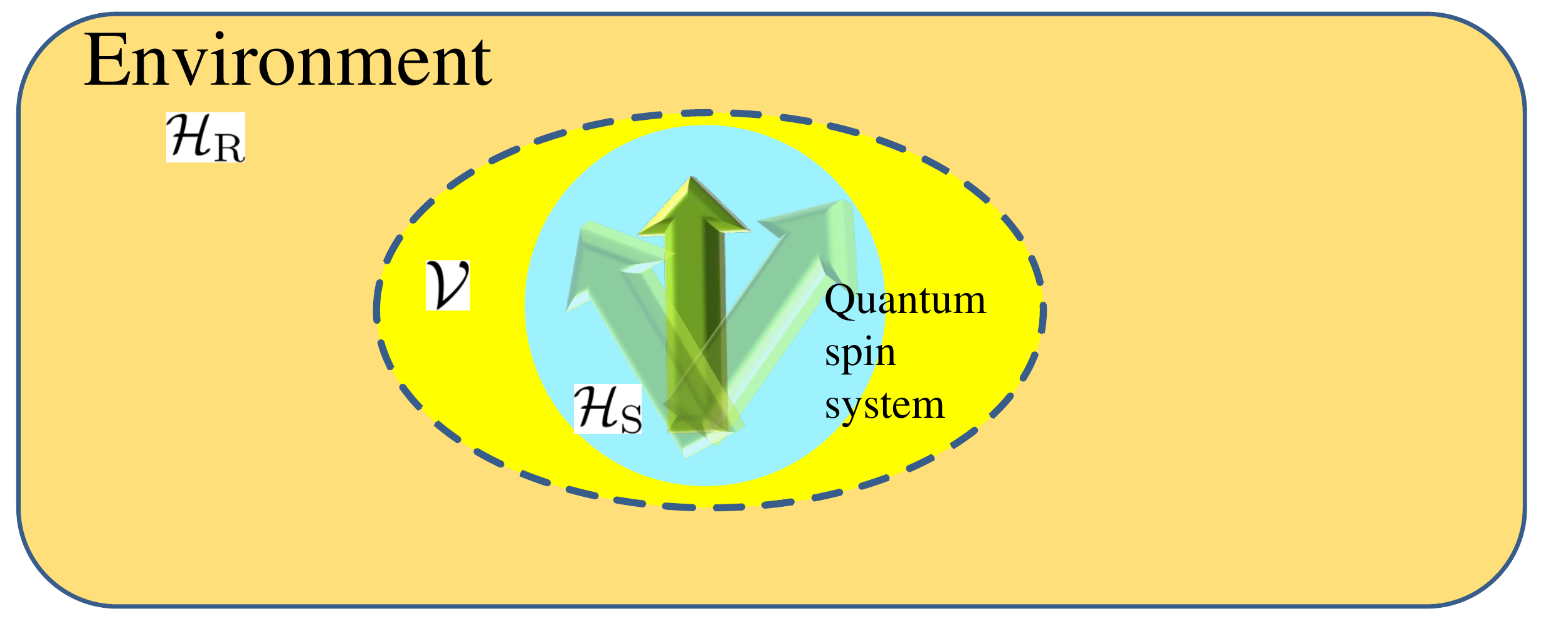}         
  \end{center}
  \caption{Scheme of an {\em open quantum system}. The central idea is to split the whole system into three: a small (and analytically or numerical treatable) quantum spin system with Hamiltonian ${\cal H}_S$, which can be studied when isolated from the rest of the universe,  the environment ${\cal H}_R$, which comprises all other degrees of freedom not included in the system, and the interaction ${\cal V}$ between them (yellow region in the scheme). 
  }
\label{fig1sh}
\end{figure}

Instead of giving up,   one always restricts $H$ to the system of interest, which we label as ${\cal H}_{\rm S}$, 
and we focus on the dynamics of the restricted set of degrees  of freedom of the system ${\cal S}$ described by ${\cal H}_{\rm S}$, see Fig.~\ref{fig1sh}.
To do so, we  split the the Hamiltonian in 3 terms: 
 \begin{equation}
H= {\cal H}_{\rm S} + {\cal H}_{\rm R}+ {\cal V} ,
\label{HTOT}
\end{equation}
where ${\cal H}_{\rm R}$ is the Hamiltonian of the environment, {\it i.e.}, the  degrees of freedom explicitly excluded from ${\cal H}_{\rm S}$, and ${\cal V}$ describes the coupling between environment and the system. 
The next step is to derive a  dynamical equation to describe ${\cal S}$, including both the action of  ${\cal H}_{\rm S}$  and the influence of the environment, which will be described in an statistical  coarse grained manner.

Now we need to find a dynamical equation to describe the ${\cal S}$, including both the action of  ${\cal H}_{\rm S}$  and ${\cal V}$.  
 This is precisely the central theme in the study of the so called  {\em open quantum systems}~\cite{Breuer_Petruccione_book_2002}.   For the dynamical equation, we adopt the density matrix language to account for the quantum dynamics of a small subsystem while tracing out the degrees of freedom of the rest, which are less interesting or irrelevant for the observer, leading in general to a non-unitary evolution.  
   Importantly, as in most cases the environment consist on a system with a macroscopically large number of degrees of freedom, it  can be modelled as a {\em reservoir} or {\em bath} that remains in thermal equilibrium, neglecting the back-action of   ${\cal H}_{S}$ on the density matrix of the reservoir.  In contrast, the bath influences the dynamics of ${\cal S}$, and more specifically, it is the  ultimate responsible of the decay of the quantum coherence of the otherwise  isolated quantum system.

For any    global (system plus reservoir) quantum state $|\psi(t)\rangle$  we can define the total density operator  $\hat \rho_{\rm Tot}(t)=|\psi(t)\rangle\langle \psi(t)|$.  
In the spirit of the open quantum system approach, we  introduce the  {\em reduced density operator}~\cite{Breuer_Petruccione_book_2002} 
\begin{equation}
\hat\rho(t)={\rm Tr}_{R} \left[\hat\rho_{\rm Tot}(t)\right],
\end{equation}
where ${\rm Tr}_{R}\left[ \dots  \right]$ corresponds to the trace over the bath degrees of freedom. When represented in the basis of eigenstates of ${\cal H}_{\rm S}$, labeled by $|n\rangle$, the
 reduced density matrix (DM) has a clear statistical interpretation.    The diagonal entries of that matrix, $\rho_{nn} \equiv P_n$,  give us the probability of finding the system ${\cal S}$ in a given eigenstate $|n\rangle$. These {\em occupations} $P_n$ satisfy the normalization condition $\sum_n P_n=1$ together with $P_n\ge 0$. 
  The off-diagonal entries $\rho_{nn'}$ ($n\ne n'$) are known as {\em coherences}.  They quantify the capability of the quantum system to be in a superposition state that combines  two different eigenstates $|n\rangle$ and $|n'\rangle$.

In general,  the dynamical equation for the reduced DM can be written down as: 
\beq
\partial_t \hat\rho(t)=-\frac{i}{\hbar}\left[{\cal H}_{S},\hat\rho(t)\right]+  {\cal R}\left[ \hat\rho(t);t \right].
\label{general}
\eeq 
The first term on the right hand side would describe the coherent dynamics if we neglect the coupling of the system to the reservoir.   The DM of a pure quantum state $\hat{\rho}= |\psi\rangle\langle\psi|$ satisfies the identity $\hat{\rho}^2=\hat{\rho}$, a condition preserved by the first term.  The second term accounts for the influence of the reservoir and it induces the non-unitary dynamics of interest here.  
 This term is responsible of both, relaxation and decoherence. In fact,  in the language of DM, relaxation is associated to the decay of any departure of the diagonal elements (populations)  on a time scale $T_1$, while the decoherence corresponds to the decay of the off-diagonal terms on a time scale $T_2$.

\subsubsection{Coherence as a basis dependent quantity}

  At this point, we must point out that  the notion of coherences is basis-set dependent.  In the context of spins, there are two natural basis sets. 
  First,  the basis defined by the eigenstates of the quantum system ${\cal H}_S$, where the thermal equilibrium  density matrix acquires a simple diagonal form.  
   Therefore,  coherence between eigenstates is a transient phenomena, and is bound to fade away. This is the phenomenon of decoherence. 
    However,    there is  a second natural choice of basis set to express the density matrix,  the basis of eigenstates of a spin observable, say $\hat S_z$,   convenient due to the existence of specific probes to measure this observable.   Whereas in some cases ${\cal H}_S$ commutes with $\hat S_z$,   so that it is possible to choose a basis set that diagonalizes both, in other instances this is not the case,  and we can have a density matrix that is diagonal in one basis, but not in the other.  Thus, coherence and decoherence, as defined above, depend on the choice of basis set, or at a more pragmatic level, on the type of experiment used to probe the system.  For this reason, whenever we have such an ambiguity, hereafter we will use a subindex in the density matrix to label the basis set in which the matrix is represented. 
  
  The  fact that coherence depends on the basis set becomes particularly tricky when there is a degenerate spectra. In that case, there are infinite possible choices of basis sets. Interestingly, the coupling to the environment is not neutral regarding this choice, which leads to the definition of the so called {\em pointer states}. We shall discuss this below.

 It must be pointed out that coherent superpositions  can be found in rather trivial situations, even at the macroscopic scale. For instance, the quantum description of the classical motion of a harmonic oscillator, given by the  Glauber quantum states,  entails coherent superpositions of the so called Fock states $|\nu\rangle$, eigenstates of the harmonic oscillator Hamiltonian $H_{\rm HO}= \hbar \omega \left(\hat{n}+\frac{1}{2}\right)$~\cite{Cohen_Roc_book_1989}.  In contrast,   having a coherent superposition of macroscopically different classical states, is a much harder task.


\subsubsection{Coherence in a two level system (TLS)\label{CohTLS}}
The simplest case in which we can illustrate the concepts of decoherence is a two level system (TLS).  In addition, the TLS approximation can be often used to describe   magnetic atoms and nanostructures.   Let us consider  a $S=1/2$ spin in the presence of a magnetic field along the $z$ axis, with eigenstates  $S_z=\pm \frac{1}{2}$.   The representation of the  DM in the basis set of eigenstates $\uparrow$ and $\downarrow$ reads as:
\begin{equation}
\hat\rho(t) =
\left(\begin{array}{cc}
P_{\uparrow}(t)  & C_{\uparrow\downarrow}(t)\\
C_{\downarrow\uparrow}(t)
 &P_{\downarrow}(t) \end{array}
\right)
\label{DM01}
\end{equation}
or, writing it in terms of the Pauli matrices 
$\vec{\tau}=\left(\hat\tau_x,\hat\tau_y,\hat\tau_z\right)$ and the identity $I$~\cite{Haroche_book_2006}:
\begin{eqnarray}
\hat\rho=\frac{1}{2}\left( I + \vec{P}\cdot\vec{\tau}\right)
\label{DM1}
\end{eqnarray}
where
\begin{equation}
\vec{P}=\Big( 2 {\rm Re}(C_{\uparrow\downarrow}), 2 {\rm Im}(C_{\uparrow\downarrow}),P_\uparrow-P_\downarrow\Big).
\label{P}
\end{equation}

We can now easily compute  the expectation value of any operator  $\langle \hat A(t) \rangle= {\rm Tr}[\hat A\hat\rho(t)]$  using the well known result for ${\rm Tr}\left[ \hat\tau_a \hat\tau_b\right] =2 \delta_{ab}$, where $a$ and $b$ denote the cartesian components.  We thus
 see that $\langle \hat S_a \rangle= P_a/2$, i.e., $\langle \hat S_z \rangle= (P_{\uparrow}-P_{\downarrow})/2$ and 
 $\langle \hat S_x (t)\rangle=2{\rm Re}\left[ C_{\uparrow\downarrow}(t)\right]$. 
Thus, the longitudinal magnetization is governed by the occupations, whereas the transverse magnetization is governed by the coherences.   

In a free induction decay experiment,  typical of NMR experiments, spins  are driven out of their  equilibrium magnetization  along the direction of a static field  (the $z$ direction),  by means of a  pulse of $ac$ transverse magnetic field.  The  re-establishment of the equilibrium situation is described by the equations: 
\begin{equation}
\langle S_z \rangle(t)- \langle S_z\rangle|_{\rm eq} \propto e^{-t/T_1} 
\end{equation}
\begin{equation}
\langle S_x\pm i S_y \rangle(t) \propto e^{-t/T_2} e^{\pm i \omega_0 t}
\end{equation}
where $\hbar\omega_0$ is the Zeeman  splitting.  These equations serve also to define $T_1$ and $T_2$, as the spin relaxation and spin decoherence time, respectively.  In addition to these two 
timescales, in the case of ensembles it is often convenient to introduce the dephasing time $T_2^*$ associated to the purely coherent dynamics of an ensemble of spins (or two level systems in general),   whose energy splitting $\Delta$   is not exactly the same for all of them.   In an ensemble measurement,  both $T_2$ and $T_2^*$ result in a decay of the  transverse spin signal. However, using spin-echo techniques,  the effect of $T_2^*$ can be reverted~\cite{Suter_Alvarez_rmp_2016}, which permits to measure  $T_2$.

Of course, Eqs. (\ref{DM1}-\ref{P}) are only  meaningful if the basis set used to represent the DM is specified, i.e., a change of basis will lead to a different matrix, yet the physical state would be the same.  The DM can describe either 
a pure state $|\psi\rangle$, in which case  $\hat{\rho}=|\psi\rangle\langle\psi|$,  or mixed states in which our knowledge about the state of the system is partial.  
For a pure state it is trivial to verify that $\hat\rho^2=\hat\rho$, which leads to $P=|\vec P|=1$. In the opposite limit we have the DM with $P=0$, with a 50 percent occupation for each state and no coherences.  
In order to  quantify the purity of a given DM we can use the so called Von Neumann entropy
\begin{equation}
S=-{\rm Tr}\left[\hat\rho \ln\hat\rho\right]= -\sum_{s=\pm1 }\frac{1+s P}{2}\log_2\left(\frac{1+s P}{2}\right).
\label{entropy}
\end{equation}
For a pure state, $P=1$ and  the entropy vanishes $S=0$, while for $P=0$ the entropy is maximized, with $S=1$.

An important result is that the property $P=0$ is preserved under any unitary change of basis. In other words, the expression of the fully decohered density matrix is basis independent.
When coupled to a reservoir for a sufficiently long time, the DM is expected to evolve towards the equilibrium DM regardless of the initial state, 
given by $\hat{\rho}=e^{-\beta {\cal H}_{\rm S}}/Z$, with $Z$ the canonical partition function and $\beta=1/k_BT$. The only exception for this rule arises if there is some symmetry that prevents the system to change the occupations of some energy levels, as discuss below.

\subsubsection{Decoherence as entanglemenet with  the bath \label{Imry}}

We now provide a very simple example of how decoherence can arise in a TLS due to its coupling to a dissipative bath in the simplest case of pure dephasing, i.e.,  when transitions between the eigenstates of the TLS are strictly forbidden, so that the diagonal terms in the density matrix stay constant. 
Here we reproduce an argument given by Stern, Aharonov and Imry~\cite{Stern_Aharonov_pra_1990}  
to show the relation between decoherence and the effect of the system on the wave function of the environment, $|\eta(t)\rangle$.

Let us assume that the system and the environment remains decoupled from each other for $t\le 0$.  The wave function for the system plus environment can be then written as: 
\begin{equation}
|\Psi(t=0)\rangle= \left(\cos \frac{\phi}{2} |C_1\rangle + \sin \frac{\phi}{2} e^{i\xi}|C_2\rangle \right)
\otimes
|\eta(0)\rangle,
\end{equation}
where $\xi$ and $\phi$ are real numbers characterizing
 completely the state of the system, modulo a phase.  
 At $t=0$ the interactions between them are turned on and we assume that the only effect of this interaction is to change the wave function of the reservoir, which acts as a witness of which path, either $C_1$ or $C_2$, is being taken by the system. 
 Let $\hbar \omega_{21}$ be the energy difference between states $|C_1\rangle$ and $|C_2\rangle$.
 Then, the wave function at $t\geq 0$ satisfies:
  \begin{eqnarray}
|\Psi(t)\rangle&=& \cos \frac{\phi}{2} |C_1\rangle \otimes |\eta_1(t)\rangle +\nonumber
\\ &+& \sin \frac{\phi}{2} e^{i\xi} e^{i\omega_{21}  t} |C_2\rangle \otimes |\eta_2(t)\rangle.
\label{psiI}
\end{eqnarray}
We see that as long as $|\eta_1\rangle\neq |\eta_2\rangle$,   state (\ref{psiI})  is no longer a product state, and it is thus said that the reservoir and the system are now entangled. 
The DM operator associated to this state is $ \hat \rho_\Psi(t)\equiv
|\Psi(t)\rangle\langle\Psi(t)|$.  If we define the reduced DM  $\rho(t) ={\rm Tr}_{R}\left[
 \hat{\rho}_{\Psi}(t)\right]$, 
we get that
\begin{equation}
\hat\rho(t) =
\left(\begin{array}{cc}
\cos^2 \frac{\phi}{2} &  c_0(t)  {\cal S}_{12}(t)\\
 c_0^*(t)  {\cal S}_{21}(t)
 &\sin^2\frac{\phi}{2} \end{array}
\right),
\label{rho2l}
\end{equation}
where we have introduced the bare coherence $c_0(t)=\cos\frac{\phi}{2}  \sin\frac{\phi}{2} e^{i\xi} e^{i\omega_{21} t}$ and the time dependent overlap
\begin{equation}
{\cal S}_{ij}={\rm Tr}_{R} \left[ 
|\eta_i(t)\rangle\langle\eta_j(t)|\right]
= \langle \eta_i(t)|\eta_j(t)\rangle.
\label{S}
\end{equation}
Of course, we have   ${\cal S}_{11}={\cal S}_{22}=1$ and  ${\cal S}_{1,2}\leq 1$.
In terms of the polarization vector defined in Eq. (\ref{P}), we get that
\beq
P=\sqrt{ \cos^2\phi+\sin^2\phi \left|{\cal S}_{12}(t)\right|^2} .
\eeq 
Therefore,  the effect of the environment mimics a which-path detector and it leads 
to a depletion of the  coherence between the $|C_1\rangle$ and $|C_2\rangle$ states that is given exactly by the overlap between the alternative reservoir wave functions: the more  sensitive the reservoir is to the state of the system,  the smaller the overlap ${\cal S}_{12}$ and the larger the decoherence.  Complete decoherence occurs when the alternative reservoir wave functions become orthogonal.  This can also be seen from an information perspective.
 Given that  the wave function of any given system  encodes the most complete knowledge that we can afford, decoherence increases the Von Neumann entropy (\ref{entropy}) at a rate 
controlled by the amount of {\it which-path} information is stored in the environment.  This complies with the rule of thumb that quantum systems are able to stick to a linear superposition state as long as nobody is watching.

\subsubsection{Decoherence as phase uncertainty induced  by a stochastic field\label{sto}}

We now discuss a second  simple picture for decoherence,   
also analyzed by Stern and coworkers~\cite{Stern_Aharonov_pra_1990}. Following the notation of Slichter~\cite{slichter2013principles}, we consider a model Hamiltonian for a spin $S=1/2$ interacting with a static field $B_z$ and a random stochastic field $\vec{b}(t)$: 
\begin{equation}
{\cal H}= g\mu_B B_z S_z + g\mu_B  \vec  b(t) \cdot  \vec S.
\end{equation}
The stochastic field is characterized by a null average $\overline{b_a(t)}=0 $ ($a=x,y,z$) and a steady noise: 
\begin{equation}
\overline{b_a(t) b_b(t+\tau)} =   \overline{b_a^2} \delta_{a,b} e^{-t/\tau} \Theta(t),
\label{whiteN}
\end{equation}
where  $\Theta(t)$ is the Heaviside function.
Here the  $\overline{b_a}$ symbol  represents an average over realizations of the stochastic classical field $\vec{b}$, $\overline{b^2}$ represents the amplitude of the noise associated to the magnetic field and $\tau$ is the so called correlation time, that characterizes the pace at which the random field fluctuates.  In the IS system, this noise has units of ${\rm T}^2 / {\rm Hz}$.   This model applies to the spin relaxation caused by the stochastic magnetic field coming from the thermal fluctuations of the current in a conductor~\cite{Jones_Vale_prl_2003}, as we discuss below.

Using the Bloch-Redfield theory discussed below,  the  relaxation time $T_1$ and the decoherence time $T_2$  can be written down as~\cite{slichter2013principles} :
\beqa
\frac{1}{T_1}&=&  \left(\frac{g\mu_B}{\hbar}\right)^2\left[ k_{xx}(\omega_0) + k_{yy}(\omega_0) \right]
\label{T1sto}
\\
%
\frac{1}{T_2}&=&  \frac{1}{2T_1}+\left(\frac{g\mu_B}{\hbar}\right)^2  k_{zz}(0)   ,
\label{T2sto}
\eeqa
where $\hbar \omega_0= g\mu_B B$ and 
\begin{equation}
k_{ab}(\omega)= \frac{1}{2}\int_{-\infty}^{\infty} \overline{b_a(t) b_b(t+\tau)}
 e^{-i\omega t} 
dt 
\label{kdef}
\end{equation}
is the Fourier transform of the correlation function. 
Expression (\ref{T2sto}) helps us to introduce several related concepts frequently used in fields like quantum optics or electron/nuclear spin resonance (ESR/NMR). The single spin decoherence time $T_2$, also called {\em dephasing time} or {\em transversal relaxation time} in the context of ESR/NMR, involves two types of processes, as revealed by Eq. (\ref{T2sto}). First, scattering processes that implies a population transfer between the the two states, which accounts for the $2T_1$ contribution in Eq. (\ref{T2sto}), its maximum value ($T_2\ge 2T_1$). In the context of quantum systems coupled to a reservoir, this is the so called {\em nonadiabatic} contribution~\cite{Cohen_Grynberg_book_1998}. 
Second, scattering processes that do not involve population transfer between the system states, which for our stochastic field, are proportional to $k_{zz}(0)$. They constitute the {\em adiabatic contribution} to decoherence. 
%

Equations (\ref{T1sto}-\ref{T2sto})  
imply that  there is a generic linear  relation between the dissipative rates  and the spectral function of the stochastic field. This result still holds when we consider  spins  coupled to quantum operators of a reservoir, in which case the spin relaxation times are proportional to the dynamic response functions of the reservoir.  Thus, $T_1$ and $T_2$ provide a local probe for the properties of the reservoir, a central notion in magnetic resonance imaging~\cite{Pykett_Newhouse_rad_1982} and in the field of quantum sensors~\cite{Giovannetti_Lloyd_science_2004,Jones_Karlen_science_2009,Tanaka_Knott_prl_2015}.

 Equations (\ref{T1sto}-\ref{T2sto})  also show how  
the stochastic field can produce population transfer between the two states only when the transverse components $\bar{b_x^2}$ or $\bar{b_y^2}$ are finite  (otherwise  $S_z$ is a constant of motion), whose efficiency is proportional to the weight of the noise spectral function at the transition energy, which relates to the conservation of the total energy.     Decoherence occurs when there are population scattering events and, in addition, it can also occur even if $\bar{b_x^ 2}=\bar{b_y^ 2}=0$. This last case is the pure dephasing case, seen above, but from a different perspective:  stochastic fluctuations of the energy splitting are straightforwardly equivalent to a loss of the phase coherence.  Finally,
 Eqs. (\ref{T1sto}-\ref{T2sto}) illustrate that 
  population scattering invariably entails decoherence, with a pure dephasing contribution proportional to $k_{zz}(0)$.

Using expressions (\ref{whiteN}) and (\ref{kdef}), we get that $k_{aa}(\omega)= \overline{b_a^2} \tau_0/(1+\omega^2 \tau_0^2)$, which leads to
\begin{equation}
\frac{1}{T_2}=  \left(\frac{g\mu_B}{\hbar}\right)^2 \left[
\frac{\tau_0}{1+\omega^2 \tau_0^2}
\left[ \overline{b_x ^2} + \overline{b_y ^2} \right]+  \overline{b_z^2}\tau_0\right].
\label{T2sto2}
\end{equation}
In the pure dephasing limit, where $\overline{b_x ^2}=\overline{b_y ^2}=0$, we have $T_2\propto \tau_0^{-1}$, which is the so called motional narrowing.  This result can be understood within the following picture. Let us consider a stochastic field that can assume only 2 values $b_\pm=\pm( \overline{b_z^2})^{1/2 }$. The phase acquired during the time interval $\tau_0$ in which the field stays active is $\delta \phi = \pm \tau_0g\mu_B b_{+}/\hbar$.  After $n$ such intervals,  the spread of the phase, governed by a binomial distribution,    equals  $\overline{\Delta \phi^2}= n (\delta \phi)^2=n\left(\tau_0 g\mu_B b_+/\hbar\right)^2 $. Now, the number of such intervals in a time $t$ is $n=t/\tau_0$.  We estimate $T_2$ as the time it takes for the phase spread to be equal to 1 radian: 
\begin{equation}
\frac{1}{T_2} = \left(\frac{g\mu_B }{\hbar}\right)^2 \tau_0\overline{b_z^2}.
\end{equation}
This result could of course be obtained  from  Eq. (\ref{T2sto2}) in the pure dephasing limit ($\overline{b_x ^2} = \overline{b_y ^2}=0$), 
where spin-flip transitions are forbidden.

\subsection{Bloch-Redfield perturbative approach to the dissipative dynamics \label{BRF}}
In this section we briefly review the Bloch-Redfield (BR) master equation theory to deal with the dissipative dynamics of quantum systems weakly coupled to a reservoir. 
The biggest advantage of this approach is that it can be applied to a great variety of systems, not restricted to individual magnetic atoms,  and permits dealing with situations where both pure dephasing and population scattering are present. On the down side, the Bloch-Redfield treats the system-reservoir coupling to second order in perturbation theory, and it can only provide a description of the dynamics on a coarse-grained
time scale larger than the correlation time of the reservoir $\tau_c$, the surface conduction electrons  or lattice vibrations in our case.  These limitations are not severe,
and this technique have been successfully used to study the spin dynamics of a single or a few magnetic atoms adsorbed on top of a monolayer of insulating material grown on a conducting substrate~\cite{Loth_Bergmann_natphys_2010,Delgado_Palacios_prl_2010,
Oberg_Calvo_natnano_2013,Spinelli_Bryant_natmat_2014,Delgado_Loth_epl_2015},  and also for few-atom clusters on metals~\cite{khajetoorians2013current}.
An excellent introduction to  the formalism can be found in Ref.~\cite{Cohen_Roc_book_1989}, in the context of quantum optics, and also in Ref.~\cite{Breuer_Petruccione_book_2002}.

The starting point of the BR formalism is the Hamiltonian of Eq. (\ref{HTOT}), ${\cal H}= {\cal H}_{\rm S} + {\cal H}_{\rm R}+ {\cal V}$. 
In the case of magnetic adatoms,  ${\cal H}_{\rm S}$  describes the atomic spins and  usually accounts for the local magnetic anisotropy, inter-spin interactions,  and Zeeman interactions. If required, one can also include the nuclear spins  in the system degrees of freedom~\cite{Delgado_Rossier_prl_2011,Delgado_Aguado_apl_2012}, in which case ${\cal H}_{\rm S}$ should include the hyperfine interactions as well.  In the basis of eigenstates of ${\cal H}_{\rm S}$, we can write   ${\cal H}_S=\sum_n E_n|n\rangle \langle n|$. 
 
 The  ${\cal H}_{\rm R}$ term  describes the environment Hamiltonian, which could corresponds to the electronic bath of conduction electrons in the metallic substrate,  a phonon bath,  photons,  or the bath of surrounding nuclear spins.  Finally, the  ${\cal V}$  term represents the interaction between the system of quantum spins and the degrees of freedom of the bath. For most systems, this interaction can be written in the form
\beq
{\cal V}=\sum_\alpha R_\alpha\otimes {\cal S}_\alpha,
\label{intera}
\eeq
where $R_\alpha$ are reservoir operators and ${\cal S}_\alpha$ spin operators, to be determined depending on the nature of the interaction. The composite index $\alpha$ will contain information about all the bath quantum numbers and, if the system is coupled to more than one bath, also the bath label.
  We remark here that the BR tensor contains only second order in ${\cal V}$  corrections to the dynamics of the quantum system due this coupling.
  In addition to condition (\ref{intera}), the BR also assumes a zero-average of the system-bath coupling, i.e., ${\rm Tr}_R\left[\hat \rho_R{\cal V}\right]=0$, where $\hat\rho_R$ is the thermal equilibrium density matrix of the reservoir and the trace is over the reservoir degrees of freedom. Notice that in cases where this last condition is not satisfied, one can always reinsert this trace into a renormalized system Hamiltonian and interaction such that the new problem has a zero average interaction.\footnote{In the case of magnetic systems coupled to an electronic bath, this
  zero-average condition implies that, whenever the bath is spin polarized, one should proceed to remove the average ${\rm Tr}_R\left[\hat \rho_R{\cal V}\right]$~\cite{Cohen_Grynberg_book_1998}. }

 If we introduce the basis of eigenstates of the system Hamiltonian, $| n\rangle$,  the  markovian evolution of the reduced DM to second order in ${\cal V}$  can be written as: 
\beq
\label{BR}
\partial_t \rho_{nm}(t)=-i\omega_{nm}\rho_{nm}(t)+ \sideset{}{'}
\sum_{nn'}{\cal R}_{nm,n'm'} \rho_{n'm'}(t),
\eeq
where $\omega_{nm}=(E_n-E_{m})/\hbar$  and ${\cal R}_{nm,n'm'}$ is the Redfield tensor. 
From the hermiticity of the density matrix, one gets ${\cal R}_{mn,m'n'}^* ={\cal R}_{nm,n'm'}$. 
 The prime over the sum in Eq. (\ref{BR}) implies that only the terms whose energies satisfy $|\omega_{nm}-\omega_{n'm'}|\ll 1/\delta t$ are included, where $\delta t=\hbar/(k_B T)$ is the coarse-grain time scale mentioned above. This is called the {\em secular approximation}~\cite{Cohen_Grynberg_book_1998}.
 
 Thus, the BR tensor  depends both on the matrix elements of the system operators,  $ S_{\alpha}^{mn} \equiv \langle m|S_{\alpha}|n\rangle$,
 as well as on the reservoir operator correlator, $g_{\alpha \beta } (t)\equiv \langle R_{\alpha}(t) R_{\beta}(0) \rangle|_{\rm eq}$
where the brackets stand for statistical average over the reservoir equilibrium density matrix and
 $R_{\alpha}(t)= e^{i{\cal H}_R t/\hbar}  R_{\alpha}(0) e^{-i{\cal H}_R t/\hbar}$.  We provide here  general expressions, decomposing
 the tensor entries  in a sum of two terms, ${\cal R}_{nn'mm'}={\cal R}^+_{nn'mm'}+{\cal R}^-_{nn'mm'}$, where~\cite{Delgado_Rossier_inprep16} ,
\begin{eqnarray}
{\cal R}^+_{nn'mm'}&=&
\sum_{\alpha,\beta}
\frac{1}{\hbar^2}
\left( 
g_{\alpha\beta}(\omega_{nm}) S_{\alpha}^{m'n'}    S_{\beta}^{nm}-
\delta_{m'n'}\sum_{n''} g_{\alpha\beta}(\omega_{n''m})S_{\alpha}^{nn''}  S_{\beta}^{n'm}\right)
\label{Rplus}
\end{eqnarray}
and
\begin{eqnarray}
{\cal R}^{-}_{nn'mm'}&=&
\sum_{\alpha,\beta}
\frac{1}{\hbar^2}
\left( g_{\beta\alpha}^*(\omega_{n'm'})S_{\alpha}^{m'n'}   S_{\beta}^{nm}
- 
\delta_{nm}\sum_{n''} g_{\beta\alpha}^*(\omega_{{n''}m'})S_{\alpha}^{m'n''}  S_{\beta}^{n''n'}
\right) 
\label{Rminus}
\end{eqnarray}
where 
$g_{\alpha\beta}(\omega)\equiv \int_0^{\infty} dt e^{-i\omega t} g_{\alpha\beta}(t)$.

The BR equation (\ref{BR}) describes the evolution of both the occupations $\rho_{nn}$ and coherences $\rho_{nn'}$ ($n\ne n'$).  Although in  general their dynamics are coupled, in some instances   
it is possible to write down an equation for the occupations, $P_n\equiv \rho_{nn}$ and the coherences separately.  The resulting equation for the occupations is the so called  Pauli master equation: 
\begin{equation}
\frac{dP_n}{dt}= \sum_{n'} \Gamma_{n'n} P_{n'}-\left(\sum_{n'}\Gamma_{nn'}\right)P_n,
\end{equation}
where  $\Gamma_{nn'}$ stands for the the scattering rates from state $n$ to $n'$.  This equation has a transparent physical interpretation:  coupling to the reservoir results in scattering events that transfer weight from some states to others. When the reservoir is at thermal equilibrium, the steady occupations $P_n$ are given by the Boltzmann distribution. 

The scattering rates $\Gamma_{nm}\equiv 1/T_1$ can be written down in terms of the Redfield coefficients~\cite{Breuer_Petruccione_book_2002}
\beqa
\label{ratesBRF0b}
\Gamma_{nm}=
\frac{2}{\hbar^2}\sum_{\alpha,\beta}  {\rm Re}\left(g_{\alpha,\beta}(\omega_{mn})\right)
 S_{\alpha}^{nm}  S_{\beta}^{mn}.
\eeqa
This general formula accounts for instance, for the $T_1$ obtained for a spin $S=1/2$ system coupled to a stochastic magnetic field, Eq. (\ref{T1sto}), when we identify the indexes $\alpha,\beta$ with the cartesian coordinates and the reservoir correlator  $g_{\alpha,\beta}$ with the magnetic field noise function $k_{\alpha,\beta}$.

In the case of the Kondo interactions discussed in this review, the quantum operator $S_{\alpha}$ is an atomic spin operator while the reservoir operator $R_{\alpha} \equiv \vec s(\vec r_l)$ 
is  the fermionic  spin density, and  their coupling is the isotropic spin  interaction $ \sum_{l,a} J_l  S_a(l) s_a(\vec r_l)$. 
In this case, the  reservoir correlator $g_{\alpha,\beta}$   is related to the spin susceptibility of the bath,  due to the fluctuation dissipation theorem.

Equation (\ref{BR}) also describes the evolution of coherences in the basis of eigenstates of the system Hamiltonian.  In the simplest case when the transition energy between a given pair of states is different from that of every other pair of eigenstates,  which automatically occurs for a two level system, the equation of motion for the coherence is independent from the rest: 
\beq
\partial_t \rho_{nm}(t)=-i\left(\omega_{nm}+\delta\Delta_{nm}\right) \rho_{nm}(t)- \gamma_{nm} \rho_{nm}(t),
\label{BR2}
\eeq
where we have split the RB tensor into its real and imaginary parts
${\cal R}_{nm,nm}=-\gamma_{nm}-i\delta\Delta_{nm}$.
Equation (\ref{BR2}) has clear physical interpretation. The coupling to the reservoir produces two effects on the dynamics of the coherence.  First, it renormalizes the energy difference between the energy levels.
Second, and central for the purpose of this review,  the coupling to the reservoir results in a damping term, that results in a decay rate of the coherence given by
\begin{equation}
\frac{1}{T_2}= \gamma_{nm}.
\end{equation}
Importantly, the decoherence rate has two types of contributions,  the  {\em adiabatic} and {\em nonadiabatic}~\cite{Cohen_Grynberg_book_1998}.
The later involve transitions between the levels $n$ and $m$ governed by the same type of rates and microscopic processes than $T_1$:
\beq
\gamma_{nm}^{\rm nonad.}=\frac{1}{2}\left( \sum_{n'\ne n}\Gamma_{n,n'}+\sum_{n'\ne m}\Gamma_{m,n'}\right).
\label{gnonad} 
\eeq
Their interpretation is transparent: phase coherence is lost when population scattering occurs.  
      The adiabatic contribution to decoherence is more subtle, as it can happen even in the absence of population scattering, and it connects directly with the ideas discussed in Secs. \ref{Imry} and \ref{sto}.   The adiabatic decoherence takes the form: 
\begin{eqnarray}
 \gamma_{nm}^{\rm ad}
 =
\frac{1}{\hbar^2}
\sum_{\alpha,\beta}
{\rm Re}( g_{\alpha\beta}(0))
\left( 
S_{\beta}^{mm}-
S_{\beta}^{nn}
\right)
\left(
S_{\alpha}^{mm} -  S_{\alpha}^{nn}
\right).
\label{adiabaticbis4}
\end{eqnarray}
The physical interpretation of Eq. (\ref{adiabaticbis4}) is the following. If we have two states, $n$ and $m$, such that the expectation value of a given operator is different, and the environment is sensitive to which of the two states the system stays at,  even if it is not able to induce scattering between them,  decoherence will occur. To be more specific, let say $n=+S$ and $m=-S$ for a quantum spin.   A magnetic coupling to the environment will create decoherence between $n$ and $m$ even if there is no population transfer.  The adiabatic rate of such process would be:
$ \gamma_{S_z,-S_z}^{\rm ad}
 \simeq 4S^2
{\rm Re}( g_{zz}(0))/\hbar^2$, which again connects with the result (\ref{T2sto2}).   

As illustrated by Eq. (\ref{BR2}), the coupling to the bath also produces a frequency shift  $\delta\Delta_{nm} $ of the original energy levels that can be decomposed as $\delta\Delta_{nm}=(\omega_n- \omega_m)$. Using the general expressions ( \ref{Rplus}) and (\ref{Rminus}), this frequency shift can be written as 
\beqa
\omega_m=
\frac{1}{\hbar^2}
\sum_{\alpha,\beta}
\sum_{n' }  
{\rm Im}\left(g_{\alpha,\beta}(\omega_{n'm}\right)
 {\bf S}^{\alpha}_{mn'} {\bf S}^{\beta}_{n'm} .
 \label{shiftE}
\eeqa
 A well known example of this renormalization is the Lamb shift of the Hydrogen spectrum.  In the case of magnetic adatoms, the associated variation of the magnetic anisotropy due to the Kondo exchange coupling with the substrate itinerant electrons has been recently reported~\cite{Oberg_Calvo_natnano_2013}. Here we should remark that result (\ref{shiftE}) is completely general, i.e., the only BR tensor elements contributing to the energy shifts are those of the form $R_{nm,nm}$~\cite{Breuer_Petruccione_book_2002}.

In summary, the BR theory accounts for the dissipative dynamics of the reduced density matrix of a quantum system weakly coupled to a reservoir.  The influence of the reservoir is included up to second order in the system-reservoir interaction, and the correlation functions of the reservoir are assumed to have a very short memory, or in a more technical jargon,  we adopt the Markovian approximation.  The BR theory accounts for 3 types of effects:
 \begin{enumerate}
\item The  occupations of the quantum states (diagonal part of the density matrix) evolve in time, experiencing transitions between the originally decoupled eigenstates of the isolated system.
\item Decoherence.  When the reservoir is in thermal equilibrium, the steady state solution of the BR is the equilibrium density matrix, which is diagonal in the basis of eigenstates of the isolated system ${\cal H}_S$. Therefore, coherences in this basis are fragile and decay in time.\footnote{The reduced density matrix can deviate from the thermal equilibrium distribution when there are some symmetries that prevents transitions between eigenstates of ${\cal H}_S$. Moreover, in the case of a degenerate spectrum it is always possible to choose a basis in which coherences does not decay with time.} 
\item Renormalization of the energy levels.  
In the case of nanomagnets, the shift of the energy levels translates into renormalized magnetic anisotropies and they can play an important role. 
For instance, the quantum spin tunneling splitting of an isolated spin, which protects quantum coherence,  can be quenched due to coupling to the environment.  
\end{enumerate}

\subsection{Bloch equation for a 2-level system\label{2levelD} }
Most of the experiments measurements of the coherence time of a system involve driving the system with an {\em ac} signal.  Therefore,  it is important to have a  dissipative theory for the density matrix where the dynamical driving term is included.   Here we review the Bloch approach following Cohen-Tannoudji {\em et al.}~\cite{Cohen_Grynberg_book_1998}.
The right hand side of the BR equation (\ref{BR}) contains two terms. The first is the commutator of the reduced density matrix with the  Hamiltonian ${\cal H}_S$ of the system, whereas the second accounts for the dissipative coupling to the environment.  This derivation was done assuming that ${\cal H}_S$ is time-independent.  

The derivation of Bloch equation for the dissipative dynamics of a driven two-level system consists, basically, in replacing ${\cal H}_S$ by ${\cal H}_S+{\cal V}_1(t)$ in the first term of the right hand side of Eq. (\ref{BR}), where ${\cal V}_1(t)$ is the driving term.
  This approach is definitely justified  in two limits. In the absence of dissipation,  the equation of motion of the density matrix is given by the commutator with the time-dependent Hamiltonian. Second,  in the absence of the driving term, we recover  the BR theory.  Basically, this approximation amounts to assume that the coherent and incoherent dynamics are additive. 
  
  For an energy-split TLS, with energies $E_a$ and $E_b$, the resulting equation of motion for the density matrix reads as:
\beeq
  \dot{\rho}_{aa}(t)=-\dot{\rho}_{bb}(t)
  = \frac{1}{2T_1} 
  \left( 
  \rho_{bb}(t)-\rho_{aa}(t)-(\rho_{bb}^{\rm eq}-\rho_{aa}^{\rm eq})
  \right) + \frac{i}{\hbar}\left(\rho_{ab} {\cal V}_1^{ba}(t)-\rho_{ba} {\cal V}_1^{ab}(t)\right)
  \label{B1}
  \eeq
and
 \beeq
\dot{ \rho}_{ab}(t)=\dot{ \rho}_{ba}^*(t)=
-\frac{\rho_{ab}(t)}{T_2}
+\frac{i}{\hbar}\Delta\rho_{ab}(t) -  \frac{i}{\hbar}
  \left( 
  \rho_{aa}(t)-\rho_{bb}(t)\right){\cal V}_1^{ab}(t),
  \label{B2}
\eeq
where ${\cal V}_1^{ba}(t)= \langle b|{\cal V}_1(t)|a\rangle$ and $\Delta=E_b-E_a\ge 0$.  Equations (\ref{B1}) and (\ref{B2}) are the so called Bloch equations.  Here we assume that the time-dependent driving field has the periodic form
${\cal V}_1^{ba}(t)=\hbar \Omega \cos\omega t$, where $\Omega$ is the {\em Rabi frequency}.

The Bloch equations in the  context of electron/nuclear spin resonance are usually written in a slightly different way.
In order to transform these equations into the most common form, we first
introduce the effective average spin operators
\beqa
\langle {\cal S}_+(t)\rangle &=& e^{-i\omega t} \rho_{ba}(t)    
\crcr
\langle {\cal S}_-(t)\rangle &=& e^{i\omega t}\rho_{ab}(t)       
\crcr
\langle {\cal S}_z(t)\rangle &=&\frac{1}{2}\left(\rho_{aa}(t)-\rho_{bb}(t)   \right),
\label{defSa}
\eeqa
and, as for the usual angular momentum operators, the {\em x} and {\em y}-components $\langle {\cal S}_x\rangle= (\left\langle {\cal S}_+\rangle+\langle {\cal S}_-\rangle\right)/2$ and 
$\langle {\cal S}_y\rangle= \left(\langle {\cal S}_+\rangle-\langle {\cal S}_-\rangle\right)/(2i)$.
With the definition (\ref{defSa}), one gets that the transversal terms $\langle {\cal S}_{x,y}\rangle$ will contain two types of oscillations, provided $|\hbar\omega-\Delta|/\Delta \ll 1$: fast oscillating terms, of the form $e^{\pm i(\omega+\Delta/\hbar)t}$, and slow oscillations of the form $e^{\pm i(\omega-\Delta/\hbar)t}$. 
Under the rotating-wave approximation described in Ref.~\cite{Cohen_Grynberg_book_1998}, the fast rotating terms 
are neglected. This is equivalent to  describe the evolution of this effective magnetic moment in a rotating frame such that the magnetization vector $\vec M(t)\equiv (\langle {\cal S}_x(t)\rangle,\langle {\cal S}_y(t)\rangle,\langle {\cal S}_z(t)\rangle)$  precesses at the effective Larmor frequency $\Delta/\hbar$: 
\beqa
\left\{
\begin{array}{l}
\langle \dot{{\cal S}_x}\rangle = \delta
 \langle {\cal S}_y\rangle
-\frac{1}{T_2}\langle {\cal S}_x\rangle,
\\
\langle \dot{{\cal S}_y}\rangle = -\delta
\langle {\cal S}_x\rangle
+\Omega \langle {\cal S}_z\rangle
-\frac{1}{T_2}\langle {\cal S}_y\rangle,
\\
\langle \dot{{\cal S}_z}\rangle = 
-\Omega \langle {\cal S}_y\rangle-\frac{1}{T_1}\left(\langle {\cal S}_z\rangle-\langle {\cal S}_z\rangle_0 \right)
\end{array}
\right.
\label{BRF_BE}
\eeqa
where $\delta=\Delta/\hbar-\omega$ is the detuning between the TLS splitting and the driving frequency and
$\langle {\cal S}_z\rangle_0 $ is defined as in Eq. (\ref{defSa}) with the initial occupations.
Notice that to avoid unnecessary complications in the notation, we have just dropped the argument $(t)$ to indicates that these quantities are given in the {\em rotating frame}.

Equations (\ref{BRF_BE}) are formally identical to the phenomenological Bloch Equations describing the dynamics of a macroscopic magnetic moment $\vec M(t)$. Thus, these equations can be used to describe the dynamics of a localized quantum spin (or an array of localized spins behaving as a TLS) driven by a classical radiofrequency field of frequency $\omega$ and Rabi frequency $\Omega$ coupled to a bath, which induces relaxation on a time scale $T_1$ and decoherence on $T_2$.

The system of differential equations (\ref{BRF_BE}) admits a steady state solution of the form (see Appendix~\ref{AppBE} for the details): 
\begin{equation}
\frac{\langle{\cal S}_z\rangle-\langle{\cal S}_z\rangle_0}{\langle{\cal S}_z\rangle_0}=
-\frac{T_1T_2\Omega^2}{1+T_2^2\delta^2+T_1 T_2 \Omega^2}
\label{SzB}
\end{equation}

\begin{equation}
\frac{\langle{\cal S}_x\rangle}{\langle{\cal S}_z\rangle_0}=-T_2\delta \frac{\langle{\cal S}_y\rangle  }{\langle{\cal S}_z\rangle_0}=
\frac{T_2^2\delta \Omega}{1+T_2^2\delta^2+T_1 T_2 \Omega^2}.
\label{SxB}
\end{equation}

\subsection{Decoherence as a limit for spectral resolution in magnetic resonance }
Equation (\ref{SzB}) provides the basis for magnetic resonance experiments. The deviation from the equilibrium of the longitudinal  magnetization can be controlled changing the detuning $\delta$ between the precession frequency $\Delta/\hbar$ and the frequency $\omega$ of the driving field.    
When sweeping over $\delta$,  Eq. (\ref{SzB}) describes a resonance curve whose maximum is given by
\begin{equation}
\left(
\frac{T_1T_2\Omega^2}{1+T_1 T_2 \Omega^2}\right).
\label{SzBmax}
\end{equation}
Notice that  the maximal deviation scales linearly with  $\Omega^2$ when the dimensionless  parameter $ x=T_1 T_2 \Omega^2$ is small.  The full width at half maximum is then given by:
\begin{equation}
\delta_{\rm FWHM}^2= \frac{1}{T_2^2}\left( 1+ T_1T_2\Omega^2\right).
\end{equation}
In the limit when the  driving amplitude $\Omega$ is small enough, or more precisely, when $x\ll 1$,   the width of the resonance is  $1/T_2$.    In the opposite limit,  the full width at half maximum is given by
 $\Omega\sqrt{T_1/T_2}$.   Thus, in both cases increasing $T_2$ results in a reduction of the FWHM of the resonance curve. 
 As we discuss in Sec. 6, experiments of spin resonance on an individual magnetic atom using STM have been recently reported~\cite{Baumann_Paul_science_2015,natterer2016}, 
demonstrating the possibilities of single spin as a magnetometer.  The accuracy of this quantum sensor is  determined by the FWMH and thereby, is ultimately limited by the decoherence rate.

\subsection{The quantum to classical transition and the spin-boson model}
The BR theory shows how a quantum state that is a linear superposition of two eigenstates with different energies 
decays with a characteristic time scale $T_2$ due to the interaction with the environment.
We now discuss   another important aspect of decoherence. Take a quantum system whose ground state is a linear combination of two eigenstates of an operator $\hat{A}$.  For instance, $\hat{A}$ could be a projection of the spin operator, $\hat S_z$, or a pseudospin operator $\hat \tau_z$  in a two level system, such a diatomic molecule or a double quantum well.    Quantum mechanically, the eigenstates of the Hamiltonian can perfectly be linear superpositions of states with different eigenstates of the operator $\hat{A}$.   
 To be more specific, let us examine  a relevant and simple example provided by an anisotropic integer spin system, governed by the Hamiltonian
\begin{equation}
{\cal H}= D S_z^2 + E(S_x^2-S_y^2),
\label{Hde}
\end{equation}
with a strong easy axis anisotropy ($-D\gg E>0$).
We discuss the simplest non-trivial case, $S=1$, which describes, for instance,  iron phthalocyanine deposited on an oxidized copper surface~\cite{Tsukahara_Noto_prl_2009}. 
Thus, the ground state doublet (for $E=0$) is formed by the eigenstates of $\hat S_z$ with  $S_z=\pm1$. As the in-plane anisotropy term is turned on, the degeneracy of the doublet is lifted,  resulting in the so called quantum spin tunneling splitting~\cite{Gatteschi_Sessoli_book_2006}  $\Delta_{\rm QST}=E_2-E_1$. Importantly, the resulting eigenvectors are linear combinations of the states with $S_z=\pm 1$: 
\begin{eqnarray}
|Q_1\rangle &=&\frac{1}{\sqrt{2}}\left( |+1\rangle + |-1\rangle \right)
 \nonumber\\
|Q_2\rangle  &=&\frac{1}{\sqrt{2}}\left(|+1\rangle - |-1\rangle \right).
\label{Qbasis0}
\end{eqnarray}
Hence, the eigenstates of the  Hamiltonian have a built-in coherence between the states with well defined $S_z$.  
In the weak-coupling BR theory,  when dissipation is included,  the steady state density matrix for this system would be given by $\hat\rho=\sum_n P_n^ {\rm Eq.}  |Q_n\rangle \langle Q_n|$, where $P_n^ {\rm Eq.}$ are the Boltzmann factors. When $k_BT \ll |D|$,  the energy gap that separates the  $S_z=\pm 1$ doublet from the $S_z=0$ state,  this equilibrium density matrix can be expressed in the $\left\{|+1\rangle,|-1\rangle\right\}$ basis as 
\begin{eqnarray}
\rho_{S_z}= \frac{1}{2}\mathbb{1}  + \frac{1}{2}{\rm Tanh}\left(\frac{\Delta_{\rm QST}}{2k_BT}\right)\left(\begin{array}{cc}  0& 1\\ 1& 0 \end{array}\right).
\label{rhoSz}
\end{eqnarray}
Thus, in thermal equilibrium the DM conserves a finite coherence of order ${\rm Tanh}(\Delta_{\rm QST}/2k_BT)$ when expressed in the basis set of eigenstates of $\hat S_z$. Interestingly, the density matrix (\ref{rhoSz}) also describes 
integer spins characterized by Hamiltonian (\ref{Hde}) with arbitrarily large $S$.  This implies having coherence between states with opposite (and arbitrarily large) magnetic moment, definitely at odds with classical systems.  
Of course, the QST splitting
rapidly decreases as $S$ increases, reducing the temperature range for which  quantum coherence  
is predicted. In particular, for the Hamiltonian (\ref{Hde}), it is basically given by $\Delta_{\rm QST}\propto E\left(E/D\right)^{S-1}$. Thus, if we take for instance the Fe$_{8}$ molecular magnet, where $S=10$, $D=-0.295$ K and $E/|D|\approx 0.19$, a $\Delta_{QST}\sim 7$ nK has been predicted~\cite{Barra_Laure_cej_2000}, a temperature very difficult to reach experimentally.
Hence, in the weak coupling approach, the coupling to the reservoir  seems to be ineffective to destroy coherence in the $S_z$ basis, as long as $\Delta_{\rm QST}/2k_BT$ is not too small.


Given the perturbative nature of the result (\ref{rhoSz}), it seems pertinent to  ask ourself what happens when the strength of the coupling to the environment increases and, in particular, if there another mechanism quenching the coherences not accounted for in (\ref{rhoSz}).
For that matter, we consider the coupling of a TLS,  described by  ${\cal H}=\frac{\Delta_{\rm 0}}{2}\hat\tau_x$
in the $S_z$-basis, to a bosonic environment that is sensitive to the value of $S_z$:
\begin{equation}
{\cal H}_{\rm SB}=\frac{\Delta_{\rm 0}}{2}\hat\tau_x+  
 \hat\tau_z\sqrt{\alpha}
\sum_{0<k<k_c} g_k
\left(b_k^\dag+b_k\right)
+ \sum_{0<k<k_c} \hbar v_F k b_k^{\dag} b_k,
\label{SBQ}
\end{equation}
where $v_F$ is the Fermi velocity and $g_k=\hbar v_F\left(\pi k/L\right)^{1/2}$ ($L$ is a quantization box length). 
This is the well known spin-boson model, proposed to study the competition between the quantum tunnelling, driven by the $\frac{\Delta_{\rm 0}}{2}\hat \tau_x$ term, and the coupling to the reservoir, that is sensitive which state $\hat \tau_z$ the system is at. 
For $\Delta_{\rm0 }=0$ the model can be solved exactly~\cite{Leggett_Chakravarty_rmphys_1987}, and it predicts a vanishing coherence even at $T=0$, in full agreement with the argument of Sec.~\ref{Imry}. 
 For a finite value of $\Delta_0$, the spin-boson model can be solved exactly in some specific limits. 
A central result of this model is that, due to the coupling to the environment,  the QST splitting becomes renormalized, according to the following equation~\cite{Leggett_Chakravarty_rmphys_1987}:
\beqa
\frac{\Delta}{\Delta_0}\approx\left(1-\Theta (\alpha)\right)
		       \left(\frac{\Delta_0}{\hbar \ome_c}\right)^{\frac{\alpha}{1-\alpha}},
\label{deltaSB}		       
\eeqa
%
where $\Theta$ is the step function,  and $\ome_c$ is a cut-off frequency. 
 For $\alpha>1$ the QST splitting vanishes strictly, which   removes the protection of coherence discussed above.  For $\alpha<1$ the QST is severely renormalized except for very small $\alpha$.   Interestingly, at weak coupling ($\alpha\ll 1$) the renormalization of the  QST can also be captured by the BR theory~\cite{Delgado_Hirjibehedin_sc_2014}.

The results of the spin-boson model acquires special relevance in the context of nanomagnets  coupled to the itinerant electrons through a Kondo exchange interaction, a system that in many instances can be mapped to the spin-boson model described above~\cite{Delgado_Loth_epl_2015,Prokoev_Stamp_rpp_2000}.
 In any event, the coupling to the environment destroys the spin coherence between states with opposite $S_z$ by renormalizing the QST splitting.

\subsection{Other approaches to the relaxation and decoherence of spins\label{OMethods}}
In this section we have presented a general overview of the decoherence problem from both a formal point of view, where decoherence and relaxation is seen basically as the effect of the coupling to the environment, an a practical point of view, with two methodologies to analyze it. In addition, in Sec.~\ref{KondoGen} we shall further particularized to the problem of Kondo induced decoherence in spin systems.  
However, there are many other possible approaches to treat the problem of spin decoherence.
Within the linear response theory, one can study the dynamical effects by looking at the behavior of the dynamical susceptibility. By using the Kubo formula~\cite{Mahan_book_1990}, one obtains the (complex) frequency dependent magnetic susceptibility, which can be related to the retarded correlation function. 
The first derivative of its imaginary part with respect to the frequency 
is related to dissipation (relaxation and decoherence), while a first derivative of 
the real part corresponds to the shifts of the energy levels, the effects already derived from the Bloch-Redfield theory \ref{BRF}.  First frequency derivative of the dynamic 
susceptibility has a frequency-dependence similar to the susceptibility and can be written as the product of two Green functions, making a separation of the shifts/dissipation possible for small couplings.

In addition to the particular treatment of the dissipative dynamics, there is a second implicit assumption in our whole treatment: the quantum system can be treated as a spin system, i.e, fluctuations of other degrees of freedom, such as charge, can be neglected. This does not need to be the case 
 in the strong fluctuation regime where the local moments are fully suppressed or in the intermediate Hund's impurity regime~\cite{Khajetoorians_Valentyuk_natnano_2015}, where charge fluctuations occurs as a result of the strong hybridization, but the local magnetic moment still survives. Furthermore, when the structural changes of the spin array modifies the local magnetic anisotropy of each spin due to the surface rearrangements~\cite{Ruiz_Stepanyuk_jpc_2015}, a more complete description including the orbital degrees of freedom may be required, as it happens in the case of Co chains on Cu$_2$N/Cu(100) surface~\cite{Bryant_Toskovic_nanol_2015}. Another examples where the spin description may fail are magnetic molecules on surfaces where conformational changes may occur either induced by charging effects or by the field of the tip~\cite{Heinrich_Ahmadi_nanol_2013,Heinrich_Braun_nanol_2015}. This is likely to occur when transport takes place close to a resonance, i.e, the addition (or removal) energy of an electron of the molecule is close to the chemical potential of the electrodes~\cite{Chen_Fu_prl_2008,Fu_Zhang_prl_2009,Delgado_Rossier_prb_2011}.

In these cases, a more complete description of the electronic transport is required, taking into account not only the spin degrees of freedom but also the charge. One possible treatment of this problem is through a generalized multiorbital Anderson model. Under this description, the correlation between electrons in the magnetic impurities are fully taken into account. For instance, in the case of transition metal adatoms, the 
electrons in the $d$-levels of the adatoms are treated as a many body system, with the effects of the surroundings already accounted for in the crystal and ligand fields~\cite{Ferron_Delgado_njp_2014,Ferron_Lado_prb_2015}. Itinerant electrons can thus   
 hop in and out of these orbitals, and their dynamical effects may be treated within the Green function formalism~\cite{Haug_Jauho_book_1996}. Within this formalism, the effect of the electronic reservoirs is summarized in two central quantities, the  on-site energy levels of the $d$-electrons, and the hybridization function. In fact, these two quantities can be estimated from ab-initio calculations~\cite{Korytar_Lorente_jpc_2011,Dang_Santos_prb_2016}. Of course, depending on the regime of parameters, one can analyze the properties of this multiorbital Anderson problem using different approaches: from the perturbation method in Ref.~\cite{Delgado_Rossier_prb_2011} to more refined numerical techniques like the non-crossing~\cite{Korytar_Lorente_jpc_2011} and one-crossing approximations~\cite{Oberg_Calvo_natnano_2013}, or numerical renormalization group analysis~\cite{Zitko_Pruschke_njphys_2010} among others.

\section{Spin Hamiltonian for magnetic adatoms}
 In this section we review the Hamiltonian describing  magnetic atoms on a surface, including both the part that, within the open quantum system picture,  we consider the ``system", as well as its coupling to other degrees of freedom, the ``baths". This includes single spin Hamiltonian and the spin-spin interactions, relevant for   engineered nanostructures such as spin chains.   

\subsection{Single spin Hamiltonian}
The low energy physics of individual quantum spins, such as magnetic atoms and molecules, can be very often described with an effective single spin Hamiltonian that describes the  magnetic anisotropy and the Zeeman coupling within the subspace of the ground state spin $S$.   The ground state multiplet of  open-shell isolated atoms has, ignoring spin-orbit coupling (in the range of $50$~meV for  $3d$ transition metals),  $(2L+1)(2S+1)$ states.  When the coupling to the surface is included, most often the orbital momentum is quenched, so that the low energy manifold has only $2S+1$ levels. 
 In that case, the single ion  Hamiltonian can be written as an even  function of the spin operators $\hat S_a$  that describes the magnetic anisotropy preserving time-reversal symmetry, plus the Zeeman interaction with the external applied field.
Quite often, this effective spin Hamiltonian is written in terms of a complete set of high-order spin operators 
\beqa
{\cal H}_{\rm S}=\sum_{k=2,4,6}\sum_{q=-k}^{k} B_k^q\hat O_k^q\left(S\right),
\label{Hani}
\eeqa
where $\hat O_k^q\left(S\right)$ are the (tesseral tensor) Stevens operators~\cite{Abragam_Bleaney_book_1970} and $B_k^ q$ are real coefficients. 
The Hamiltonian coefficients $B_k^q$ 
are determined by the symmetry of the
the spin system,  the crystal field,  hybridization and, importantly, the  spin-orbit interaction.  
Several groups have tried to  determine the anisotropy  parameters from first principles calculations, although in general this is a hard problem.   
This technique has been applied to transition metal atoms  adatoms~\cite{Zitko_Pruschke_njphys_2010,Lin_Jones_prb_2011,Etzkorn_Hirjibehedin_prb_2015,Ferron_Lado_prb_2015} or even rare earth like Ho on Pt(111)~\cite{Miyamachi_Schuh_nature_2013}. 	
 Multiplet calculations with rescaled Coulomb integrals where the atomic spin-orbit interaction is used~\cite{Cowan_book_1981,Laan_Thole_prb_1991}
have also been used to compute $B_k^ q$. This method requires to calculate spatial averages $\langle r^ n\rangle$ over the atomic wavefunctions, but most frequently they are taken as fitting parameters~\cite{Baumann_Paul_science_2015,Baumann_Donati_prl_2015,Donati_Rusponi_science_2016}.

A second approach to determine the single spin anisotropy coefficients consist on fitting the IETS spectra~\cite{Hirjibehedin_Lin_Science_2007,Otte_Ternes_natphys_2008,Oberg_Calvo_natnano_2013,
Bryant_Spinelli_prl_2013,Yan_Choi_nanolett_2015}
For this matter,  it is important to account for the adatom crystal field symmetry.
For instance, for magnetic adatoms, the dominant (quadratic) uniaxial term is usually written as  $D\hat S_z^2$ ($D=3B_2^0$), 
while the quadratic in-plain anisotropy, typical of adsorption sites with symmetry $C_{2v}$, is written as $E\left(\hat S_x^2-\hat S_y^2\right)$, as in Eq. (\ref{Hde}), where $E=B_2^2$. This is the case of the transition metal atoms studied on  Cu$_2$N/Cu(100)~\cite{Choi_Gupta_jp_2014},  MgO/Ag(100)~\cite{Rau_Baumann_science_2014,Baumann_Donati_prl_2015,Baumann_Paul_science_2015},   or h-BN/Rh(111)~\cite{Jacobson_Herden_natcom_2015}. Whereas in some cases the $z$-axis is the off-plane direction, in others the $z$-axis may lie in the surface plane~\cite{Hirjibehedin_Lin_Science_2007}. 
Another example of interest in the field of magnetic atoms on surfaces is the adsorption on sites with $C_{3v}$ symmetry, such as the Pt(111) surface~\cite{Steinbrecher_Sonntag_natcom_2016,Miyamachi_Schuh_nature_2013}, where the lowest order transversal term takes the form $B_4^3\hat O_4^3=B_4^3/2\left\{S_z,S_+^3+S_-^3\right\}$,
 i.e., it mixes states with spin projection $S_z$ differing in three units~\cite{Miyamachi_Schuh_nature_2013}. Higher order term may also appear in molecular magnets due to their high intrinsic spins~\cite{Gatteschi_Sessoli_anchemint_2003}.

In some high symmetry  instances orbital momentum is not fully quenched. This is the case of Fe on top of MgO.
 In this unquenched case, the low energy levels are described by~\cite{Baumann_Paul_science_2015,Baumann_Donati_prl_2015}:
\begin{equation}
{\cal H}_{\rm S}= D \hat L_z^2 + F_0 (\hat L_x^4+\hat L_y^4) +  \lambda\vec{L}\cdot\vec{S} +
 \mu_B\vec{B}\cdot\left(\vec{L}+g\vec{S}\right)
\end{equation}
where $L=2$ and $S=2$.  The most important consequence of this finite orbital momentum is that the spin-orbit coupling 
induces a splitting of the energy levels linear in $\lambda$, resulting in an enhanced magnetic anisotropy.
Thus, for the Fe/MgO, the low energy spectrum of this effective Hamiltonian is a two level system with approximate quantum numbers $S_z\approx 2L_z\approx \pm 2$. 
This doublet lies 14 meV below the next excited state, so that for many practical purposes, the system behaves like  a  two level system.  

\subsubsection{Integer vs half integer spins\label{intvshi}}

\begin{figure}[t]
  \begin{center}
       \includegraphics[width=1.\linewidth,angle=0]{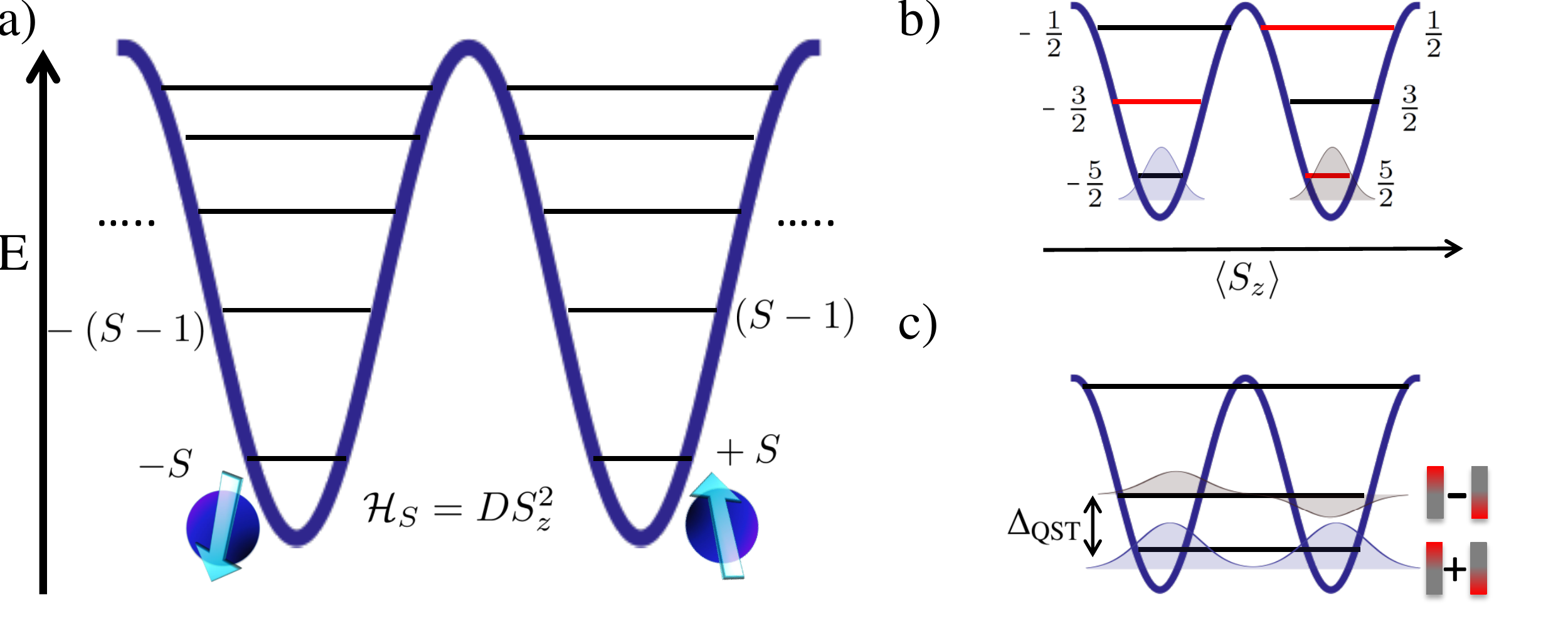} 
  \end{center}
  \caption{ Effects of the magnetic anisotropy on quantum magnets. (a) Energy levels scheme of a nanomagnet with an uniaxial, easy axis ($D<0$ ) anisotropy. The two degenerate ground states correspond to the classically oriented states along the easy axis. (b) Effect of the transversal $E$ terms ($\propto (S_x^2-S_y^2)$ ) on an exemplifying $S=5/2$ half-integer spin, involving mixing between states represented with the same color. (c) Idem to (b) but for an integer spin ($S=1$ in the example displayed).  The new mixed states are bonding and antibonding  linear combinations of the classical states, split in energy by $\Delta_{QST}$. 
  }
\label{figAni}
\end{figure}

The behavior of integer and half-integer spins, described with the same Hamiltonian,  can be radically different.  According to Kramers' theorem,  half-integer spins have at least doubly degenerate spectrum, on account of time reversal symmetry, whereas integer spins can be non-degenerate at $B=0$.  In the case of strong uniaxial anisotropy,  where the dominant term in the Hamiltonian is ${\cal H}_{\rm S} = D \hat S_z^2$,  with $D<0$,  the ground state doublet is made of the two states with $S_z=\pm S$, see Fig.~\ref{figAni}.  The effect of the the remaining terms in the Hamiltonian depends dramatically on the parity of $2S+1$.
To be specific  we discuss here the case of Hamiltonian (\ref{Hde}), but the discussion applies to general spin Hamiltonians as well~\cite{Prada_arXiv_2016}. 

 In the case of half-integer spins, as well as some integer-spins where there are symmetry protected degeneracies~\cite{Karlewski_Marthaler_prb_2015,Donati_Rusponi_science_2016,Prada_arXiv_2016},
   there is a strict degeneracy of the ground state doublet at zero magnetic field~\cite{Kramers_paa_1930},  and we can always choose the eigenvectors of ${\cal H}_{\rm S}$ such as the uniaxial term $D\hat S_z^2$ is diagonal in the $E=0$ limit: 
 \begin{eqnarray}
|C_1\rangle &\propto & |-S\rangle 
+ {\cal O}\left(\varepsilon^{S-1}\right) |\delta \phi_1\rangle
 \nonumber\\
|C_2\rangle &\propto & |+S\rangle 
+ {\cal O}\left(\varepsilon^{S-1}\right) |\delta \phi_2\rangle,
\label{Cstates}
\end{eqnarray}
where ${\cal O}\left(\varepsilon^{S-1}\right) |\delta \phi\rangle$  is the  contribution to the wave function coming from states with  smaller $|S_z|$ and  it is controlled by 
$\varepsilon$, a small number that in the case of the Hamiltonian (\ref{Hde}),  is  given by $\varepsilon=E/|D|$.\footnote{If $3E>D$ we relabel the Hamiltonian, so that the new easy axis is again $z$.} Since this type of situation is compatible with the conventional classical picture of a magnet with two equivalent ground  states with a finite and opposite average magnetization $M_z\simeq \langle C_1 |S_z|C_1\rangle= -\langle C_2 |S_z|C_2\rangle$, we refer to these spins as   {\em type C} or {\em classical type of spins}.  
  We stress, however, that since these two wave functions correspond to degenerate states, any linear combination of $|C_1\rangle$ and $|C_2\rangle$  
will also be a valid choice to describe the ground state doublet.  However, the coupling to the environment will select the
$|C_1\rangle$ and $|C_2\rangle$ states as the most stable, a process discussed in Sec.~\ref{basisCh}.

By contrast, the zero-field spectrum of integer spins with finite transversal anisotropy $E$ is non-degenerate due to quantum spin tunneling, so that $E_2=E_1+\Delta$, where $\Delta\propto E\varepsilon^{S-1}$ in the case of Hamiltonian (\ref{Hde}). It is then apparent that, as $S$ increases, $\Delta$ decreases.  As long as $\Delta\neq 0$,  the wave functions of the doublet are uniquely defined, modulus a global phase, and approximately given by~\cite{Gatteschi_Sessoli_book_2006}:
\begin{eqnarray}
|Q_1\rangle &\propto & |S\rangle +(-1)^{S-1} |-S\rangle 
+ {\cal O}\left(\varepsilon^{S-1}\right) |\delta \phi\rangle \nonumber\\
|Q_2\rangle &\propto & |S\rangle +(-1)^S |-S\rangle + {\cal O}\left(\varepsilon^{S-1}\right) |\delta \phi\rangle ,
\label{Qbasis}
\end{eqnarray}
where the last term is a small contribution coming from states with $|S_z|<S$.
Thus, the wave functions are approximately given by the bonding and antibonding combination of the states with $S_z=\pm S$, while the bonding-antibonding nature of the ground state wave function alternates as $S$ increases.  We refer to these states as {\em type Q} or {\em quantum type} of states because they are quantum superpositions of the two classical ground states, and have thereby very different properties, such as the vanishing  expectation values of the magnetic moment~\cite{Delgado_Loth_epl_2015}.

\subsection{Pseudo-spin $1/2$  approximation  \label{TLT}}
In those instances where there is a gap between the ground state doublet  and the rest of the energy levels, one can truncate the Hilbert space and treat the anisotropic  spins as two level systems.  This occurs for the Hamiltonian (\ref{Hde}) when $-D\gg E, k_BT$.   This condition holds, for instance, in the case of Fe or Co on Cu$_2$N~\cite{Hirjibehedin_Lin_Science_2007,Otte_Ternes_natphys_2008}.\footnote{The 2-level approximation will be valid to describe the spin dynamics as long as excitation of higher energy states, either by thermal fluctuations or by scattering with transport electrons under a finite bias voltage, are negligible.} 
 When projected over this subspace,  the spin rotational invariance inherent to the Heisenberg coupling between magnetic adatoms, $\vec{S}(l)\cdot\vec{S}(l')$, or the Kondo interaction with the surface electron spin density, $\vec{S}(l)\cdot\vec{s}(\vec{r}_l)$,  is broken on account of the single ion  magnetic anisotropy.  To see this,  we represent the spin operators $\hat S_a(l)$ in the subspace of dimension 2 of the ground state doublet. The resulting matrices of dimension 2 can be written down as linear combinations of Pauli matrices $\hat\tau_a$ acting on the space of the TLS:
\begin{equation}
\hat S^a(l)= \sum_{b=x,y,z} \kappa_{ab}(l) \hat \tau_b.
\label{S2tau}
\end{equation} 
Specifically, we write this up for the cases of a single spin described by Hamiltonian (\ref{Hde}). The expressions are then quite different for the half-integer and integer spin.  In the case of half-integer, using the wave functions (\ref{Cstates}), we  have 
$\kappa_{ab}\propto\delta_{ab}$, with 
\begin{eqnarray}
\kappa_{zz}\approx  S+{\cal O}(\varepsilon^{2S-1}) \nonumber\\
\kappa_{xx},\kappa_{yy}\sim S\varepsilon^{2S-1}.
\label{kappahalf}
\end{eqnarray}
Thus, in the limit of pure uniaxial anisotropy with $S>1/2$,  the only operator with non-zero matrix elements in the  ground state manifold is
the $\hat S_z$ operator, and the Heisenberg coupling $\vec{S}\cdot\vec{s}(0)$ takes the form of an Ising interaction 
$ \hat S_z  s_z(0)\simeq S \hat \tau_z  s_z(0)$.     
Another interesting case occurs for half-integer spins where $D>0$, such that the ground state doublet has $S_z=\pm 1/2$.   This is relevant for instance for Cobalt on Cu$_2$N ($S=3/2$)~\cite{Toskovic_berg_natphys_2016}.   In this case  the operators $\hat S_x$ and $\hat S_y$ have non-vanishing matrix elements within the $S_z=\pm 1/2$ doublet, but their strength is larger than in the $\hat S_z$ term.
For $S=3/2$,  the representation of the $\hat S_x$ matrix in the subspace of the $S_z=\pm 1/2$ doublet give matrix elements twice as large as those obtained for a real $S=1/2$ spin.  In contrast, the representation of the $\hat S_z$ operator gives the same matrix elements in both cases.  The resulting exchange interactions in the $S=3/2$ case are thereby anisotropic.

We now consider the representation of the spin operator in the $2\times 2$ space within the Q-basis (\ref{Qbasis}), where~\cite{Delgado_Loth_epl_2015}

\beqa
\vec S \equiv
  \hat z \langle Q_1|\hat S_z|Q_2\rangle \left(
  \begin{array}{cc}
  0 & 1\\
  1 & 0\\
  \end{array}
  \right).
\eeqa
In other words, the  atomic spin-flip operators $\hat S_x$ and $\hat S_y$  have no effect in the sub-space of states $|Q_1\rangle $ and $|Q_2\rangle$, and only the $\hat S_z$ operator has a non-vanishing contribution whose effect is to {\em induce transitions} among these states.  
It is important to note that for systems with a degenerate ground state, it is a matter of choice whether we use the $Q$ or the $C$ basis.  

The important take home message here is that the conventional spin-isotropic interactions, working in the subspace of low energy selected by the large magnetic anisotropy, result in anisotropic effective spin Hamiltonians. This is of course a resource that might be used to do quantum simulations~\cite{Toskovic_berg_natphys_2016}.

\subsection{Hamiltonian for multi-spin systems\label{multi}}
One of the most appealing features of surface magnetism is the possibility to assemble magnetic structures, adding atoms one by one.  In addition to the single spin anisotropy, these atoms interact with each other through different kind of sin-spin interactions.  
A broad class of magnetic atoms on surfaces can be described with the spin Hamiltonian
\begin{equation}
{\cal H}_{\rm S}= \sum_{l\leq N}  {\cal H}_0(l) + \sum_{l\neq l'}^N \sum_{a,b=x,y,z} J_{a,b}(l,l') S_a(l)S_b(l')
\label{HSpin}
\end{equation}
where the first term describes the single ion  magnetic anisotropy, given by Eq (\ref{Hani}),  while the second one represents a spin-spin exchange interaction.  Several  atomically engineered  nanostructures, such as spin chains of Fe  and Mn on Cu$_2$N~\cite{Hirjibehedin_Lutz_Science_2006,Otte_Ternes_prl_2009,Loth_Baumann_science_2012,
yan2016non,choi2016magnetic,
Bryant_Spinelli_prl_2013,Spinelli_Bryant_natmat_2014,Toskovic_berg_natphys_2016},
  can be modelled assuming that spin-spin interaction is rotational invariant (Heisenberg type,  $J_{a,b}= J_H \delta_{a,b}$) and neighbor isotropic, either antiferromagnetic (AF)~\cite{Hirjibehedin_Lutz_Science_2006,Loth_Baumann_science_2012,Toskovic_berg_natphys_2016}  or ferromagnetic (FM)~\cite{Bryant_Spinelli_prl_2013,Spinelli_Bryant_natmat_2014}, while the second neighbor interaction is either negligible, or much smaller~\cite{Rossier_prl_2009,Rudenko_Mazurenko_prb_2009}.  Dzyaloshinskii-Moriya (DM)  interaction coupling, where the inter-spin interaction takes the form
  $\sum_{ab}J_{ab}S_a(l) S_b(l')\equiv \vec  D_{ll'}\cdot \left( \vec S(l)\times \vec S(l')\right)$,  with $\vec D_{ll'}$ a vector whose orientation is given by some high-symmetry direction of the surface, is known to be relevant for magnetic atoms on top of heavy metal surfaces, such as Pt~\cite{Honolka_Lee_prl_2009} and Ir~\cite{Menzel_Mokrousov_prl_2012}, resulting in non-collinear broken symmetry states, such as  spin spirals for Fe chains on Ir~\cite{Menzel_Mokrousov_prl_2012} or Mn on W~\cite{Ferriani_Bergmann_prl_2008,Serrate_Ferriani_nature_2010}, and skyrmions for monolayers of Fe on Ir~\cite{Heinze_Bergmann_natphys_2011,Romming_Hanneken_science_2013}.  Both indirect coupling mediated by the surface electrons and super-exchange mediated by non-magnetic surface atoms are believed to contribute to these interactions~\cite{Schweflinghaus_Zimmermann_prb_2016}.  In the following we only consider first neighbor Heisenberg interactions $\vec{S}(l)\cdot\vec{S}(l\pm 1)$, the dominant coupling between magnetic atoms adsorbed on a decoupling layer such as Cu$_2$N, MgO or AlO.

Depending on the competition between the single ion anisotropy and the Heisenberg interaction, the geometry of the structure (chain, ladder,  cluster),  and the spin parity,  the ground state of the system can have very different nature.
From the experimental information, obtained using both  IETS and time resolved magnetization switching using spin polarized STM, together with the information from the model Hamiltonian, one can infer  the different types of ground states. 
 Even if we limit the discussion to chains and ladders formed with transition metals on Cu$_2$N, the following ground states have been reported:  
\begin{enumerate}
\item{ Antiferromagnetically correlated spin ground state, without broken symmetries,  for Mn chains ($S=5/2$) with up to $N=10$ atoms~\cite{Hirjibehedin_Lutz_Science_2006}. In the case of even (odd) the spin of the ground state  is $S_G=0$ ($S_G=5/2$).  The exchange interaction for the Mn dimer is $J_H\simeq 6$ meV, much larger than the single ion anisotropy $D\approx -0.04$ meV. }   
\item {Broken symmetry  AF ground states (Ising type classical N\'eel states), for Fe chains along the $N$ rich direction,  and $N \geq 3$~\cite{Loth_Baumann_science_2012}.  For $N=2$  no signal of broken symmetry is observed ~\cite{Loth_Bergmann_natphys_2010}. } 
\item{ Broken symmetry  FM ground states (Ising type),   for Fe chains along the
 Cu rich direction,  and $N \geq 3$~\cite{Spinelli_Bryant_natmat_2014}.  }
\item{ Distributed Kondo effect in hybrid spin Fe-Mn$_N$ spin chain~\cite{Choi_Robles_arXiv_2015}.   The Fe-Mn dimer ($N=1$) have a strong AF interaction that leads to a ground state  spin  $S_G=1/2$, that results in a zero bias Kondo peak. As $N$ increases, staying odd, the Kondo effect is preserved, but it is more prominent in the side of the structure opposite to the Fe-Mn, reflecting the non-local and rather counterintuitive nature of the  ground state of the chain.}
\item{ Correlated pseudo-spin $1/2$ XXZ  model~\cite{Toskovic_berg_natphys_2016}.  A chain of Cobalt atoms with $S=3/2$ and strong uniaxial anisotropy that  favors the $S_z=\pm 1/2$ doublet realizes the  $XXZ$  model. }

\end{enumerate}

This list illustrates that the competition between uniaxial anisotropy,   exchange interaction and Kondo coupling can result in a large variety of possible ground states.  This can be understood, in part, when the anisotropy is large enough to permit a truncation of the single spin Hamiltonian, keeping only two levels. This approximation leads then to anisotropic pseudo-spin $1/2$ interactions.  For instance, for an integer spin chain governed by the single ion anisotropy Hamiltonian (\ref{Hde}) and Heisenberg  exchange $J_H$,  and provided $D<0$ with $|D|\gg E, J_H$,   we can retain the doublet with maximal $S_z$ at each atomic spin. 
The resulting effective Hamiltonian reads~\cite{Delgado_Loth_epl_2015}:
\begin{equation}
{\cal H}_S\approx \sum_n \frac{\Delta}{2} \hat\tau_z(n) + j\sum_{n} \hat \tau_z(n)\hat \tau_z(n+1)
\label{QIM}
\end{equation}
where $\Delta$ is the quantum spin tunneling splitting of an individual adatom and $j\equiv J_H|\langle G_1|\hat S_z(l)|G_2\rangle|^2$. Of course, Hamiltonian (\ref{QIM}) is nothing but the quantum Ising model in a transverse field (QIMTF).  This is approximately the case of the FM~\cite{Spinelli_Bryant_natmat_2014} and AFM~\cite{Loth_Baumann_science_2012} Fe chains on Cu$_2$N/Cu(100). 
In  contrast, in the case of $D\gg E,\;k_BT 0$,  the  two level truncation that keeps only the $S_z=\pm 1/2$ doublet leads to the XXZ model~\cite{Toskovic_berg_natphys_2016}. We thus see that  atomically engineered magnetic structures with atomic spins with $S>1/2$  can be used to realize effective spin $1/2$ models and might be used for quantum simulation~\cite{Toskovic_berg_natphys_2016}.

 \section{Decoherence due to Kondo exchange\label{KondoGen}}
 In this section we treat the problem of decoherence of   magnetic atoms on a surface  due  to the Kondo exchange interaction with the itinerant electrons of  the underlying conductor.   We treat in detail the case of an individual magnetic atom and, later on, we briefly address the problem of spin decoherence of finite size chains of exchanged-coupled magnetic atoms.   
 
 We assume that at $t=0$ some quantum demon has been able to prepare the system  in  a linear superposition of two states that have very different magnetic properties. In the case of an individual quantum spin, the wave function would be:  
\begin{equation}
|\psi(t=0)\rangle=\frac{1}{\sqrt 2}\left(|S_z=+S\rangle + |S_z=-S\rangle\right).
\label{t0}
\end{equation}
Our goal is to determine for how long this coherent superposition can be maintained, taking into account that the spin is exchange coupled to the electron gas of the surface. This characteristic decoherence time will be denoted as $T_2$.

\subsection{Kondo exchange interaction}
The Kondo exchange interaction with the surface electrons can be written down as: 
\beqa
{\cal V}_K=  \sum_l^N J_l\vec S_l\cdot \vec s(\vec r_l),
\label{HTUN}
\eeqa
where $\vec S_l$ is the spin of the $l$-magnetic adatom and $\vec s(\vec r_l)$ is the surface  spin density evaluated at the position $\vec r_l$ of the $l$ magnetic atom,
\beqa
\vec s(\vec r_l)=\sum_{\vec k\vec k'\sigma\sigma'}e^{i\left(\vec k-\vec k'\right)\cdot \vec r_l}\frac{\vec \tau_{\sigma\sigma'}}{2}
c^{\dagger}_{\vec k,\sigma} c_{\vec k'\sigma'},
\label{Sdensity}
\eeqa
 where $c_{\vec k,\sigma}^\dag$ indicates the creation operator of a conduction electron with
momentum $\vec k$, and spin projection $\sigma$ along the  quantization  axis $z$. In the case of a single magnetic atom, we always choose its location at the origin, so that the phase factor in the Hamiltonian goes away.  For many atoms the phase factor can not be gauged away and it can play an important role. 
We assume that the conduction electrons can be described within the independent particle picture:
\begin{equation}
{\cal H}_{\rm surface}= \sum_{\vec{k},\sigma}\epsilon_k c^{\dagger}_{\vec k,\sigma} c_{\vec k\sigma}
\end{equation}
 For simplicity, in this work we limit the discussion to the case of a non-polarized electrodes. A similar discussion could be carried out for the coupling to a spin-polarized bath~\cite{Delgado_Rossier_prb_2010}.  Most often, the  exchange constant $J_l\equiv J$ is the same for all the spins, which is a quite reasonable assumption for magnetic adatoms adsorbed on equivalent lattice sites on the surface.
 
\subsubsection{Kondo Hamiltonian in the two-level approximation\label{ssTLA}}
As we discussed in Sec.~\ref{TLT},  in many instances we can restrict the  $(2S+1)$ Hilbert space of magnetic adatoms down to just two levels.  
Using equations (\ref{S2tau}) and (\ref{HTUN}), we can write down  the Kondo exchange Hamiltonian, projected into the subspace of the ground state doublet,  as: 
\begin{equation}
{\cal V}_K= 
 J \sum_{l\leq N} \sum_{a,b} \kappa_{ab}(l) \hat \tau_b   s_a(\vec r_l).
 \label{KondoG}
\end{equation}
where $\sum_b \kappa_{ab}(l)\hat \tau_b$  is the representation of the atomic spin operator $\hat S_a(l)$ in the basis set of the two levels. In the case of degenerate two levels, this representation is not unique.  If we choose the basis set as to diagonalize the $\hat S_z$ operator,  
for a single spin that we place at the origin of coordinates 
 ($\vec r_1=0$), the representation of the Kondo coupling in the truncated basis takes the form
\begin{equation}
{\cal V}_{K}^C= j_z \hat\tau_z s_z(0)+ j_{x} \hat\tau_x s_x(0)+ j_y \hat\tau_y s_y(0),
\label{VC}
\end{equation}
where $j_z\simeq  JS \gg j_x,j_y$.   Thus, the single ion anisotropy results in an effective anisotropic exchange.  In the case where the eigenstates of the TLS are also eigenstates of $\hat S_z$, with $|S_z|>1/2$,   we have $j_x=j_y=0$ and the coupling anisotropy is maximal as the  truncated Kondo Hamiltonian takes the form of an Ising-Kondo model~\cite{Sikkema_Buyers_prb_1996}, ${\cal V}_{K}^C= j_z \hat\tau_z s_z(0)$.
 As a consequence,
 population scattering is strictly forbidden between the states with different atomic $S_z$,  yet the coupling to the environment is able to induce decoherence, as we  describe in Sec.~\ref{Imry}.

Another interesting case is provided by a TLS  whose wave functions are given by states as those in Eq. (\ref{Qbasis}). This situation arises naturally for integer spins  described with the single ion Hamiltonian (\ref{Hde}),  with $-D \gg |E|>0$.  In that case we have
$\vec S=\langle Q_1|S_z|Q_2\rangle (0,0,\hat\tau^x)$~\cite{Delgado_Loth_epl_2015}, 
which leads to the single impurity Kondo Hamiltonian
\begin{equation}
{\cal V}_K^{Q}= j_z \hat\tau_x s_z(0)
\label{VQ}
\end{equation}
where $j_z= J \langle Q_1|S_z|Q_2\rangle$. 
In other words, the  atomic spin-flip operators $\hat S_x$ and $\hat S_y$  have no effect in the sub-space of states $|Q_1\rangle $ and $|Q_2\rangle$, and only the $\hat S_z$ operator has a non-vanishing contribution whose effect is to {\em induce transitions} among these states~\cite{Delgado_Loth_epl_2015}.

\subsection{General expressions for $T_1$ and $T_2$ due to Kondo exchange}

The general expression for a $T_1$ like transition rate between two eigenstates of a spin chain due to Kondo coupling to the reservoir is given by~\cite{Delgado_Rossier_inprep}: 
\beqa
\Gamma_{MM'}=
\frac{\pi J^2}{2\hbar^2} 
\sum_{\vec k,\vec k'}
f(\epsilon_k) \left(1-f(\epsilon_{k'})\right)
\times
\chi_{M,M'}(\vec{k}-\vec{k}')
\delta\left(\epsilon_k+E_M-\epsilon_{k'}-E_{M'}\right),
\label{dinteg}
\eeqa
where
\beq
\chi_{M,M'}(\vec{k}-\vec{k}')\equiv 2\sum_{l,l'=1}^N
e^{i(\vec{k}-\vec{k}')\cdot (\vec r_l-\vec r_{l'})}
\sum_aS^a_{MM'}(l)S^a_{M'M}(l') = \sum_a \left|{\cal S}^a_{M,M'}(\vec{k}-\vec{k}')\right|^2
\label{defchi}
\eeq
where ${\cal S}^a_{M,M'}(\vec{k}-\vec{k}')=\sum_{l} e^{i(\vec{k}-\vec{k}')\cdot  \vec r_l} S^a_{MM'}(l)$
and 
$S^a_{MM'}(l)\equiv \langle M|\hat S^a(l)|M'\rangle$ with $a=x,y,z$. The interpretation of Eq. (\ref{defchi}) is quite transparent. The rate  contains  the $f(1-f)$ factor that weights the occupation of the initial quasiparticle state and the availability of the final quasiparticle state.  The $\delta$ function ensures the overall conservation of energy, while energy exchange between system and reservoir is
allowed.   The spin structure function, $\chi_{M,M'}(\vec{k}-\vec{k}')$ accounts both for the spin conservation and for the non-local couplings that naturally arise when the sum over all the atoms in the structure is squared in order to obtain a scattering rate. 


For the decoherence rates, we obtain two contributions, very much like in the single spin case. 
The first comes from $T_1$-like  population scattering  processes~\cite{Cohen_Grynberg_book_1998}:
\beqa
\gamma^{nonad.}_{M,M'}=\frac{1}{2}\left(\sum_{N\ne M} \Gamma_{M,N}+\sum_{N\ne M'} \Gamma_{M',N}\right),
\label{gnonadv}
\eeqa
where $\Gamma_{M,M'}$ are the scattering rates defined in Eq. (\ref{dinteg}).
The adiabatic contribution  corresponds to processes that occur even in the absence of changes in populations
of the $|M\rangle$ states. It is driven
by  elastic scattering processes with the reservoir and it is often known as pure dephasing.  
 In our case,  the adiabatic decoherence rate is given by:
\beqa
\gamma_{M,M'}^{ad.}&=&\frac{\pi J^2}{2\hbar}
\sum_{\vec k,\vec k'}
f(\epsilon_k) \left(1-f(\epsilon_{k'})\right)
\chi_{M,M'}^ {ad.}(\vec{k}-\vec{k}')
\delta\left(\epsilon_k-\epsilon_{k'}\right).
\label{invT2chain}
\eeqa
The matrix elements $\chi_{M,M'}^ {ad.}(\vec{q})$ are given by (see Appendix \ref{T2gen} for details):
\beq
\chi_{M,M'}^ {ad.}(\vec{q})=\sum_{a} \left|\sum_l \left(e^{i\vec q\cdot \vec r_l}S^a_{MM}(l)-e^{-i\vec q\cdot \vec r_l}S^a_{M'M'}(l)\right)\right|^2
=
 \sum_a \left|{\cal S}^a_{M,M}(\vec{q})-{\cal S}^a_{M',M'}(-\vec{q})\right|^2
.
\label{chiadb}
\eeq

\subsection{Decoherence of a single degenerate spin   \label{T2single}}

\subsubsection{Decoherence of a single spin with degenerate spectrum: perturbative results  \label{KondoT2}}
We now  consider  the simplest case of an individual
 magnetic atom  with  spin $S$  and  uniaxial anisotropy, ${\cal H}_{\rm S}= -|D|\hat S_z^2$, which leads to a degenerate ground state. 
 We choose as a basis set the eigenstates of $\hat S_z$,  dubbed as classical, see Eq. (\ref{Cstates}). 
 
Within the perturbative BR theory, the decoherence rate has two types of contributions,  adiabatic and nonadiabatic.  
The latter  are due to $T_1$-like process that change the population of the two states 
 $|+S\rangle$ and $|-S\rangle$, see Eq. (\ref{gnonadv}). The Kondo interaction can only induce transitions with $\Delta S_z=\pm 1$.  Therefore, for $S>\frac{1}{2}$,  the nonadiabatic channel implies inelastic transitions from the states of the ground state doublet to an excited state,  separated in energy by $\Delta= |D| (2S-1)$.  Energy conservation implies that {\em a thermally excited electron-hole pair across the Fermi energy has to be annihilated}.
 The scattering rate for these $T_1$-like processes excitation of the spin is: 
\begin{equation}
\frac{1}{T_1}=
 \frac{\pi}{2\hbar} \left(\rho J\right)^2 S\frac{\Delta}{e^{\beta \Delta}-1}
\label{T1Kondosimplest}
\end{equation}
where $\rho$ is the density of states at the Fermi level of the surface electrons.
As the temperature goes to zero, the density of thermally excited electron-hole pairs vanish exponentially, and
excitation rate   $1/T_1\propto e^{-\beta\Delta}$ vanishes altogether.   
 
 The relaxation rate of the $S_z=S-1$  excited state (or $Sz_=-S+1$) to the ground state by {\em spontaneous emission} of an electron-hole pair across the Fermi energy can be obtained from Eq. (\ref{T1Kondosimplest}) by  
reversing the sign of $\Delta$. In the low temperature limit ($\beta \Delta \gg 1$), this relaxation rate scales like 
\begin{equation}
\frac{1}{T_1}\Big|_{\rm relax}\equiv \gamma^{\rm ad.}\simeq   \frac{\pi}{2\hbar} \left(\rho J\right)^2 S\Delta
\label{T1relax}
\end{equation}
The linear relation between the transition energy $\Delta$ and the relaxation of excited states has been observed experimentally for 
Fe adatoms on Cu(111) surface~\cite{Khajetoorians_Lounis_prl_2011} and also for an antiferromagnetically coupled chain of $N=3$ Fe atoms on Cu$_2$N~\cite{yan2016non}

For the nonadiabatic decoherence of the ground states, the relevant rate is given by Eq. (\ref{T1Kondosimplest}).  Thus, 
at sufficiently low temperatures this contribution is suppressed exponentially, and we are left with the adiabatic contribution, whose   rate is given by~\cite{Delgado_Rossier_prl_2012,Shakirov_Shchadilova_arxiv_2016}: 
\begin{equation}
\frac{1}{T_2^{\rm ad}} \equiv \gamma^{\rm ad.} =\frac{\pi}{2\hbar} \left(\rho J\right)^2 S^2 k_BT.
\label{pureT2}
\end{equation}
This result  shows that the bigger $S$, the faster the decay of the coherent superposition of the states $+S$ and $-S$. 
It also shows that even without inelastic scattering, the phase coherence of the superposition state is fragile. 
The temperature dependence in this case comes from a phase space argument.  Elastic scattering requires the presence of an electron and a hole at the same energy. The  density of electron hole pairs scales linearly with $k_BT$.   Perturbative results work well when $\rho J\ll 1$.  For instance, taking $\rho J=0.1$, and $S=2$, and $k_BT=100$ mK,  the decoherence time is $T_2\simeq 1.2$ ns. 
 Therefore, even in the most favorable case of  a system where both spin flip and inelastic scattering are suppressed, elastic  spin-preserving Kondo interactions are  extremely detrimental for atomic spin coherence.    Baumann {\em et al.} have reported $T_2\simeq 120$ ns for $k_BT=0.6$ K for Fe on MgO.  If the decoherence rate in that system was governed exclusively by Eq. (\ref{pureT2}), and assuming $S=2$ for that system, we would infer $\rho J\simeq 3\times 10^{-3}$.  Thus, this is a good upper limit for this quantity in that system. 

The perturbative results seems to indicate that, in the limit of  $T=0$, the coupling to the environment would not be able to decohere the linear superposition. Actually, we show below that this conclusion is wrong, and elastic spin-conserving Kondo process would result in a finite decoherence rate even at $T=0$.  The failure of the BR theory relates, in this case, to the fact that the the theory only describes processes that are slow compared to the correlation time of the bath $\tau_c$. In the case of the Fermi sea, $\tau_c\approx  \frac{\hbar}{k_B T}$.  Thus, as the temperature goes to zero, the theory can only describe processes that are very slow, and eventually this prevents the proper description of $T_2$.

\subsubsection{Decoherence of a single spin with degenerate spectrum: non-perturbative results  \label{KondoT2}}
We now compute the adiabatic decoherence without using perturbation theory and without invoking the Markovian limit, both essential ingredients of the BR theory.  This is possible in this specific case, because we can  use two well established methods. First,  we perform the partial wave expansion of  the itinerant electron states of the surface. 
For a single Kondo impurity, only the $s$ wave contributes~\cite{Schotte_Schotte_prb_1969}.  This turns the original problem in a one-dimensional one, for which we can apply the bosonization description of the spin density operator as described in Appendix \ref{bosoni}.  Second, and as described in Sec~\ref{ssTLA}, we restrict the spin Hilbert space to the ground state doublet.  We end up with the following Hamiltonian for the 
TLS coupled to the electron-hole pairs of the Fermi sea, described with bosonic operators:
\begin{equation}
{\cal H}_{\rm SBC}=\hbar v_F\sum_{0\le k\le k_c} k b^{\dagger}_{k} b_{k},
+ \sqrt{\alpha} \hat\tau_z 
\sum_{0<k<k_c}g_k
(b_k^\dag+b_k).
\label{SBC0}
\end{equation}
Notice that this model is a particular version of the spin boson model when the tunneling term vanishes, see Eq. (\ref{SBQ}).  
Hamiltonian (\ref{SBC0}) is diagonalized by the following unitary transformation~\cite{Shnirman_Schon_book_2003}:
\begin{equation}
U{\cal H}_{\rm SBC} U^{-1} = h_0+\frac{\Delta}{2}\hat\tau_z
\label{UnitT}
\end{equation}
where  $U=\exp(-i\hat \tau_z \Phi/2)$ and
\begin{equation}
\Phi\equiv i \sum_{k>0}^{k=k_c} 
 \left(\frac{k}{\pi L}\right)^{1/2}\left(b_k^{\dagger}- b_k\right).
\end{equation}
This permits us to calculate the evolution of the wave function of the bosonic operators that describe the electron-hole pair excitations of the bath. Using the argument of Sec.~\ref{Imry}, we can compute the decay of the coherence, given by Eq. (\ref{S}). 
With the help of transformation (\ref{UnitT}), one arrives to ${\cal S}_{12}(t)=e^{K(t)}$~\cite{Shnirman_Schon_book_2003}, where
\begin{equation}
K(t)=\frac{4}{\pi\hbar}\int_{0}^{\omega_c}d\omega
\frac{J(\omega)}{\omega^2} {\cal F}(\omega,T,t),
\end{equation}
with
\begin{equation}
{\cal F}(\omega,T,t)=\coth\left(\frac{\hbar\omega}{2k_BT}\right)
(\cos\omega t -1) -i \sin(\omega t),
\end{equation}
and $J(\omega)$  the spectral density. 
For a Ohmic bath, the case relevant for a Fermi gas, we have $J(\omega)\propto \omega$,
and we can obtain two very interesting limiting results. For finite temperature, and not too short times,
 $t>\frac{\hbar}{k_BT}$, we have: 
 \begin{equation}
 {\cal S}_{12}(t)\approx e^{-\Gamma t} e^{-i{\rm Im}\left[ K(t)\right]}
 \label{PD1}
 \end{equation}
with
\begin{equation}
\Gamma= 2\pi\alpha k_BT/\hbar.
\label{Gamma}
\end{equation}
This results is the same that can be obtained using the Bloch-Redfield approach, outlined in the previous subsection. It basically means that, even in the absence of spin-flip interactions, the electron gas is able to decohere a ``Schr\"odinger-cat "-like state, Eq. (\ref{S}), where the spin is prepared in a superposition of the two states with opposite magnetization along the easy axis. 
  In the opposite limit, $1/\omega_c< t<\hbar/k_BT$,  which becomes specially relevant as $T$ goes to zero,  we have $K(t)\approx -2\alpha \ln(\omega_c t)$, which yields
   \begin{equation}
 {\cal S}_{12}(t)\approx  (\omega_c t)^{-2\alpha}.
 \label{PD2}
 \end{equation}
This result is interesting on two counts. First, it shows that even at $T=0$ the bath is able
to decohere the spin.  Second, it permits covering a limit that can not be addressed by
 the Bloch-Redfield approach, as we discussed in Sec.~\ref{BRF}.
 
Equations (\ref{PD1}) and( \ref{PD2}) can be reinterpreted as follows. Since they clearly show a transition from an exponential decay law in Eq. (\ref{PD1}) to a power law in Eq. (\ref{PD2}), one can define a transition temperature $T^*=\hbar \omega_c/k_B\exp (-1/2\alpha)$ below which the power-law decay dominates over the later exponential one~\cite{Shnirman_Schon_book_2003}. Thus, we can distinguish between a low temperature regime ($T< T^*$) where the dephasing rate is temperature independent and approximately given by $k_bT^*/\hbar$, and a high temperature one ($T>T^*$) with the dephasing rate $\Gamma$ given by Eq. (\ref{Gamma}).

\subsubsection{Quasiparticle phase shift  as the origin of the pure dephasing
\label{phaseshiftsec}}
Both the perturbative BR approach and the non-perturbative method based on the exact solution of the bosonized Hamiltonian show that  the Kondo interaction of an anisotropic spin with a Fermi gas is able to decohere the linear superposition state of Eq. (\ref{t0}).  This occurs with scattering events that  preserve  both the energy and the angular momentum  of the two particles involved in the scattering, the quasiparticles and the  magnetic atom,
 and although linear momentum is transferred,  this information is averaged out.  Thus, 
and obvious question arises: how does the the interaction modify the environment wave function, which is the ultimate responsible of the decoherence in the absence of scattering? 

For the spin-conserving interaction considered here,  the environment actually is formed by   two independent reservoirs, the Fermi gas for $\uparrow$ electrons and the Fermi gas for $\downarrow$ electrons.   If we focus on the $\uparrow$ reservoir, it is apparent that the interaction with the atomic spin in the $+S$ state results in a phase shift different from the one when the atomic spin is in the $-S$ state.  This is trivially seen in the case of one dimensional Fermions interacting with a delta function $ V_0 \delta(x)$, for which the phase shift transmission coefficient is given by $t(\epsilon)= \frac{1}{1+ i V_0 \rho(\epsilon)}$.   For the Kondo-Ising problem, we can write down $V_0= J S \sigma$, where $S$ and $\sigma=\pm 1/2$ are  the atomic and quasiparticle spins  respectively.  In the weak coupling limit, the phase shift  is:
\begin{equation}
\delta_{\sigma}(S)  \simeq \rho J S \sigma  
\end{equation}
where $\rho=\frac{m}{\hbar^2 k(\epsilon)}$ is the density of states at the energy of the quasiparticle, that we omit from the arguments for simplicity.  Importantly, for a fixed quasiparticle spin, the phase shift  depends on whether the atomic spin is in the $+S$ or $-S$ state.  Using this result, the pure dephasing (\ref{pureT2}) can be written down as: 
\begin{equation}
\frac{1}{T_2^{\rm ad}} =\pi \frac{k_B T}{2\hbar} \sum_{\sigma} \left(\delta_{\sigma}(S)-\delta_{\sigma}(-S)\right)^2  .
\label{pureT2phaseshift}
\end{equation}
 So, it is natural to think that the environment reads the information about the system via the spin dependent phase shifts.  This is in line with the standard results of the quantum impurity  phenomena,  such as the Fermi edge singularity~\cite{Anderson_prl_1967} and the Kondo effect~\cite{Kondo_ptp_1964}, where some important results can be expressed in terms of the quasiparticle scattering phase shift~\cite{Nozieres_Dominicis_pr_1969,Anderson_prl_1967}.

The connection between spin decoherence and the quasiparticle phase shift opens up an interesting venue of research: in the case of spin structures,  the scattering with multiple point scatterers can results in dramatic reductions of the phase shifts, that would lead to enhanced coherent lifetimes for the atomic spins.

\subsubsection{Decoherence for the degenerate doublet: the choice of basis set \label{basisCh}}
The discussion above has considered decoherence for spins with uniaxial anisotropy, such that $S_z$ is a good quantum number. Thus, the Kondo interaction  with the surface electrons  was {\em not} capable of inducing transitions from $S_z$ to $-S_z$, on account of the conservation of angular momentum (provided that $2S>1$). 
Hence, in that situation,  the choice of the eigenstates of  $\hat{S}_z$ 
as basis set is quite natural, although not unique. However this situation where scattering is forbidden is exceptional and, in general, for two level systems both decoherence and population scattering can occur leading to coupled dynamical equations. 
 In this case,  typical of half-integer spin at zero field, the distinction between $T_1$ and $T_2$ becomes ill defined, or at least it has to be referred to a specific choice of the basis set. 
 
 Interestingly,    sometimes  the coupling to the environment is fully responsible of selecting the final states among the infinite possible choices of the isolated system. The  states  with  this property receives the name of  pointer states~\cite{Zurek_revmodphys_2003}. This is known as einselection process~\cite{Zurek_revmodphys_2003} and it bears obvious resemblance  with the quantum measurement problem.

In the context of half-integer  anisotropic spins, Kramers' theorem ensures  at least a double degeneracy of the energy spectrum at zero magnetic field.  Thus,  the transverse anisotropy Hamiltonian $E(S_x^2-S_y^2)$ can not lift the degeneracy of the $S_z=\pm S$ doublet when $S$ is half-integer. However, it does modify the wave functions.  The dimensionless parameter that controls the mixing is $\varepsilon=E/|D|$, that in the following is assumed to be small. 
We illustrate the process of eigenselection taking the case of a $S=3/2$ spin, but the discussion could be easily generalized.  
The wave functions of the ground state doublet can be written as: 
\beqa
\left\{
\begin{array}{l}
|C_1\rangle = \cos\frac{\theta}{2} |+3/2\rangle + \sin\frac{\theta}{2} |-1/2\rangle \\
|C_2\rangle = \cos\frac{\theta}{2}|-3/2\rangle - \sin\frac{\theta}{2} |+1/2\rangle \\
\end{array},
\right.
\label{basispm}
\eeqa
where 
  $\theta \approx \sqrt{3}\varepsilon+{\cal O}(\varepsilon^2)$. Thus, the mixing angle $\theta$ is imposed by the magnetic anisotropy.  Given that Kondo interactions can only connect states that differ in at most one unit of angular momentum $(\Delta S_z =\pm 1, 0)$, having $\varepsilon=0$  makes the scattering rate between states $|C_1\rangle$ and $|C_2\rangle$  impossible.

States (\ref{basispm}),  which hereafter we call {\em classical} states and thus the notation  $|C_1\rangle$ and $|C_2\rangle$,
have  a well defined magnetization along the easy axis of the system~\cite{Delgado_Loth_epl_2015}. 
However, any other pair of states linear combination of $|C_1\rangle$ and $|C_2\rangle$, such as  
 \beqa
\left(\begin{array}{c}
       | \hat{\Omega}_1\rangle \\
       | \hat{\Omega}_2\rangle 
      \end{array}
\right)=
\left(
\begin{array}{cc}
 \cos \frac{\phi }{2} & \sin \frac{\phi }{2}e^{i\xi} \\
 -\sin \frac{\phi }{2}e^{-i\xi} & \cos \frac{\phi }{2} \\
\end{array}
\right)
\left(\begin{array}{c}
       | C_1\rangle \\
       | C_2\rangle 
      \end{array}
\right),
\label{basisC}
\eeqa
where $(\phi,\xi)$ are spherical coordinates  in the unit sphere, is an equally valid choice of quantum states for the doublet. 
Notice that the angles $\phi$ and $\xi$ are a matter of choice, in contrast with the angle $\theta$, that is given by the Hamiltonian.
We now study how the decoherence induced by the  Kondo coupling to a metallic substrate depends on $\phi$. 
   Since the only role of the angle $\xi$ is to interchange the real and imaginary part of the coherences, in the following we just take $\xi=0$.

As noticed in Sec.~\ref{CohTLS}, the density matrix for a TLS has only 3 independent real quantities  that can be encoded in  the vector $\vec{P}$, see Eq. (\ref{P}). The BR master equation of a degenerate TLS can be then written as
\beqa
\frac{\partial\vec P(t)}{\partial t}=M\cdot\vec P(t), 
\label{difeqs}
\eeqa
where 
the matrix $M$  
is a functional of the Redfield coefficients, as shown in the Appendix \ref{apendixap}. For the  matrix $M$ it is convenient to introduce the notation
\beqa
M=\left[\begin{array}{ccc}
-\gamma/2 & M_{r,i} & M_{r,p} \\
M_{i,r} &-\gamma'/2 & M_{i,p}\\
M_{p,r} & M_{p ,i} & -\Gamma
\end{array}
\right],
\label{Mmatrix}
\eeqa
where $p$, $r$ and $i$ stand for populations, real and imaginary part of the coherences, respectively.\footnote{As we are interested in the relaxation and decoherence rates, here we neglect the induced energy shift (imaginary part of ${\cal R}_{12,12}$). This could be incorporated as a renormalized energy difference $\tilde \Delta$. } 
 Thus, in this notation $\Gamma$ stands for the transition rate between the eigenstates of ${\cal H}_S$ while $\gamma,\gamma'$ denote decoherence rates. From Eqs. (\ref{difeqs}) and (\ref{Mmatrix}), it is clear that the evolution of the occupations imbalance and real and imaginary parts of the coherences are fully decoupled from each other whenever $M_{k,j}=0$.  
In the case of the dissipative dynamics of the anisotropic $S=3/2$ spin that we are considering here, we obtain the following closed expressions for the entries of (\ref{Mmatrix}):
\beqa
\Gamma(\phi)
&\approx & 
\frac{9\pi k_B T}{4\hbar}  (\rho J)^2
\left[ 
\left(1-\cos (2 \phi )\right)+ \varepsilon^2 \left(1+3 \cos (2 \phi )\right)
\right]
\label{popscatt}
\\
\gamma(\phi) & \approx &
  \frac{9\pi k_B T}{2\hbar}  (\rho J)^2
\left[
 \left(1+  \cos (2 \phi )\right) + 
\varepsilon^2 \left(1-3  \cos (2 \phi )\right)
 \right]
\\ 
\gamma'(\phi) &\approx&  \frac{9\pi k_B T}{\hbar}  (\rho J)^2
 \left(1+\varepsilon^2\right)
\\ 
M_{p,r} &= & 4M_{r,p} \approx -\frac{9\pi k_B T}{2\hbar}  (\rho J)^2
  \left(1-3 \varepsilon^2\right) \sin (2 \phi ) .
\label{offdiag}
\eeqa 
The rest of the  $M_{ij}$  entries are zero for this system.    From these equations,
  we see that the {\em classical basis}  ($\phi=0$) is special on three counts: 
\begin{enumerate}
\item It  minimizes the elastic population scattering (\ref{popscatt}). 
\item It  decouples the evolution of coherences and occupations 
\item It maximizes the decoherence rate. In other words, the environment is particularly efficient  destroying
coherent superpositions  established between the states $|C_1\rangle$ and $|C_2\rangle$. 
\end{enumerate}

For these reasons,  the choice of states $|C_1\rangle$ and $|C_2\rangle$ as a basis set to study this class of systems  is very convenient. Of course, other choices of basis set are also legitimate, but  make the description of the dynamics more complicated.

\subsection{Non-degenerate two level systems\label{nondg}}
Now we turn our attention to the type-Q systems,  for which the spectrum is non-degenerate and the ground state is a linear combination of two states with opposite magnetic moment.    The representation of
the density matrix in the basis of eigenstates of the spin Hamiltonian is thus unique. Within the BR picture, the coupling to the reservoir will lead to a steady state density matrix that is diagonal in the eigenstate basis, and thereby it will have build-in coherences when represented in the $C$ basis, as long as $k_BT$ is not much larger than the energy splitting $\Delta_{\rm QST}$.  It na\"ively looks like, within the BR theory, the coupling to the reservoir is not able 
to prevent the coherence, i.e., the linear superposition of two states with opposite magnetization.     In the following we see how this is not really true,  even within the BR picture.   

\subsubsection{Renormalization of the QST splitting \label{ShiftsSB}}

So far we have considered in some detail 
 how the coupling to the reservoir changes the dynamical evolution of the reduced density matrix.  There is another aspect of this coupling that plays an important role in the case of type-$Q$ ground state, namely, the renormalization of the energy differences in the spectrum.    This type of effect  was first observed in the context of atomic physics:
 the coupling of the hydrogen atom to the vacuum photon field results in the so called Lamb shift, a renormalization of the $1s-2p$ transition whose treatment requires to take care of the singularities
 and played a major role in the development of quantum field theory. 
 
 In the context of magnetic adatoms, the notion of renormalization of the energy levels was used to explain the correlation between the changes in the transition energy  between the $S_z=\pm 1/2$ ground state doublet and the $S_z=\pm 3/2$ excited states, on one hand, and the intensity of the zero bias Kondo peak on the other.   It was argued that~\cite{Oberg_Calvo_natnano_2013,Delgado_Hirjibehedin_sc_2014}, given that the  height of the Kondo peak depends on $\rho J$ and it was correlated to the inelastic transition energy, this energy shift had also to depend on $\rho J$. The perturbative BR theory would naturally account for this energy shifts.  The same theory can also be applied to integer spins with quantum spin tunneling.  For instance, for a spin $S=1$ described with Hamiltonian (\ref{Hde})   with $D<0$  and $E\neq 0$, the splitting is renormalized according to~\cite{Delgado_Hirjibehedin_sc_2014}:  
\begin{equation}
\Delta_{\rm 0,1}= \Delta_0 \left(1-  \frac{3}{2} (\rho J)^2  \ln \left(\frac{2 W}{\pi k_B T}\right) \right)
\label{renorS1}
\end{equation}
where $W$ is the bandwidth of the itinerant electrons that appears as an ultraviolet cut-off in the theory~\cite{Oberg_Calvo_natnano_2013}.

An obvious consequence of Eq. (\ref{renorS1}) is that there is a critical value of $\rho J$ at which the splitting vanishes. However, the BR perturbative approach used  to get this results may break down before this critical value, which together with the presence of the logarithmic
 term  calls for a critical questioning of the validity of the perturbative approach.   Non-perturbative numerical calculations based on the One Crossing approximation for a multi-orbital  Anderson model that includes magnetic anisotropy  confirmed the validity of the perturbative calculation both for half-integer~\cite{Oberg_Calvo_natnano_2013} and integer~\cite{Jacob_Rossier_arXiv_2015}. In the last case,  the QST splitting goes to zero as $\rho J$ increases above a threshold value.    

A more elegant nonperturbative description of this phenomenon can be obtained from the bosonization procedure described in Appendix \ref{bosoni}.  This permits us to map the problem of an individual quantum spin exchanged coupled to an electron gas into the spin-boson model~\cite{Delgado_Loth_epl_2015}.  
The mapping between the two models then leads to the following identity between the coupling of the TLS and the baths~\cite{Delgado_Loth_epl_2015}: 
\beq
\alpha \equiv \left|\langle Q_1|S_z|Q_2\rangle\right|^ 2 (\rho J)^2.
\label{alphaM}
\eeq
Making use of the well known results for the  spin boson model,  Eq. (\ref{deltaSB}), one can see that the QST vanishes as  $\alpha$ goes beyond one~\cite{Delgado_Loth_epl_2015}.  In the limit of small $\alpha$, 
we can  expand Eq. (\ref{deltaSB}) in a Taylor series to yield
\beeq
\Delta= \Delta_0 \left(1-\alpha\ln \frac{\hbar \omega_c}{\Delta_0} \right).
\label{DeltaTayl}
\eeq
The result  (\ref{DeltaTayl}) basically reproduces the perturbative result (\ref{renorS1}), with the main difference that $k_BT$ within the Log function  has now been replaced by $\Delta_0$, the bare zero-field-splitting.
The main and completely general result is that the coupling to the environment  reduces the quantum spin tunneling splitting and, if sufficiently strong, it  completely cancels it.

\subsubsection{Perturbative Dynamics of the non-degenerate TLS \label{dynND}}
In Sec.~\ref{KondoT2} we have already studied  the dynamics of the reduced density matrix describing an effective two-level degenerate spin interacting with an electron gas via a Kondo interaction.  Now we tackle this  
question in the case of a non-degenerate TLS, with an energy splitting $\Delta\neq0$.  
In particular, by using the BR theory, we will analyze in detail the situation when the relaxation rate $\Gamma$ is comparable to the splitting $\Delta$.

We focus in the case when $\Delta$ arises from quantum spin tunnelling. 
Without loss of generality, we assume here that it corresponds to an anisotropic spin described by Hamiltonian (\ref{Hani}).
The dissipative dynamics is generated by the Kondo coupling,  given by Eq. (\ref{VQ}), where $\hat \tau_x$ induces transitions between the two eigenstates of the TLS.    Importantly, in the $|C\rangle$ basis Kondo scattering does not produce scattering.  In the limit when $\Delta$ is very small,  it is convenient to keep in mind these two complementary points of view, as we discuss below. 

Since we are considering a non-degenerate system, 
 coherences and occupations evolves in a fully independent way.  For the specific decoherence mechanism considered here, 
  the whole DM dynamics is determined uniquely by two parameters, the splitting $\Delta$ and the relaxation rate $\Gamma$.

\begin{figure}[t]
  \begin{center}
    \includegraphics[height=0.8\linewidth,angle=-90]{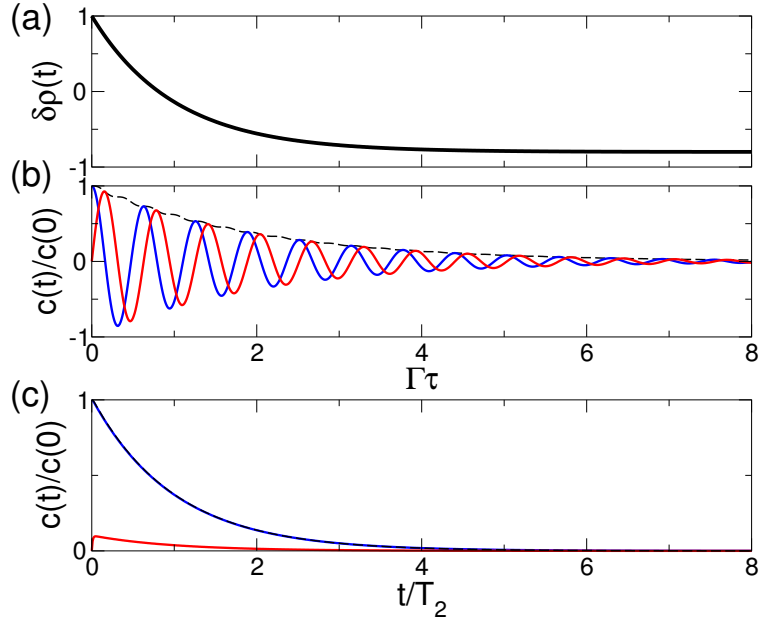} 
  \end{center}
  \caption{Time evolution of the density matrix of an effective two level system corresponding to an integer spin. (a) Occupations difference for $\Gamma_{1,2}=0.1\Gamma$ and $\Gamma_{2,1}=0.9\Gamma$. (b) and (c)  Real (blue) and imaginary (red) parts of the coherence $c(t)$ in the underdamped ($\Delta=10\hbar\Gamma$) and overdamped ($\Delta=0.1\hbar\Gamma$) regimes respectively.  Notice that in panel (c) the time axis has been scaled by $T_2 \approx 100/\Gamma$.  In all cases, $\delta\rho_0=1$, $c(0)=1/2$.
  }
\label{fig1}
\end{figure}

We first derived the evolution of the occupation. 
Simple analytical expressions for the dynamics of $\hat\rho(t)$ can be obtained for this particular case. If the initial occupation imbalance is $\delta\rho_0$, the population difference $\delta\rho=P_{\uparrow}-P_{\downarrow}$ satisfy
\beqa
\delta \rho (t)&=& e^{-t\Gamma}\left(\delta\rho^0+\frac{\delta\Gamma}{\Gamma}\right)
-\frac{\delta\Gamma}{\Gamma},
\label{evodP}
\eeqa
where $\delta\Gamma={\cal R}_{22,11}-{\cal R}_{11,22}$. 
This simple time evolution is depicted in Fig.~\ref{fig1}a).
  As expected, the steady state occupations is only determined by the ratio between Boltzmann factors, ${\cal R}_{22,11}/{\cal R}_{11,22}$. Notice that the evolution tends to the one of a degenerate TLS in the classical basis as $\Delta\to 0$.

We now consider the solution of Eq. (\ref{difeqs}) for the coherences, which 
can behave in a quite different way depending on the relation between $\Delta$ and $\Gamma$. This gives rise
 two different regimes: an {\em underdamped oscillatory regime} for $\hbar^2\Gamma^2-4\Delta^2<0$ and an  {\em overdamped oscillatory regime} for $\hbar^2\Gamma^2-4\Delta^2>0$.

\vspace{0.3cm}
{\em Underdamped regime: $\hbar^2\Gamma^2<4\Delta^2$}\\
In this regime the splitting $\Delta$ is the largest energy scale. Coherences $C_{\uparrow\downarrow}$ are given by
\beqa
\frac{C_{\uparrow\downarrow}(t)}{C_0} &=& \frac{ e^{-t\Gamma/2}}{\vartheta} \Big[\sin \frac{\vartheta  t}{2} (\Gamma e^ {-i\beta} -2i\frac{\Delta}{\hbar}  e^ {i\beta})
+\vartheta  e^ {i\beta}  \cos \frac{\vartheta  t}{2}\Big],
\label{rhoR}
\eeqa 
where we have defined   $C_{\uparrow\downarrow}(0)=C_0 e^{i\beta}$ with $C_0>0$ and we have introduced the  rate 
$\vartheta^2=|\Gamma^2-4\Delta^2/\hbar^2|$.
We thus see that the coherence oscillates with an amplitude that  decays with  a characteristic time
\begin{equation}
\frac{1}{T_2}= \frac{\Gamma}{2}= \frac{1}{2T_1}.
\end{equation}
The  oscillation  period $T=4\pi/\vartheta$ is  different from $\frac{\hbar}{\Delta}$,  a renormalization that also 
 happens in an underdamped classical harmonic oscillator. 
This situation is illustrated o Fig.~\ref{fig1}b) where, in addition to the real and imaginary part of $C_{\uparrow\downarrow}(t)$, we plot the envelope $|c(t)|$.

\vspace{0.3cm}
{\em Overdamped regime: $\hbar^2\Gamma^2>4\Delta^2>0$}\\
We now consider the case where $\Gamma$ is the dominant energy scale. 
For $t\gg 2/\vartheta$,  we can write
\beqa
\frac{C_{\uparrow\downarrow}(t)}{C_0}
\approx   \vartheta^{-1} e^{-t /T_2} \left[
\Gamma e^ {-i\beta}+e^ {i\beta}\left(\vartheta -2\Delta/\hbar\right)\right],
\label{overdamp}
\eeqa
with
\beqa
\frac{1}{T_2}\equiv \left(\frac{\Gamma-\vartheta }{2} \right)
\label{T2rate}
\eeqa
The coherence no longer oscillates and,  importantly,  the timescale for the decoherence depends both on $\Gamma$ and on $\Delta$. 
In the limit of small $\Delta$,  $\hbar \Gamma \gg \Delta$ we can write down
\beqa
\frac{1}{T_2}
\approx \frac{\Delta}{\hbar}\left(\frac{\Delta}{\Gamma \hbar}\right)\ll \Gamma.
\label{T2rate}
\eeqa
 Therefore, a strong suppression of decoherence rate $1/T_2$ occurs in this regime, as depicted in Fig.~\ref{fig1}c). This rather counterintuitive results can be better understood if we reinterpret the results in the classical basis as explained in the next section.
 
\subsubsection{Density matrix in the classical basis for the split TLS}

Equations (\ref{evodP}-\ref{rhoR},\ref{overdamp})  describe the diagonal ($\delta \rho$) and off-diagonal ($C_{\uparrow,\downarrow}$)  parts of the  density matrix for the split TLS  expressed in the basis of eigenstates (i.e., the basis (\ref{basisC}) with $\phi=\frac{\pi}{2}$).   We can write down the same density matrix in   the basis set of the classical states: 
 \begin{eqnarray}
\hat\rho_C(t)=\left[\begin{array}{cc} \frac{1}{2}+{\rm Re}[C_{\uparrow,\downarrow}(t)]& \frac{\delta\rho(t)}{2}+i{\rm Im}[C_{\uparrow,\downarrow}(t)] \\ \frac{\delta \rho(t)}{2}-i{\rm Im}[C_{\uparrow,\downarrow}(t)]  & \frac{1}{2}-{\rm Re}[C_{\uparrow,\downarrow}(t)]\end{array}
\right]_{C},
\label{initrho2}
\end{eqnarray} 
where the subindex ``$C$" denotes that $\hat\rho$ is written in the classical basis of states $|C_{1,2}\rangle$. 
This equation shows how   population scattering in the Q-basis ($\phi=\frac{\pi}{2}$), which leads to the decay of $\delta \rho$,  results in a decay of the off-diagonal terms of the DM  expressed in the classical basis.  
  Analogously,  decoherence in the Q-basis  can be interpreted in terms population transfer in the $C$ basis.  
 This result permits us
to  re-interpret the Eqs. (\ref{evodP}-\ref{overdamp}) of the dynamics of the split TLS obtained.
  For the system considered here,  angular momentum conservation prohibits direct Kondo scattering between the classical states. On the other hand,  the coherent evolution of the system permits, via quantum spin tunneling $\Delta$, connecting the classical states, giving rise to  Rabi oscillations in the diagonal of the density matrix (\ref{initrho2}).  
  In the overdamped regime these oscillations are suppressed, and result in an incoherent population transfer governed by a  exponential decay  with $T_2$ given by Eq. (\ref{T2rate}).   This mechanism for population transfer combines thus the coherent quantum spin tunneling and the dissipative coupling to the environment, and has been proposed by  
   Gauyacq and Lorente~\cite{Gauyacq_Lorente_jp_2015} to explain the switching between N\'eel states observed for AF chains of Fe atoms on Cu$_2$N~\cite{Loth_Baumann_science_2012}.     Following  Ref.~\cite{Gauyacq_Lorente_jp_2015}, we refer to this mechanism as the decoherence assisted switching , although perhaps it would be more accurate to refer to it as quantum tunneling assisted switching.

\subsubsection{Non-perturbative derivation of the decoherence assisted switching
\label{Nico}}

We now provide an independent derivation for the decoherence assisted switching mechanism. For that matter, we use again the mapping of the Kondo model for the split TLS to the spin boson model of Eq. (\ref{SBQ}).    Within this model, 
the time-dependent spin autocorrelation of the system is given by $P(t)=\langle \hat \sigma_z(t)\rangle$,  where $\sigma_z=\pm 1$ labels the two classical states. We  assume that at $t=0$ the system is prepared in the $\sigma_z=+1$ state. 
 Using the noninteracting-blip approximation~\cite{Leggett_Chakravarty_rmphys_1987}, one gets $P(t)= g(t)e^{-t/\tau}$ where $g(t)$ is an oscillating function (constant) of order unity for $\alpha<1/2$ ($\alpha\ge 1/2$), and
\beqa
\tau^{-1}=(\Delta^2/\hbar^2\omega_c)
\frac{\sqrt{\pi}}{2}\frac{\tilde\Gamma(\alpha)}{\tilde\Gamma(\alpha+1/2)}
\left(\frac{\pi k_B T}{\hbar\omega_c}\right)^{2\alpha-1},
\label{tausb}
\eeqa
where $\tilde\Gamma(x)$ is the Gamma function. Expression (\ref{tausb}) is valid whenever $\alpha\ge 1$, for which $\Delta$ is renormalized to zero [see Eq. (\ref{deltaSB})],  or, for $\alpha<1$, if 
\beqa
\alpha k_B T\gg \Delta\left(\frac{\Delta}{\hbar \omega_c}\right)^{\alpha/(1-\alpha)}.
\eeqa
In both cases, $\Gamma\gg \Delta$, so that we are in the overdamped limit discussed above.  
Equation (\ref{tausb}) already shows that the switching time $\tau$ is proportional to $ \Delta^2$. This result is in line with the perturbative result in Sec. \ref{genpertnd}. In fact, when the coupling constant $\alpha$ goes to zero, we get that Eq. (\ref{tausb}) takes the simple form
\beqa
\tau^{-1}\approx \frac{\Delta^2}{2\hbar\pi \alpha k_bT},
\label{itau}
\eeqa
which taking into account that $\Gamma=2\pi \alpha k_bT/\hbar$, reproduces the perturbative result (\ref{T2rate}) to lowest order in the coupling $\alpha$.

\subsection{Relaxation and decoherence in spin chains and ladders  \label{genpertnd}}

We now address the problem of spin relaxation and decoherence of finite size spin chains due to their Kondo coupling to a nearby electron gas.   Many of the results obtained in the previous sections will be useful. 
  An important  concept to keep in mind is that  decoherence affects quantum states, rather than the spins. Whereas in the single spin case this distinction if less important, in the case of multi-spin chain it does play a key role. For instance,  a single scattering event between two states might flip  the spin of all the atoms in the chain.
The problem of spin relaxation of a Heisenberg-coupled spin array due to Kondo exchange  is  formally connected  with the so called Kondo lattice model, where otherwise independent itinerant fermions are exchanged coupled to a lattice of quantum spins~\cite{Zachar96,Hewson_book_1997,neto2000non}.   Even if direct exchange between the local spins is sometimes not explicitly written down in the  the Kondo lattice Hamiltonian,  indirect exchange interactions emerge, which can compete with the Kondo correlations~\cite{neto2000non}.  The Kondo lattice model is a rich many-body problem, with several different types of electronic order~\cite{Zachar96}.    Here we treat the Kondo coupling as a perturbation within the BR theory. In addition, the systems of interest have a rather small number of spins. 
These two simplifications  make the problem   easier than the Kondo lattice model  and tractable by means of numerical diagonalization.  
On the other hand,  in order to model magnetic adatoms we have to include the effect of magnetic anisotropy, sometimes overlooked in the Kondo lattice model.  As we discussed above,    spin chains and ladders do behave very differently depending on the interplay between magnetic anisotropy of individual spins, exchange interactions, the number $N$ of spins in the system,
and the parity of $N$,    as discussed in Sec. \ref{multi}.    This  affects both $T_1$ and $T_2$.

We now discuss two different problems.  First, we discuss  how the spin decoherence times $T_2$  of an Ising spin chain depends on the properties of the spin chain, such as the number of atoms $N$, and the atom-atom distance $d$. Second 
we discuss the  role played by Kondo induced decoherence in the  emergence of classical behavior in spin chains made of atoms that,  when isolated,  have finite quantum spin tunneling splitting.

\subsubsection{$T_2$ for broken symmetry states in spin chains}
In the context of magnetic adatoms, the BR theory has been applied to study the spin relaxation $T_1$, and to a lesser extend $T_2$,  of a variety of finite size spin chains and ladders  due to Kondo exchange~\cite{Delgado_Rossier_prb_2010, Gauyacq_Lorente_prb_2013,Delgado_Loth_epl_2015,Gauyacq_Lorente_jp_2015}.     We discuss here the decoherence time of Ising chains without spin-flip terms in the Hamiltonian, 
so that $S_z$ is a good quantum number, and the ground state is doubly degenerate. We consider the case of AF coupling, so that the lowest two states, which we label as $1$ and $2$,
 correspond to  the two  classical N\'eel state.  Hence, we can write the matrix elements of the local spins operators as $S^a_{11}(l)=-S^a_{22}(l)=S(-1)^l$.    Following our recent work~\cite{Delgado_Rossier_inprep},  and using Eq. (\ref{invT2chain}), the relevant form factor from Eq. (\ref{chiadb}) for a chain of atoms lying along the $x$ axis is given by:

\beq
\chi_{1,2}^ {ad.}(\vec{q})=S^2 \left|\sum_n (-1)^n  \left(e^{i q_x na}+e^{-i q_x na}\right)\right|^2.
\label{structure1DAF}
\eeq 
The resulting pure decoherence rate is given by~\cite{Delgado_Rossier_inprep}:
\beq
\frac{1}{T_2^*} 
 \approx 
\frac{\pi \left(\rho {\cal J}\right)^2}{8\hbar}
k_BT \Lambda^{\rm AFM}(k_Fd,N).
\label{T2AFM}
\eeq
where ${\cal J}=J|S^z_{1,2}|^2$ and $\Lambda^{\rm AFM}(k_Fd,N)$ is a dimensionless function that represents the average of the structure factor (\ref{structure1DAF}) over the Fermi surface, and it is therefore different for fermions in $D=1,2,3$ dimensions.  For $D>1$ and  $k_F d>1$, i.e., in the limit where the interatomic distance is larger than the Fermi wavelength,
relevant for metals, the function $\Lambda^{\rm AFM}(k_Fd,N)\simeq N$  with some small oscillations as a function of $k_F d$~\cite{Delgado_Rossier_inprep}.  Thus, {\em the decoherence rate of an Ising spin chain with $N$ spins is $N$ times quicker than the single spin, Eq (\ref{pureT2})}.

The origin of the oscillations of the decoherence time as a function of $k_F d$ is interesting in itself~\cite{Delgado_Rossier_inprep}.  Within the BR theory, it is apparent that both decoherence and energy shifts in an open quantum system are two sides of the same coin: the reactive and dissipative response of the system to the coupling with a reservoir.  The reactive coupling results in a shift of the energy levels. In the case of a spin chain coupled to an electron gas, this shift of the energy levels is nothing but the RKKY interactions~\cite{Delgado_Rossier_inprep}, which is known to oscillate as a function of $k_Fd$.  Therefore, the oscillations in $T_2$ are expected, being mathematically related to the RKKY interaction.   They can also be understood in terms of phase shifts of the quasiparticles. As we discussed  in Sec.~\ref{phaseshiftsec},   the pure dephasing rate  is proportional to the variation  in the quasiparticle scattering phase shift  for the two states of the spin(s).  Elastic scattering between waves with wavenumber $k_F$ and a structure with period $d$ is expected to depend in an oscillatory manner on $k_F d$, on account of multiple scattering interference.  

This type of effect depends crucially on   the phase factor $e^{i\left(\vec k-\vec k'\right)\cdot \vec r_l}$ that appears in the Kondo Hamiltonian,  overlooked by some previous work in the context of magnetic adatoms.    It's omission is equivalent to  ignore the local nature of Kondo coupling, $\sum_n \vec{S}_n\cdot \vec s(\vec{r}_n)$ and  replace it by a coupling with the total spin $\vec s(0)\cdot\sum_n \vec{S}_n$.   This coupling commutes with the total atomic spin operator, $\vec{S}_T\equiv \sum_n \vec{S}_n$, and it is thereby  not capable of producing transitions between eigenstates with different $S$.   As  a result, the Kondo interaction without the phase factors can not induce spin transitions between chain states with different total spin $S_T$.  

\subsubsection{The quantum to classical transitions in spin chains}
In the seminal work of Loth {\em et al.}~\cite{Loth_Baumann_science_2012}, by using spin-polarized STM experiments, it was shown how chains of $N>6$ Fe atoms at $T\sim 1$ K deposited on Cu$_2$N  would acquire spontaneous local,  forming the N\'eel state,  expected on account of the AF interaction of Fe atoms along the nitrogen rich direction. This AF character was further confirmed by an accurate exploration of the Fe dimer~\cite{Bryant_Spinelli_prl_2013}.  
In addition, a time-resolved tracking of the magnetization made it possible to observe random telegraph noise with two states, revealing switching between the two N\'eel states. As noted by Loth and coworkers,  the system provided a unique opportunity to study the transition between the classical behavior, observed for chains,  and the quantum behavior of the single Fe atom in the same surface~\cite{Delgado_Loth_epl_2015}  inferred from the IETS spectroscopy~\cite{Hirjibehedin_Lin_Science_2007}.

In this system, both the local magnetic anisotropy~\cite{Hirjibehedin_Lin_Science_2007} as well as the Fe-Fe exchange~\cite{Bryant_Spinelli_prl_2013} are well determined from IETS. This provides an accurate description in terms of the Hamiltonian~\cite{Delgado_Loth_epl_2015} 
\begin{equation}
{\cal H}= \sum_l D \hat S_z(l)^2 + E\left(\hat S_x(l)^2 -\hat S_y(l)^2\right)  + J \sum_{l=1,N_1} \vec{S}(l)\cdot\vec{S}(l+1).
\end{equation}

\begin{figure}[t]
  \begin{center}
       \includegraphics[width=0.5\linewidth,angle=0]{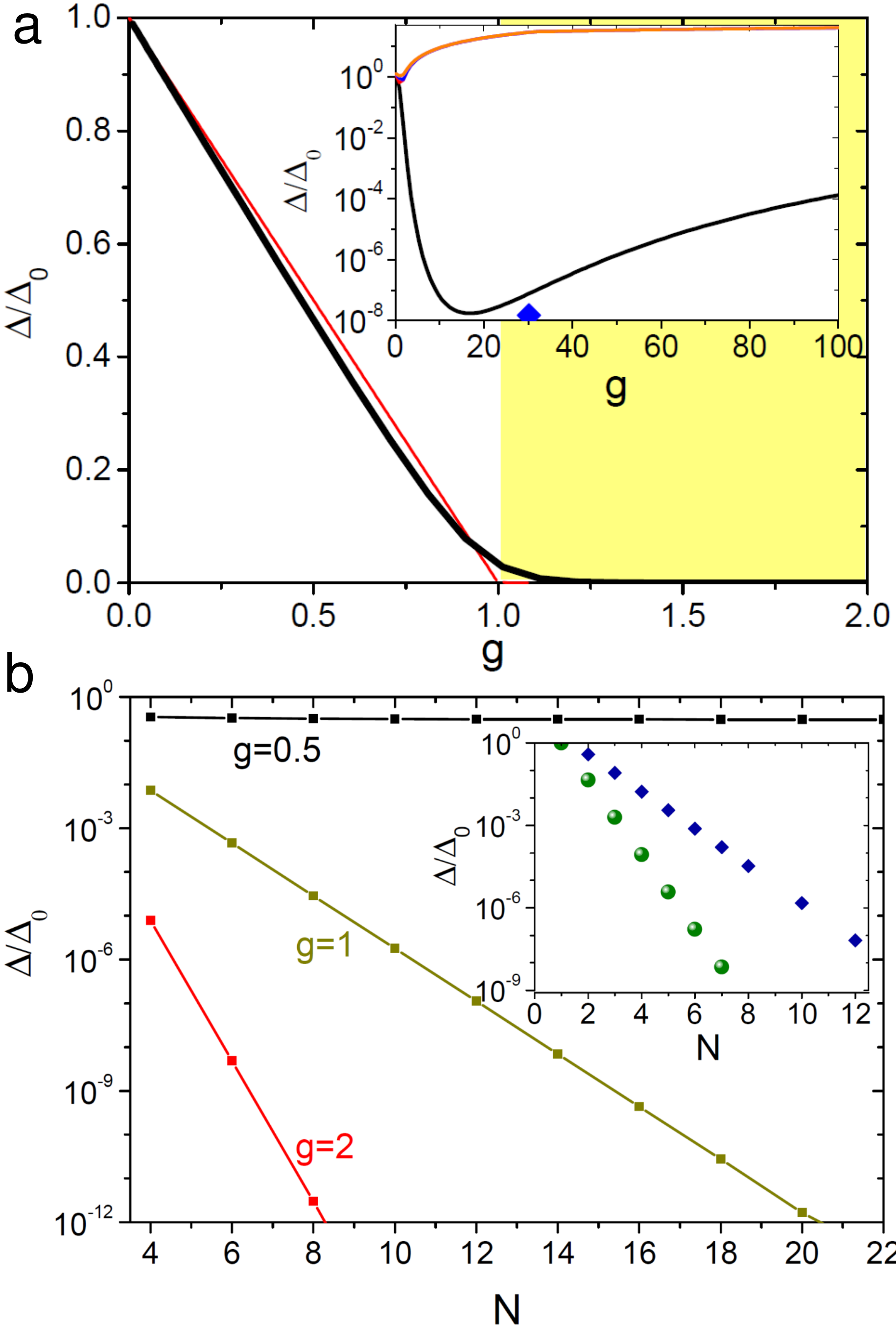}  
  \end{center}
  \caption{Reprinted with permission from F. Delgado {\em et al.}, Europhysics Letters {\bf 109}, 57001 (2015) (a) QST splitting of the Ising chain vs. $g=2|j_H|/\Delta_0$ for a $N = 20$ chain (black line) and the infinite chain (red line). $g_c=1$ marks the quantum phase transition. Inset: QST splitting of the $S = 2$ Heisenberg spin chain together with higher-energy excitations (orange lines) vs. $g$ for the Fe chains with $D=-1.5\ \text{meV}$ and $E=0.3\ \text{meV}$ (the diamond marks the experimental condition~\cite{Loth_Baumann_science_2012,Bryant_Spinelli_prl_2013} with $g\approx 27$ ). (b) Chain size dependence of $\Delta$ in the QIMTF for $g=0.5<g_c$ (weak size dependence), and $g=1,\;2$ (exponential dependence that leads to a type-C ground state for large $N$). Inset: size dependence of $\Delta$ for the experimental parameters, both for the AF~\cite{Loth_Baumann_science_2012,Bryant_Spinelli_prl_2013} and FM chains~\cite{Spinelli_Bryant_natmat_2014}, showing an exponential dependence.
  }
\label{figEPL}
\end{figure}

In the $J=0$ limit,  this interaction describes an ensemble of independent anisotropic spins. For $S=2$, the case relevant for Fe on Cu$_2$N,  the ground state is unique, and the expectation value of the atomic spin operators, $\vec{M}(l)\equiv \langle \vec{S}(l)\rangle$ is strictly  zero.  In the opposite limit of very large $J$, we can ignore the anisotropy, and we have again a unique ground state, with $S=0$, and null $\vec{M}(l)$ for all atoms in the chain.    In order to reproduce the phenomenology observed experimentally, with two equally likely  N\'eel states, with finite $\vec{M}(l)$,  we would need a doubly degenerate ground state. Mathematically, this situation appears for instance in the limit $E=0$ and $|D|\gg J$.   For the values of $D$, $E$ and $J$ obtained from IETS,   we can compute  the energy difference 
$\Delta_0$ between the ground state and first excited state, and we can study their wave functions.     Interestingly, we obtain that 
$\Delta_0$ decreases exponentially as a function of  $N$, as shown in the Fig.~\ref{figEPL}. Moreover, the wave functions of the ground and first excited states are bonding and anti-bonding combinations of N\'eel states.  Thus, the combination of single-ion anisotropy and exchange interactions  shrink $\Delta_0$. From the numerical diagonalization with the best fitting parameters we obtain that $\Delta_0(N)\approx  \Delta_{\rm Fe} e^{-N/0.74}$~\cite{Delgado_Loth_epl_2015}, where $\Delta_{\rm Fe}\approx 0.15$ meV is the zero field splitting of the single Fe.

Thus,  the energy scale that protects the quantum behavior of finite size spin chains decreases exponentially as the number of atoms in the chain increases, but it does not cancel. This can be seen in the case of the quantum Ising model in a transverse field.  Even when the AF interaction dominates, leading to the doubly degenerate ground state in the thermodynamic limit,  for finite size chains there is a small splitting~\cite{Delgado_Loth_epl_2015}.  However,  when the splitting is small enough,  we can argue that the  classical behavior for the spin chain takes over, on account of the dissipative Kondo coupling with the surface electrons.  This switching can be better understood in the N\'eel basis  as we discussed in Sec. \ref{Nico}.   There we showed
that the decoherence (of the N\'eel states) in this limit is $N$ times faster the one of an individual atom. For instance, $T_2(N=8)$ is in the 100 ps range, whereas the Rabi time $\hbar/\Delta(8)$ is well above microseconds.  Thus, we are in the limit $\hbar T_2^{-1} \gg \Delta(N)$.   This accounts for the observed random telegraph noise with switching times slower than milliseconds~\cite{Gauyacq_Lorente_jp_2015}. 

\section{Other spin relaxation mechanisms: photons, phonons and nuclear spins}

We now consider  additional sources of spin relaxation and decoherence affecting spins intrinsic to magnetic atoms surfaces.   In the previous section we have considered a mechanism that  is characteristic from metals, namely, Kondo interactions.  Given that in many instances the underlying metal and the  magnetic atom  are separated by an atomically thin insulating material,   we briefly discuss the dominant spin relaxation mechanisms for magnetic atoms in insulating hosts:  spin-phonon coupling and hyperfine interactions with nearby nuclear spins.  In addition to these, there is another mechanism that is known to result in spin decoherence: a conductor creates a stochastic magnetic field due to the unavoidable thermal fluctuations of the current around its zero average (Johnson noise).   These mechanisms  could be particularly relevant when the Kondo interaction is suppressed, something that might be achieved using superconducting substrates~\cite{Heinrich_Braun_natphys_2013,Nadj_Perge_science_2014}, or decoupling layers with less transparent tunnel barrier, achieved either with a thicker spacer or a wider band-gap material, such as MgO~\cite{Rau_Baumann_science_2014,Baumann_Paul_science_2015,Donati_Rusponi_science_2016,
natterer2016}  or h-BN~\cite{Jacobson_Herden_natcom_2015}.

\subsection{Spin-phonon coupling \label{SPh}}
Spin-phonon interaction ${\cal V}_{\rm s-ph}$ is an important  source of spin relaxation for paramagnetic centers in insulating materials where the density of itinerant carriers is negligible, blocking Kondo exchange.  It can be the dominant mechanism when the density of spin centers is small, so that dipolar coupling is negligible. 
The spin-phonon (SP) interaction also affects the parameters in the static
spin Hamiltonian, giving rise to shifts in the $g$-value, the fine structure
splitting, and the hyperfine interaction~\cite{Abragam_Bleaney_book_1970}. In addition, phonons may also induce
a spin-spin interaction between ions. Such effects arise even at zero temperature from the zero-point vibrations of the lattice, and though they increase at finite temperatures, they are often rather small.

Relaxation processes mediated by phonons involve the emission or absorption of a
quantum by the spin system, which should be absorbed or emitted by the lattice vibrations.
Thus, a transition between two spin levels will be driven efficiently if the lattice can produce such an
oscillatory electromagnetic field. 
One possible mechanism is the one proposed by Waller~\cite{Waller_zp_1932}: 
the local dipolar magnetic field created by neighboring ions, which depends on the ions distance, fluctuates due to the lattice vibration. 
A second process, more relevant in diluted magnetic centers, consists of modulation
of the crystal electric field or ligand field through motion of the electrically
charged ions under the action of the lattice vibrations~\cite{Abragam_Bleaney_book_1970}, essentially a dynamic
orbit-lattice interaction, which influences the spin levels through the
spin-orbit coupling.

Spin relaxation due to one-phonon emission scales with the density of states of phonons at the spin transition energy $\Delta$:\footnote{Being a small perturbation, one can apply the Bloch-Redfield approach of Sec.~\ref{BRF} to study the relaxation and decoherence processes induced by  ${\cal V}_{\rm s-ph}$.}
\beqa
1/T_1\propto \rho_{ph}(\Delta)=\Delta  n_B(\Delta) \Sigma(\Delta),
\eeqa
 where $\Sigma(\Delta)$ is the number of phonon modes per unit volume in a frequency
range $\Delta/\hbar,(\Delta+d\epsilon)/\hbar$ and $n_B(\Delta)$ the Bose-occupation factor. 
For the phonon radiation bath one has, taking into account that there are two transverse polarizations an one longitudinal wave motion:
\beqa
\Sigma_\Delta=\frac{3\Delta^2}{\hbar^2\pi^2}\left(\frac{2}{v_t^3}+\frac{1}{v_l^3}\right),
\eeqa
where $v_t$ and $v_l$ are the transverse and longitudinal 
wave velocities.

The SP induced relaxation rates involves matrix elements of the crystal and ligand electric field perturbations. Thus, the specific form of ${\cal V}_{\rm s-ph}$ depends in general on the particular phonon mode and the symmetry of the environment of the magnetic atoms~\cite{Abragam_Bleaney_book_1970}. 
It always contains an even power of atomic spin operators, like the zero field single ion Hamiltonian,   on account of time reversal symmetry, and it is linear in the atomic displacement operator.
 Chudnovsky and coworkers~\cite{Chudnovsky_Garanin_prb_2005} made
a particularly elegant derivation of a universal spin-phonon coupling for the transverse phonons.
As very often the matrix elements also scale with the energy, a $T_1^{-1}\propto \Delta^3 /v^{5}$  is commonly found~\cite{Abragam_Bleaney_book_1970,Chudnovsky_Garanin_prb_2005}, with $v$ the 
wave velocity.
As typically the ratio between the transversal and longitudinal wave velocities satisfies $v_t/v_l \lesssim 0.7$~\cite{Abragam_Bleaney_book_1970}, the contribution from longitudinal modes can be usually neglected. 
A particularly simple an beautiful result found by Chudnovsky and coworkers~\cite{Chudnovsky_Garanin_prb_2005}  is that the SP relaxation rate at zero applied field is given by
\beqa
\frac{1}{T_1^{ph}}=\frac{\left|\Xi\right|^2}{12\pi\hbar^4}\frac{\Delta^5}{\rho_m v_t^5 },
\eeqa
where $\rho_m$ is the mass density and $\Xi$ is a dimensionless matrix element of the spin operator.\footnote{Here we have defined the matrix elements $\Xi$ as dimensionless parameters, contrary to Ref.~\cite{Chudnovsky_Garanin_prb_2005}  }

The application of the BR theory for spin relaxation of a spin center due to spin-phonon coupling  has been implemented
by Leuenberger {\em et al.}~\cite{Leuenberger_Loss_prb_2000}, which considered the case of  uniaxial molecular magnet  crystals. The same analysis could be applied for individual magnetic atoms with the single ion Hamiltonian (\ref{Hde}).
The quadratic spin operators that enters into the  general form (\ref{intera})
can be written in that case as:
\beqa
{\cal S}_\alpha \equiv q_1(\alpha) S_+^ 2+q_2 (\alpha)S_-^ 2+q_3(\alpha) S_+ S_z+ q_4(\alpha) S_-S_z +h.c.,
\label{sphGen}
\eeqa
where $q_i(\alpha)$ are numerical coefficients that depends on the single ion parameters $D$ and $E$. By using
a Bloch-Redfield master equation to study the spin dynamics, they obtained an excellent agreement with experimental data in a Mn$_{12}$-acetate crystal~\cite{Leuenberger_Loss_prb_2000}. 
Ganzhorn {\em et al.}~\cite{Ganzhorn_Klyatskaya_natcom_2013} also study the effects of the SP coupling of a TbPc$_2$ molecular spin  with a longitudinal phonon mode  in  a  carbon nanotube and suggested that it could induce the suppression of quantum tunnelling of magnetization, similar to the quenching of the QST by Kondo exchange explained in Sec.~\ref{ShiftsSB}.

%
%
%

It is particularly interesting to consider the effect of the spin-phonon coupling on the low energy doublet for the dominant uniaxial term $D  S_z^ 2$. In that case, given the form (\ref{sphGen}), one gets that after the projection in the subspace spanned by $|Q_1\rangle$ and $|Q_2\rangle$, ${\cal V}_{\rm s-ph}=0$, i.e., the spin phonon coupling can not induce relaxation or decoherence between the low energy states $|Q_i\rangle$.
Similarly, in the case of a half-integer spins with Hamiltonian (\ref{Hde}), the pure decoherence between the classical states $|C_1\rangle$ and $|C_2\rangle$ also cancels.  
In other words, the spin-phonon coupling does not provide an efficient pure-decoherence mechanism for these spins.  Of course, spin-phonon coupling can still produce decoherence via inelastic events, but these can be thermally suppressed.

\subsection{Hyperfine interactions}

Hyperfine interactions  account for the spin coupling between electronic and nuclear degrees of freedom.
It has two main components:  the so called Fermi contact interaction term,  that is only non-zero whenever the electronic spin density is non-vanishing at the nuclear site, i.e., for electrons with a finite $s$-wave component.   In addition, there is the 
dipole-dipole coupling, that can give a contribution for  the hyperfine interaction of $p$, $d$ and $f$ electrons with the nuclear spin in the same atom.  

Spin decoherence due to nuclear spins is known to play a major role in the case of electrons confined in semiconductor quantum dots~\cite{Awschalom_Loss_book_2013}. In that case, a single electron visits thousands of nuclei.
For the atomic spins considered here, contact interaction is not particularly relevant because the magnetic moment lies mostly on $d$ electrons, for which the contact term vanishes, and also because the local moment is localized in just one atom.  Hyperfine interaction with the same-atom nuclear spin will result in a small splitting of the energy levels, the so called  hyperfine structure. 
 For instance, for the $d$ electrons of $^{51}$Mn, one has $A\approx 0.3-1\;\mu$eV~\cite{Walsh_Jeener_pr_1965},  while for the $I=9/2$ nuclear spin of the $^{209}$Bi embedded in Si  $A\approx 6.1\;\mu$eV~\cite{George_Witzel_prl_2010}.
These splittings  might be resolved with IETS at extremely low temperature~\cite{Delgado_Rossier_prl_2011} and are definitely   within reach with the spectral resolution achieved with STM-ESR~\cite{Baumann_Paul_science_2015}.  In the case of Fe, only $^{57}$Fe has a nuclear spin $I=1/2$, and a natural abundance of 2 percent. 
The hyperfine coupling for $^{57}$Fe$^{3+}$ in bulk MgO~\cite{Abragam_Bleaney_book_1970} is $A\approx 0.1\;\mu$eV.

The dipolar coupling of a single electronic spin with an ensemble of surrounding spins could lead to decoherence. For an electronic spin $\vec{S}$ located at the origin, the dipolar interaction with 
the nuclear spins $\vec{I}_i$, located at $\vec{r}_i$, takes the form
reads:
\begin{equation}
{\cal H}_{\rm dip}=-\frac{g_I \mu_N g_s \mu_B}{4\pi}\sum_i \frac{1}{r_i^3}
\left(
3\left(\vec S\cdot \hat r_i\right)
 \left(\vec I_i \cdot \hat r_i\right)-\vec S\cdot \vec I_i
\right) .
\label{Hdipol}
\end{equation}
If we treat the nuclear spins as classical moments, we could write down the dipolar coupling Hamiltonian for the electronic spin as ${\cal H}_{\rm dip}=g\mu_B\sum_a S_a H_a$, where $H_a$ would be the $a$ component of the nuclear spin field. If we assume this field takes random values, associated to the statistical thermal fluctuations of the nuclear spins, it is possible to estimate the electronic spin decoherence time, using Eq. (\ref{T2sto}).  The magnetic field created by a proton at 3$\AA$ is $40\;\mu $T.  
Thus, we can estimate a lower limit for $\frac{g\mu_B }{\hbar}\simeq  6$ kHz.  If we assume that the inverse  nuclear spin correlation time $\tau_0$ is also in that range,  this will give decoherence times in the milliseconds time scale. Of course, this is a very rough estimate of the order of magnitude.

\subsection{Spin relaxation and decoherence due to Johnson noise}
 
 We have just seen that nuclear spins, as sources of random magnetic fields, can induce decoherence of remote spins.   Conductors are known to be sources of random magnetic fields, so that is not surprising that they 
 have been identified as possible sources of relaxation and  decoherence for cold atom spins~\cite{Scheel_Rekdal_pra_2005}, electronic spin qubits~\cite{Morello_pla_nature_2010}, spins in quantum dots~\cite{Poudel13}, Phosphorus donors in Si~\cite{Muhonen_Dehollain_natnano_2014}, 
  and for $S=1$ NV centers on surfaces~\cite{kolkowitz2015}.
  
   The random magnetic fields close to a conductor are created by the thermally induced statistical fluctuations of the current, the so called Johnson noise~\cite{johnson1928,nyquist1928}.
   In a quantum language, the electronic currents in the conductor emit photons that interact with the remote spins.    
    Modulations of the transverse (longitudinal) component of the magnetic field result in relaxation (decoherence), as inferred from the general equations (\ref{T1sto}) and (\ref{T2sto}). 
 The spin relaxation and decoherence rates are thus directly proportional to the amplitude of the transverse and longitudinal components of the magnetic noise, at the spin-transition frequency $\Delta/\hbar$  in the case of $T_1$, and at zero frequency for $T_2$.  
When the motion of the electrons in the conductor is diffusive ($k_BT \gg \Delta$),  and 
when the conductor is relatively far from the spin, so that the far field approximation is valid,\footnote{This condition holds when the skin depth of the metal $\delta_e$ is  much larger than the distance $d$  between the metal and the magnetic impurity. For instance, for the Ag surface used in Ref.~\cite{kolkowitz2015}, this corresponds to $d\ll 1\;\mu$m, an easily achievable condition.} 
an expression for the magnetic noise spectral density $S_B^z$ can be easily computed~\cite{kolkowitz2015}:
 \begin{equation}
 S_B^z=\left(\frac{\mu_0}{4\pi}\right)^2\frac{k_BT\sigma_e}{d},
 \end{equation}
 where $\sigma_e$ is the electric conductivity.  Applying Fermi's golden rule one can arrive to a rather simple analytical result for the $T_1$ relaxation due to Johnson magnetic noise~\cite{Poudel13,kolkowitz2015}:  
\begin{equation}
 \frac{1}{T_1}=\frac{3g^2 \mu_B^2}{2\hbar} S_B^z.
 \end{equation}

A controlled experiment with a wedge shape conductor has been carried out,  where
the $T_1$ of NV centers  could be measured as a function of their distance to the conductor, finding
spin relaxation rates in the range of  $1/T_1\simeq 1/3\; {\rm ms}^ {-1}$ for $d=50$ nm,
results in good agreement with this theory .  A naive scaling of this result to distances of $d \approx 0.5$ nm, would lead to $T_1 \sim 3\;\mu s$ at the same temperature.  However, at that point the far field expression no longer applies and a more sophisticated treatment of the evanescent components of the field becomes necessary~\cite{Scheel_Rekdal_pra_2005,Poudel13}.
Qualitatively, the correlation time of the magnetic noise in this regime is determined by the ballistic time of flight of electrons through the relevant interaction region, resulting  
in a saturation of the noise spectral density (and the spin relaxation rate) as either the NV approaches the surface or the mean free path becomes longer at lower temperatures, with its limiting value given by~\cite{kolkowitz2015}:  
 \begin{equation}
 S_B^z=\frac{2\mu_0^2 k_B T }{\pi} \frac{ne^2}{m_e v_F}
 \end{equation}
 where $m_e$ is the electron effective mass and $v_F$ is their Fermi velocity.   Again, good agreement with the experiments was found in this limit  when  $NV$ centers are at $d\approx 4$ nm, with $T_1$ in the range of $0.5$ ms  at $T=27$ Kelvin.   Whereas a more detailed analysis is probably necessary to assess how this mechanism affects 
 magnetic adatoms,  perhaps including the quantum  effects in the Johnson noise, these experimental results provide a good starting point for the order of magnitude.  In addition,  the shot noise associated to current flow across the STM-surface junction will also contribute to the magnetic noise mechanism.

\section{Experimental methods}

In this section we briefly overview the different methods that are used to probe atomic spins on surfaces  with STM, with emphasis in spin dynamics. 

\subsection{Single spin Inelastic Electron Tunnelling Spectroscopy}

The technique of inelastic electron tunnelling spectroscopy was first applied in tunnel junctions back in the sixties~\cite{Jaklevic_Lambe_prl_1966,Lambe_Jaklevic_pr_1968}.  Three decades  later it was implemented with an STM, first  for single molecule  vibrational excitations~\cite{Stipe_Rezaei_science_1998},  and a few years later,  for the spin excitations of individual magnetic atoms~\cite{Heinrich_Gupta_science_2004,Hirjibehedin_Lutz_Science_2006,Hirjibehedin_Lin_Science_2007}.   In all cases the working principle is the following: 
when   the excess bias voltage energy, $eV$,  of a transport electron is larger than the transition energy $\Delta$ of some other localized degree of freedom in the barrier, an inelastic transport channel opens in which   the electron  tunnelling is accompanied by the  inelastic excitation of the local degree of freedom.   The experimental fingerprint is a step $\Delta G$ in the $dI/dV$ 
  at the  bias voltages $V=\pm \Delta/e$. Quite frequently, the second derivative $d^2I/dV^2$ is plotted, which offers an increased contrast. The inelastic transitions are then identified as characteristic peaks (dips) at $eV=\Delta$  ($eV=-\Delta$).

A shift of the excitation energies as a function of an applied magnetic field provides 
a clear fingerprint of the magnetic nature of a given excitation~\cite{Heinrich_Gupta_science_2004,Hirjibehedin_Lutz_Science_2006,Hirjibehedin_Lin_Science_2007}.  
Interestingly, in many of these magnetic excitations, 
the magnitude  of  $\Delta G$ is of the same order of magnitude than the zero bias conductance. This can be accounted for
by the fact that both elastic and spin-flip inelastic events occur via 
 cotunelling  ~\cite{Delgado_Rossier_prb_2011}. By contrast, the vibrational steps are frequently quite small, specially for large molecules. 
Very often  the evolution of the $dI/dV$  spectra for various magnetic fields permits one to infer the effective spin Hamiltonian of a given structure~\cite{Hirjibehedin_Lutz_Science_2006,Hirjibehedin_Lin_Science_2007,  Khajetoorians_Chilian_nature_2010} providing a starting point for subsequent modelling.

The spectral resolution provided by this method is limited by the unavoidable thermal smearing of the Fermi distributions of the quasiparticles in the electrodes,  given by $5.4 k_B T$~\cite{Jaklevic_Lambe_prl_1966,Lambe_Jaklevic_pr_1968}.  In addition, the amplitude of the lock-in potential provides an additional broadening, although is normally smaller than the thermal smearing.

In a first seminal experimental work by S. Hirjibehedin {\em et al.}~\cite{Hirjibehedin_Lin_Science_2007}, it was shown that height of the inelastic step in the $dI/dV$ was proportional to $\sum_{a=x,y,z} \left|\langle i |S_a |f\rangle \right|^2$, where $i$ and $f$ are the initial and final states of the spin Hamiltonian. 
 This proportionality establishes a spin selection rule $\Delta S_z = 0, \pm 1$.
Later it was demonstrated that this spin selection rule is a consequence of the Kondo exchange interaction that dominates the inelastic scattering~\cite{Rossier_prl_2009}.

In the case of structures with several magnetic atoms, the $dI/dV$ spectra at different atoms have the steps at the same bias voltage, reflecting the collective nature of the spin excitations, but the height of the steps $\Delta G$  at atom $n$ is expected to be proportional  to 
 $\sum_{a=x,y,z} \left|\langle i |S_a(n) |f\rangle \right|^2$~\cite{Rossier_prl_2009}, which can be strongly spatially modulated~\cite{Spinelli_Bryant_natmat_2014}. This provides a tool to image the spin excitations in these structures,  that has been  used   to image the spin waves in ferromagnetically coupled Fe spin chains~\cite{Spinelli_Bryant_natmat_2014}, and in Cobalt spin chains on Cu$_2$N~\cite{Bryant_Toskovic_nanol_2015}. Dramatic variations of the intensity of the Kondo peak have also been reported in Mn-Fe$_N$ spin chains~\cite{Choi_Robles_arXiv_2015}.

\subsection{Spin polarized STM}

The technique of spin polarized STM (SP-STM), pioneered by R. Wiesendanger~\cite{wiesendanger1990,Wiesendanger_revmod_2009},  was the first spin-sensitive STM based probe. A spin-polarized tip and a magnetic surface form an atomic-scale magnetic tunnel junction.  When the relative orientation of the magnetic moments of tip and sample can be controlled independently, by application of an external magnetic field in most instances,  the conductance of the system can present changes, providing thereby a spin-dependent signal.   The technique has been used to explore a variety of surfaces with different types of magnetic order~\cite{Wiesendanger_revmod_2009} and, more relevant  to this review,  the magnetization of individual magnetic atoms deposited on non-magnetic surfaces~\cite{Meier_Zhou_Science_2008,Zhou_Wiebe_natphys_2010,Zhou_Meier_prb_2010,Khajetoorians_Lounis_prl_2011,
Khajetoorians_Wiebe_natphys_2012,Khajetoorians_Schlenk_prl_2013,Khajetoorians_Steinbrecher_natcomm_2016}.

The early approach to achieve spin polarizaton in  SP-STM  was the use of tips of magnetic materials.  More recently,  the use of non-magnetic tips with just a few magnetic atoms in the apex of the tip, picked from the surface, has been demonstrated as a viable alternative~\cite{Loth_Bergmann_natphys_2010,Loth_Etzkorn_science_2010,Loth_Baumann_science_2012,
Spinelli_Bryant_natmat_2014,Baumann_Paul_science_2015}
to obtain magnetic contrast, although this approach requires the application of a magnetic field to freeze paramagnetic fluctuations of the small magnetic cluster in the tip.

\subsection{Methods to determine $T_1$\label{T1exp}}

Several different techniques have been applied to measure, or to infer,  the spin relaxation time of individual magnetic atoms using a STM.    In the case when the intrinsic broadening $\hbar T_1^{-1}$ associated to the finite spin lifetime is larger than the thermal splitting,  it  can be extracted from the full width at half maximum of the $d^2I/dV^2$ spectra. Using this approach, the lifetime of Fe atoms on Cu(111)  was measured to be
$T_1\approx \hbar/(2\Delta E)\approx 200$ fs at $0.3$ K~\cite{Khajetoorians_Lounis_prl_2011}. In addition,  Khajetoorians {\em et al.},  observed a linear scaling of the spin relaxation rate $1/T_1$  with the excitation energy, tuned with the magnetic field,   in agreement with  Eq. (\ref{T1relax}).

A second way to infer $T_1$ from the IETS spectra is based on the experimental results by 
 Loth and coworkers~\cite{Loth_Bergmann_natphys_2010}, that observed how some inelastic steps appear when the current was increased by reducing the tip-surface distance.  The connection with $T_1$ is the following: when the pace at which the current  excites the spin, given by $I_{\rm inelastic}/e$,  is faster than the pace at which the atomic spin relaxes, $T_1^{-1}$,  a non-equilibrium occupation of the excited atomic spin state $|X_1\rangle$  builds. This makes it possible to observe new inelastic steps corresponding to transitions from $|X_1\rangle$ to higher energy excited states $|X_2\rangle$.   The inelastic current can be then estimated as $I_{\rm inelastic}/e\simeq \Delta G V$, where $\Delta G$ is the height of the inelastic step associated to the primary inelastic transition, from ground state to $|X_1\rangle$.      This method works better when the excitation energy $E_{|X_2\rangle}-E_{|X_1\rangle}$ is larger than the primary excitation $E_{|X_1\rangle}-E_{|G\rangle}$. In the case of a Mn dimer on a Cu$_2$N/Cu(100) surface, $T_1$ can be estimated around $30$ and $5$ ps for the first and second excited states~\cite{Loth_Bergmann_natphys_2010}.

The pioneering implementation, in 2010,  of electrical pump and probe measurements~\cite{Loth_Etzkorn_science_2010} with a spin polarized STM provided a   direct measurement of $T_1$ for  an individual magnetic  entity, a dimer of  Fe and Cu on top of Cu$_2$N.  The working principle is the following: an electrical bias pulse capable of driving the atomic spin out of equilibrium is applied on the tip-surface junction.  A second smaller probe is send, after a time lapse $\tau$, and the conductance is measured.  As long as $\tau<T_1$, the atomic spin is still excited, and the magnetoresistive component of the conductance is different from the one prior to the pump pulse.   
This permitted a direct access to spin-transitions occurring on a time scale of a few tens of ns~\cite{Loth_Etzkorn_science_2010}. This technique has been applied for instance to study the relaxation time of a single Fe and Co atom on MgO/Ag(100)~\cite{Rau_Baumann_science_2014,Baumann_Paul_science_2015}.
 or small arrays of magnetic adatoms on Cu$_2$N~\cite{Loth_Baumann_science_2012,Spinelli_Bryant_natmat_2014,yan2016non}.

\begin{figure}[t]
  \begin{center}
       \includegraphics[width=1.\linewidth,angle=0]{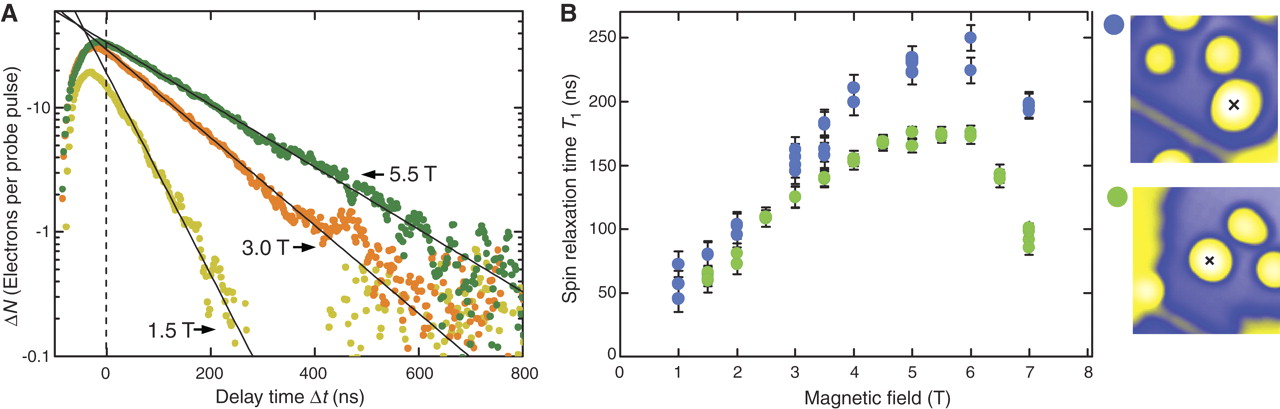}    
  \end{center}
  \caption{Spin relaxation times of an Fe-Cu dimer on a Cu$_2$N/Cu(100) substrate measured by pump-probe scanning tunnelling microscopy. From S. Loth {\em et al.}, Science  {\bf 329}, 1628 (2010). Reprinted with permission from AAAS. (A) Pump-probe measurements for different magnetic fields on an Fe-Cu dimer. (B) $T_1$ as a function of magnetic field for the two Fe-Cu dimers shown in the accompanying 5-nm by 5-nm STM topographs.} 
\label{fig2}
\end{figure}

Yet another method to determine $T_1$ is possible for magnetic structures with two magnetic states that are stable enough to resolve their random telegraph noise using a spin polarized tip~\cite{Wiesendanger_revmod_2009,Loth_Baumann_science_2012,Spinelli_Bryant_natmat_2014}. This requires that the switching time is slower than the time it takes to record the conductance, typically $1$ ms. 
This is the case of long Fe spin chains on Cu$_2$N,  coupled either AF~\cite{Loth_Baumann_science_2012} or ferromagnetically~\cite{Spinelli_Bryant_natmat_2014}.  The histogram of the switching times $\tau_s$ can be fit to an exponential function that allows one to fit the switching rate, $T_1^{-1}$.   In the case of Fe chains on Cu$_2N$, the switching rates of antiferromagnetic chains decreases dramatically for smaller chains,  making it impossible 
to resolve the switching dynamics of individual Fe atoms with conventional measurements. Remarkably,  the spin dynamics of individual Ho atoms on MgO has been recently  probed using this method~\cite{natterer2016}, thanks to the extremely long switching time of Ho on this surface.
Independent XMCD experiments for  Ho adatoms on  MgO  demonstrate the appearance of magnetic remanence for a single atom up to 30 Kelvin, with relaxation times exceeding 1500 seconds at 10 K~\cite{Donati_Rusponi_science_2016}. 
 A similar claim had been done for Ho on top of Pt~\cite{Miyamachi_Schuh_nature_2013}. 
However, later XMCD measurements revealed no evidence of magnetic stability and a different ground state~\cite{Donati_Singha_prl_2014}, while
 more recent SP-STM and IETS-STM measurements have not found any evidence of magnetic moment of Ho on  Pt(111)~\cite{Steinbrecher_Sonntag_natcom_2016}.

\subsection{Determination of $T_2$ via EPR-STM}

The implementation of electron paramagnetic resonance of an individual magnetic atom using an STM~\cite{Baumann_Paul_science_2015} has been one of the most recent dramatic developments in the field of scanning probes.  In a conventional EPR experiment,  a static field $B_0$ splits the spin levels of the target atoms, and the microwave ac magnetic field
drives spin transitions.  When the applied field $B_0$  is tuned to set a spin transition in resonance with the driving field frequency $\omega_0$,  the absorption is maximal. In c.w. experiments, the absorption of  microwaves  is probed as a function of $B_0$, resulting in spectra with narrow resonance lines that provide valuable information about the local spins.  Fitting this curve with the steady state solution of Bloch equation permits one to infer 
$T_2$~\cite{Abragam_Bleaney_book_1970}.  EPR  is extensively used in a variety of fields, including  biology, chemistry or physics.  The spectral resolution of EPR is limited by both instrumentation and by the intrinsic line width of the target spins. Spectral resolutions in the range of a few MHz permit probing the hyperfine structure~\cite{morley2005,George2013}, vastly superior to the spectral resolution of STM. The down side of conventional EPR is spatial resolution, or sensitivity: state-of-the-art EPR setups  handle volumes  larger than  $1\; \mu {\rm m}^3$, and the minimal number of spins that can be detected is in the range of $10^{7}$~\cite{blank2003high}.

In order to carry out a single spin resonance experiment with an STM it is necessary both to drive the spins with an $ac$ signal and to be able to probe their response. These are two challenging requirements for a charge sensitive instrument.  In order to drive the spin, Baumann {\em et al.} applied an ac. voltage in the 20-30 GHz range superimposed to the $dc$ bias with an STM. They applied this signal on top of an individual Fe atom on a MgO(100) layer grown on top of Ag~\cite{Baumann_Paul_science_2015}. The coupling to the spin is believed to occur~\cite{Baumann_Paul_science_2015,Lado_Ferron_arxiv_2016,Berggren_Fransson_srep_2016} via a combination of several effects: the tip electric field slightly distorts the position of the Fe atom, that in turn changes the crystal field of the atom; in combination with the spin-orbit coupling,   transitions are induced between the two lowest energy states of the system. 

The detection relies on the spin sensitivity of the tip that hosts one magnetic atom whose orientation is fixed by the large external magnetic field. For a fixed value of the dc bias,  the current of the STM-tip-surface junction is scanned as a function of the RF frequency $f$,  giving a $I(f)$ curve.   A very narrow peak, with a full width at half maximum  $\delta f \simeq 21$ MHz,  was found when $f$ is in resonance with the lowest energy excitation $f_0 = \Delta/h$. Application of a small  off-plane magnetic field tunes $f_0$, resulting in a shift of the peak.  Importantly,  as the frequency of the RF is changed,  Baumann {\em et al.}~\cite{Baumann_Paul_science_2015} varied   the power of the  external RF source in order to maintain a constant RF amplitude $V_{RF}$ at the STM tunnel junction.

  Spin contrast is verified by pump-probe experiments 
  that are used to determine $T_1$, in the range of 100$\;\mu s$, much longer than the 0.2 ps of Fe on Cu(100)~\cite{Khajetoorians_Lounis_prl_2011}, and also Mn dimers on Cu$_2$N (5-30 ps). This highlights how  the decoupling layer can increase  dramatically the spin relaxation time. 

The determination of $T_2$, together with the  driving Rabi frequency $\Omega$,   is done through fitting of the $I(f)$ curves for various amplitudes $V_{RF}$  to the steady state solution of the Bloch equation for a driven two level system [Eq. (\ref{SzB})].  Values of $T_2\simeq 210\pm 50$ ns were found, and $\Omega \simeq 2.6\pm 0.3 {\rm rad}/\mu s$~\cite{Baumann_Paul_science_2015}.     The energy resolution of this experiment is 10 nanovolts, outperforming by 3-4 orders of magnitude the one obtained with IETS.  This setup can be used as magnetometer, as we discuss in the next subsection, thanks to its  remarkable sensitivity to  tiny variations  of magnetic field.

\subsection{Quantum technologies: quantum sensors with magnetic adatoms}

The combination of all these single spin probing and manipulation techniques is opening now the possibility to devise
various sensing strategies.  In the case of the STM-EPR experiment,  the resonance curve of the Fe atom can be used as a probe for the magnetic moment of atoms nearby.  Taking advantage of the atomic manipulation capabilities of STM, it is possible to change the distance between the detection Fe atom  under the tip, and the atoms nearby.   When both the probe and target atoms are Fe on MgO,  the shift in the resonance of the probe atom is  expected to be given by
 $\mu_0  m_{Fe}^2/(4\pi d^{3})$, where $m_{Fe}$ is the magnetic moment of the Fe atom and $d$ is their distance, that can be determined with pm resolution~\cite{choi2016magnetic}.   This permits an absolute measurement of the magnetic moment of Fe on this surface. Once the probe magnetic moment is calibrated,  it can be used to measure the magnetic moments of different atoms and structures. This has been used to probe the magnetic moment of Ho atoms on MgO~\cite{natterer2016}.  In this case, the spectral resolution is limited by the intrinsic $T_2$  of Fe on MgO.

A second  example  is the magnetometer designed by  Yan and coworkers~\cite{yan2016non}, a nanoengineered
Fe spin chain on Cu$_2$N that acts as a probe, using the
linear scaling between the spin relaxation rate $T_1^{-1}$ and its transition energy $\Delta$.   The  magnetic field created by   a second nanoengineered structure, at a approximately $3$ nm,   results in a modification of the transition energy of the probe $\delta \Delta$, that in turn, modifies the $T_1$ time of the probe, measured with the pump-probe technique.   This setup can be used to detect the random telegraph switching of the source nanostructure, provided it is much longer than   the $T_1$  of the probe structure.  The working principle of this setup benefits from a rapid spin relaxation of the probe.

\section{Outlook  and conclusions\label{discc}}

A conclusion of this review is that the  exploration of the spin coherence of surface spins by means of STM is  just starting but has a promising future ahead.    The development of the field will depend crucially on the capability to increase the relaxation and coherence times
of the surface spins. Given that Kondo interactions are the dominant source of spin relaxation,  
this translates into a reduction of $\rho J$, the product of the density of electronic states at the Fermi level and the Kondo exchange constant.  The reduction of $J$ can be achieved using thicker decoupling layers with a wider gap. MgO and BN seem very promising materials in this regard. 

Another route to reduce the effect of Kondo interactions is to use superconducting substrates, for which the density of states at the Fermi energy vanishes. In fact, it has been shown that 
the spin dynamics of individual magnetic adatoms  slows down on superconducting substrates~\cite{Heinrich_Braun_natphys_2013}.   The  modification   of the  the nuclear spin relaxation follows a non-monotonic curve, well understood within BCS theory. Future theory work should address this problem in the limit when the spin excitation energy is larger than the superconducting gap,  in contrast with the  nuclear spin case, as well as the theory for $T_2$ due to the coupling with the superconductor.

  The use of graphene as a conducting substrate might be helpful to reduce $\rho$ and increase $T_1$ thereby~\cite{Cervetti_Rettori_natmat_2016}, on account of the vanishing density of states at the Dirac points, together with the small density of nuclear spins.  In addition, graphene has is own very interesting spinful point defects, such as chemisorbed hydrogen~\cite{gonzalez2016} and zigzag edges~\cite{ruffieux2016}.
 
In addition to the reduction of $\rho J$, there are other more subtle ways to reduce the effect of Kondo exchange.   The spin relaxation rate of the Kondo exchange is controlled by the symmetry of the atomic spin wave functions, that in turn depends on the symmetry of the substrate.  A proper choice of the adsorbate magnetic moment, the adsorption site and the surface can provide 
an almost full quench of the Kondo induced relaxation~\cite{Miyamachi_Schuh_nature_2013,Hubner_Baxevanis_prb_2014,Donati_Rusponi_science_2016}. Multi-spin structures  have been predicted to provide additional  opportunities to tune the spin relaxation~\cite{Delgado_Rossier_inprep16}.

Finding other magnetic atoms and surfaces that permit STM-EPR experiments,  beyond Fe on MgO, will be  be another milestone for the development of this field. From the theory side it will be important to understand in detail the mechanisms that make it possible to couple $ac$ electric fields to the spin of the surface atoms~\cite{Nussinov_Crommie_prb_2003,Balatsky_Nishijima_adphys_2012}.    Another challenge for theory is to have a realistic description of the system on two counts. First,  to have  a proper quantum spin model for a given system, inferred from experiments or derived from first principles~\cite{Ferron_Lado_prb_2015}, or a combination of both~\cite{Baumann_Donati_prl_2015}.   Second,  addressing the problem of the dissipative coupling of the surface spins to their environment.  This problem can become particularly challenging in the case of strong  Kondo coupling that lead to the Kondo effect.  On the other hand, in that  limit we expect a very strong decoherence, and from this perspective, it is a less interesting limit.  Finally,   a theory for the STM current as a full functional of the atomic spin density matrix across, including coherences,  might be necessary to describe transport in the  presence of the RF field.  

 The roadmap for the experimental development of the field of
  coherent manipulation of surface spins with STM will be inspired by the accomplishments with NV centers and P donors in silicon.   For that matter, it would be convenient  to extend the pump-probe techniques~\cite{{Loth_Etzkorn_science_2010}} to the coherent domain,  to be able to perform coherent control  experiments with pulsed RF perturbations.     Time will tell how far can we make it in the development of quantum computers based on atomic surface spins or if  the remarkable feats afforded by the exquisite quantum control of individual spins can be combined with the amazing  potential of STM to engineer spin structures.


\section*{Acknowledgements}
We acknowledge fruitful discussions and private communications ,  with
 A. Ardavan, S. Baumann,  M. A. Cazalilla A. Ferr\'on,  A. Heinrich,  C. Hirjibehedin,  N. Lorente, S. Loth, A. Morello,  
  A. F. Otte, and J. Wrachtrup.
JFR acknowledges financial supported by MEC-Spain
(FIS2013-47328-C2-2-P) and Generalitat Valenciana
(ACOMP/2010/070), Prometeo.
This work is funded by ERDF funds through the Portuguese Operational Program for Competitiveness and InternationalizationÐ COMPETE 2020, and National Funds through FCT- The Portuguese Foundation for Science and Technology, under the project ``PTDC/FIS-NAN/4662/2014" (016656).
FD acknowledges funding  by the Ministerio de Econom\'{i}a y Competitividad (MINECO, Spain), with grant MAT2015-66888-C3-2-R..  
We thank J. Sinova for financial support in the  SPICE  workshop on ``Magnetic adatoms as building blocks for quantum magnetism". FD acknowledges hospitality of the Departamento de F\'{i}sica Aplicada, Universidad de Alicante.

\appendix

\renewcommand*{\thesection}{\Alph{section}}

\section{Bloch-Redfield tensor ${\cal R}_{mm',nn'}$ \label{appendixa}}
In this Appendix we provide the general expressions for the population scattering rates ${\cal R}_{nn,mm}$ ($n\ne m$), decoherence rates ${\rm Re}\left[{\cal R}_{nn',mm'}\right]$ ($n\ne n'$ and $m\ne m'$), and energy shifts 
${\rm Im}\left[{\cal R}_{nn',mm'}\right]$.
For that matter, it is particularly convenient to introduce the reservoir correlation function 
$g_{\alpha\beta}(t)= \langle R_\alpha(t)R_\beta(0)\rangle\left. \right|_{eq} $, which can be  recast as
\begin{eqnarray}
&&g_{{\alpha}\beta}(t)\equiv \sum_{r}  P_r \sum_{r'}
R_\alpha^{rr'} R_\beta^{r'r} e^{i(\epsilon_r-\epsilon_{r'})t/\hbar},
\label{gDef}
\end{eqnarray}
where $R^{rr'}_\alpha=\langle r|R_\alpha|r'\rangle$.
Here  we introduced the reservoir eigenvectors $|r\rangle$ associated to the eigenvalues $\epsilon_r$,  together with the thermal occupations $P_r$. The different transition amplitudes and energy shifts appearing in the Bloch-Redfield tensor can be written in terms of the 
Fourier transform
\begin{equation}
g_{\alpha\beta}(\omega)=\int_{0}^ \infty dt\;e^ {-i\omega t}g_{\alpha\beta}(t)
\end{equation}
Thus, using Eq. (\ref{gDef}) together with the Sokhotsky's formula, one can write
\begin{equation}
g_{\alpha,\beta}(\omega)= \sum_{r}  P_r \sum_{r'} R_\alpha^{rr'} R_\beta^{r'r}
\left(\pi\delta\left(\omega-\omega_{rr'}\right)-i{\cal P}\frac{1}{\omega-\omega_{rr'}}\right),
\label{gDefome}
\end{equation}
where ${\cal P}$ stands for the Cauchy principal part.

\subsection{Population scattering $1/T_1\equiv \Gamma_{n,m}$ \label{T1gen}}
For $n\ne m$, one gets after the substitution in Eqs. (\ref{Rplus}-\ref{Rminus})  that
the scattering rate $\Gamma_{n,m}={\cal R}_{mm,nn} $ can be written in the form
\beqa
\label{ratesBRF0b}
&&\Gamma_{nm}=
\frac{2\pi}{\hbar^2}\sum_{\alpha,\beta} 
 \sum_{r} P_r \sum_{r'}  R_{\alpha}^ {rr'} R_\beta^ {r'r} S_\alpha^ {nm} S_{\beta}^ {mn}
 \delta\left(\omega_{mn}-\omega_{rr'}\right),
 \crcr
&&
\eeqa
or, in terms of the correlation function $g_{\alpha\beta}(\omega_{nm})$,
\beqa
\label{ratesBRF0b2}
&&\Gamma_{nm}=
\frac{2}{\hbar^2}\sum_{\alpha,\beta} {\rm Re}\left[g_{\alpha\beta}(\omega_{nm})\right]
S_\alpha^{nm} S_\beta^{mn}.
\eeqa 
Notice that this tensor elements are real and positive, as corresponds to transition rates between the eigenstates of the isolated system ${\cal H}_{\rm S}$. They reproduce the result of the Fermi Golden Rule.

\subsection{Decoherence rates $1/T_2$\label{T2gen}}
The evolution of the coherences $\rho_{nm}$  are dominated by the BR tensor components ${\cal R}_{nm,nm}$. This situation may change when there are other coherences $\rho_{n'm'}$ with degenerate Bohr frequencies, i.e., $|\omega_{n'm'}-\omega_{nm}|\ll 1/\delta t$, in which case one can use the general expressions of  ${\cal R}_{nm,n'm'}$.
For the components ${\cal R}_{nm,nm}$, one can generally write ${\cal R}_{nm,nm}=-\gamma_{nm}-i\delta\Delta_{nm}$, with $\gamma_{nm}$ ($\delta\Delta$) the real (imaginary parts). We can split the decoherence rate $\gamma_{nm}$ into a non-adiabatic component $\gamma_{nm}^{\rm nonad.}$, which involves scattering between different quantum system states, and a adiabatic one, where the system does not actually changes its state. For the non-adiabatic one gets
\beq
\gamma_{nm}^{\rm nonad.}=\frac{1}{2}\left( \sum_{n'\ne n}\Gamma_{n,n'}+\sum_{n'\ne m}\Gamma_{m,n'}\right).
\label{gnonad} 
\eeq
The genuine decoherence mechanism, responsible of the pure decoherence, is given by the adiabatic component
\begin{equation}
\gamma^{\rm ad}_{nm}= 
\frac{\pi}{\hbar^2}\sum_{r} P_r \sum_{r'}
\left|
\sum_{\alpha} R_\alpha^{rr'}
 \left(S_\alpha^{nn}-S_{\beta}^{mm}\right)
\right|^2 \delta(\omega_{rr'}) ,
\label{gammaad}
\end{equation}
which after introducing the correlator $g_{\alpha\beta}(\omega)$ can be written in the form (\ref{adiabaticbis4}).

Similarly, one can also evaluate the decoherence rate due to the coupling with other coherences. Using the general expressions (\ref{Rplus}) and (\ref{Rminus}), one gets for $n\ne n'$ and $m\ne m'$ the decoherence rates
\beq
{\rm Re}\left[{\cal R}_{nm,n'm'} \right]=
\sum_{\alpha\beta}\frac{2}{\hbar^2}
 {\rm Re} \left[  S_\alpha^{m'm} S_\beta^{nn'}
 \left( g_{\alpha\beta}(\omega_{nn'})
+g_{\alpha\beta}^*(\omega_{mm'})\right)\right].
\eeq

\subsection{Energy shifts \label{Shifts}}

The imaginary part of the BRF tensor leads to a modification of the bare Hamiltonian which can be accounted for as a new renormalized Hamiltonian. The energy shifts $\delta\Delta_{nm}\equiv \delta \omega_n-\delta \omega_m$ associated to ${\cal R}_{nm,nm}$ can be written in a similar way to the decoherence rates. Then, the general expression is given by
\beqa
\delta  \omega_m=
\frac{1}{\hbar^2}
\sum_{\alpha,\beta}
\sum_{n' }  
{\rm Im}\left[g_{\alpha,\beta}(\omega_{n'm})\right]
S_{\alpha}^{mn'} S_\beta^{n'm}.
\eeqa
In addition to this energy shifts induced by the rates ${\cal R}_{nm,nm}$,
 there can be additional terms coming from coupling with other coherences. By writing the Liouville operator as
\beqa
{\cal L}(\hat\rho) =   -\frac{i}{\hbar}\left[H_{\rm eff},\hat\rho\right]+{\cal R}^{\rm Re}(\hat\rho),
\eeqa
where ${\cal R}^{\rm Re}$ stands for the real part of the BR tensor, 
one can arrive to the following result~\cite{Breuer_Petruccione_book_2002}:
\beq
\langle n|H_{\rm eff}|m\rangle
=E_n \delta_{nm}+\sum_{\alpha\beta}\sum_{n'}{\rm Im}\left[g_{\alpha\beta}(\omega_{n,n'})\right]
S_\alpha^{nn'} S_\beta^{n'm} .
\eeq
Importantly, this effective Hamiltonian satisfies $\left[{\cal H}_S,H_{\rm eff}\right]=0$~\cite{Breuer_Petruccione_book_2002}.

\section{Bloch-Redfield tensor for the Kondo coupling \label{appendixa1}}

In this appendix we give explicit expressions for the Bloch-Redfield tensor in terms of the bath correlators.

In the case of the Kondo coupling with the conduction electrons of an electronic bath, which could be the metallic substrate or an STM tip in the case of magnetic adatoms, the bath operators $R_{\alpha}$ takes the form
\begin{equation}
 R_{\alpha} \equiv  
 \sum_{\sigma,\sigma'}   \sum_{\vec k,\vec k'} e^{i(\vec{k}-\vec{k'})\cdot\vec{r}_l}
\frac{\tau_{\sigma,\sigma'}^a}{2} c^{\dagger}_{\vec k\sigma}c_{\vec k'\sigma'} ,
\label{RKondo}
\end{equation}
where $\alpha\equiv \left(l,a \right)$, with $a=x,y,z$ and $l$ labelling the atoms while $S_{\alpha} \equiv  J(l) S_a(l)$.\footnote{When the system is coupled to more than one electronic bath at different chemical potentials, as it would be the case for adatoms subjected to a tunnel current as when studied by STM, the indexes $\alpha,\beta$ should also include an electrode index that must be summed up in ${\cal V}$. This will give place to intra-electrodes  and also inter-electrode scattering events~\cite{Delgado_Rossier_prb_2010}.}   

Considering fermionic bath states resulting from the creation of a single electron-hole pair, one can  obtain after some algebra that
\begin{eqnarray}
g_{\alpha,\beta}(\omega)&\equiv & \frac{\delta_{a,b}}{2} 
\sum_{\vec k\vec k'} 
  f(\epsilon_{\vec k}) \left(1- f(\epsilon_{\vec k'})\right)
e^{i\left(\vec k-\vec k'\right).\cdot \left(\vec r_l-\vec r_{l'}\right)}
e^{i\omega_{\vec k\vec k'}t}
\label{gKondo}
\end{eqnarray}
where $\beta\equiv (l',b)$ and $f(x)$ is the Fermi-Dirac occupation distribution.
 Here we have assumed for simplicity that the only reservoir quantum number are the spin $\sigma$ and the wavevector $\vec k$.  Notice that the assumption of null expectation value of the interaction is equivalent in this case to assume a spin-unpolarized electronic reservoir. 
Then, assuming a constant density of states $\rho$ at the Fermi level and using Eqs. (\ref{RKondo}-\ref{gKondo}), which leads to a analytical expression of the energy integrals, together with Eqs. (\ref{Rplus}-\ref{Rminus}), the real part of the BRF tensor can be then written as
\beqa
\label{ratesafbF}
{\cal R}_{nn',mm'}&=&\frac{\pi}{\hbar}\rho^2
\Big[
-\delta_{n',m'}\sum_{n''} \Lambda_{n,n'',n'',m}{\cal G}\left(\Delta_{mn''}\right)
\crcr
&&\hspace{-1.8cm}+\Lambda_{n,m,m',n'}
{\cal G}\left(\Delta_{mn}\right)
-\delta_{n,m}\sum_{n''}\Lambda_{m',n'',n'',n'}
{\cal G}\left(  -\Delta_{n''m'}\right)
\crcr
&&\quad+\Lambda_{n,m,m',n'}
{\cal G}\left( -\Delta_{n'm'}\right)
\Big],
\eeqa
with ${\cal G}(\Delta)=\Delta/(1-e^{-\beta\Delta})$ and $\beta=1/k_bT$. The spin-dependent matrix elements are defined as
\begin{eqnarray}
\Lambda_{nn',mm'}=\frac{1}{4} \sum_{l,l';a}J(l)J(l')\lambda_{ll'}(k_F)
S^a_{nn'}(l) S^a_{mm'}(l') 
\end{eqnarray}
where $S^a_{nm}(l)\equiv \langle E_n|S^{a}_l|E_m\rangle$. 
  In addition, we have introduced the factors 
$\lambda_{ll'}(k_F)$ that depends on the geometrical distribution of the spins together with the dimensionality of the electron gas and Fermi wavenumber. Assuming that in the small neighborhood of the Fermi level, where the product of the Fermi functions in Eq. (\ref{gKondo}) are non-zero, we can approximate $|\vec k|\approx k_F$, we have that
\beqa
 \lambda_{l,l'}(k_F)=\frac{1}{\Omega^2}\int d\hat k d\hat k' 
 \exp\left[\pm ik_F\left(\hat k-\hat k'\right) \cdot  \left(\vec r_l-\vec r_{l'}\right) \right],
 \label{lphase}
\eeqa 
with $\hat k=\vec k/|\vec k|$ and $\Omega=\int d\hat k$. The phase integral $\lambda_{l,l'}(k_F)$ is a function of the Fermi wavenumber $k_F$ and the dimensionality of the electron gas.
 For a linear chain of equidistant spins  one can find explicit analytical expressions of $\lambda_{ll'}(k_F)$~\cite{Delgado_Rossier_inprep16}.

\subsection{Bloch-Redfield tensor of a pseudo-spin $1/2$\label{appendixb}}
In this section we provide the explicit expressions for all the Bloch-Redfield tensor elements ${\cal R}_{nn',mm'}$ in the case of a 2-level spin system.
For convenience, we define $\Delta=E_2-E_1\ge 0$. From the hermiticity of the density matrix operator $\hat \rho$ one has that ${\cal R}_{nn',mm'}^*={\cal R}_{n'n,m'm}$.
Using Eq. (\ref{ratesafbF}), we get the pure population rates
\beqa
{\cal R}_{22,11}&=&  -{\cal R}_{11,11} =   \frac{\pi\rho^2}{\hbar} \Lambda_{1,2,2,1}
{\cal G}(-\Delta),
\crcr
{\cal R}_{11,22}&=&  -{\cal R}_{22,22}=
\frac{\pi\rho^2}{\hbar}
\Lambda_{1,2,2,1}{\cal G}(\Delta),
\cr
&&
\label{ratesRnmnm}
\eeqa
and the scattering rates between populations and coherences
\beqa
{\cal R}_{11,21}&=&   {\cal R}_{11,12}^*=
\frac{\pi\rho^2}{\hbar}
  \Big[
  -\Lambda_{1,2,2,2}{\cal G}[0]  -\Lambda_{1,1,1,2}{\cal G}[\Delta ]
  \crcr
  &&
  +\Lambda_{1,2,1,1} 
({\cal G}[0]+{\cal G}[\Delta ])\Big],
\crcr
{\cal R}_{22,21}&=&  {\cal R}_{22,12}^*=
\frac{\pi\rho^2}{\hbar} 
\Big[  -\Lambda_{2,1,11}{\cal G}[0]
\crcr
&& -\Lambda_{2,2,12}{\cal G}[-\Delta ]
+\Lambda_{2,1,2,2}({\cal G}[0]+{\cal G}[-\Delta ])
\Big],
\crcr
{\cal R}_{12,11}&=& {\cal R}_{21,11}^*=
\frac{\pi\rho^2}{\hbar}
\Big[  \Lambda_{1,1,1,2}-\Lambda_{1,2,2,2}\big]{\cal G}[-\Delta ]
\crcr
{\cal R}_{12,22}&=& {\cal R}_{21,22}^*=
\frac{\pi\rho^2}{\hbar}
\Big[ -\Lambda_{1,1,1,2}+\Lambda_{1,2,2,2}\Big]{\cal G}[\Delta].
\crcr
&&
\label{Rnmnpmp}
\eeqa
In addition, there are two different transition rates between the coherences
\beqa
{\cal R}_{12,12}&=& {\cal R}_{21,21}^*=
-\frac{\pi\rho^2}{\hbar}
\Lambda_{1,2,2,1} \left({\cal G}[-\Delta]+ {\cal G}[\Delta]\right)
\crcr
{\cal R}_{12,21}&=& {\cal R}_{21,12}^*=
\frac{\pi\rho^2}{\hbar}
\Lambda_{1,2,1,2} \left({\cal G}[-\Delta ]+{\cal G}[\Delta ]\right).
\label{puredech}
\eeqa

\subsection{Relation between the $M$-matrix and the Bloch-Redfield  tensor \label{apendixap}}
When the time evolution of the linearly independent  components of the density matrix of a degenerate 2-level system, 
encoded in the vector $\vec P$ of Eq. (\ref{difeqs}), are written in terms of the BR tensor components ${\cal R}_{nm,n'm'}$,
the different matrix elements in Eq. (\ref{Mmatrix}) are given by
\beqa
\label{Cgamma}
 \Gamma &=& {\cal R}_{11,22}+{\cal R}_{22,11},\crcr
\label{mgamma}
\gamma &=&  -2{\rm Re}\left[{\cal R}_{12,12}+{\cal R}_{21,12}\right],\crcr
\label{mgammap}
\gamma' &=&  -2{\rm Im}\left[{\cal R}_{12,12}+{\cal R}_{21,21}\right],\crcr
\label{M12}
M_{p,r} &=&  -2{\rm Re}\left[ {\cal R}_{11,12}-{\cal R}_{22,12}\right],\crcr
\label{M13}
M_{p,i} &=&  2{\rm Im}\left[ {\cal R}_{11,12}-{\cal R}_{22,12}\right],\crcr
\label{M21}
M_{r,p} &=&  -{\rm Re}\left[ {\cal R}_{12,11}-{\cal R}_{12,22}\right]/2,\crcr
\label{M23}
M_{r,i} &=&  -{\rm Im}\left[ {\cal R}_{12,12}-{\cal R}_{12,21}\right],\crcr
\label{M31}
M_{i,p}&=&  -{\rm Im}\left[ {\cal R}_{12,11}-{\cal R}_{12,22}\right]/2,\crcr
\label{M32}
M_{i,r} &=&  {\rm Im}\left[ {\cal R}_{12,11}+{\cal R}_{12,21}\right]/2.
\eeqa

\section{Bosonic representation of the excitations of the Fermi gas and the spin-boson model\label{bosoni}}

The mapping to the spin-boson model makes use of the so called bosonization technique~\cite{Emery_Devreese_book1979} that permits mapping the excitations of a one dimensional interacting  Fermi system into a theory of bosons that, in some instances, reduces to a free boson theory. 
  For 2D and 3D baths, it is still possible to use the bosonization technique  when it comes to describe non-interacting fermions coupled to a local impurity with rotational symmetry, such as the single impurity Kondo Hamiltonian.  Thus, one can use the partial the wave decomposition of the scattering states and keep only the $s$-wave states, which basically leads to a set of decoupled one dimensional channels~\cite{Schotte_Schotte_prb_1969}. Notice that, in general, the rotation invariance condition is not satisfied by the spin-phonon coupling.

 Introducing the linear dispersion $\epsilon_k=\hbar v_F |k|$, valid in a small region around the Fermi energy, the free electron Hamiltonian can be rewritten as
\begin{equation}
h_0= \hbar v_F\sum_{k\sigma} |k| c^{\dagger}_{k,\sigma} c_{k\sigma}.
\label{hfe}
\end{equation}
Next, we define the following bosonic operators
\beqa
b_k=\left(\frac{\pi}{k L}\right)^{1/2}\sum_{q\sigma}\sigma c_{q-k,\sigma}^\dag c_{q\sigma},
\eeqa
where $L$ is the length of a normalization box such that the wave vectors are quantized as $k_n=2\pi n/L$ and the limit $L\to \infty$ is finally taken. Hence, $b_k^ \dag$ basically creates a spin-flip electron hole pair. Then, ${\cal H}_R$ can be written as
\begin{equation}
h_0= \hbar v_F\sum_{k\le k_c} |k| b^{\dagger}_{k} b_{k},
\label{h0b}
\end{equation}
where $k_c$ is a momentum cut-off of the order of the bandwidth introduced to remove wavevectors beyond the scale of the Fermi wavenumber $k_F$.

The bosonization technique is particularly suitable for the Ising coupling. In fact, when the Kondo exchange coupling projected into the low energy doublet takes the form
\beq
{\cal V}_K \approx j_z \hat\tau_z s_z(0),
\eeq
the bosonized version of the Kondo Hamiltonian involves only the spin density operator~\cite{Leggett_Chakravarty_rmphys_1987}. As explained in Sec. \ref{TLT}, this is the case of the quantum type of nanomagnets, such as integer spins described by Hamiltonian (\ref{Hde}).
The Ising part of the fermion spin density can be then written down in terms of the bosonic operators as:
\begin{equation}
s_z(0)=
\sum_{0<k<k_c}
\left(\frac{k}{\pi L}\right)^{1/2}(b_k^\dag+b_k).
\label{szb}
\end{equation}

In the following we introduce two important quantities, the density of conduction electrons states at the Fermi energy, given by $\rho=(2\pi \hbar v_F)^{-1}$ and the dimensionless coupling constant 
$\alpha=(\rho j_z)^2$. Using  Eqs. (\ref{h0b}) and (\ref{szb}), we can  can write the 
Hamiltonian of the truncated TLS interacting with the electrons gas in the bosonized form. This can be easily done in the two  limiting cases in which the Kondo interaction is reduced to the Ising form. 
First, for the integer spin, described by the coupling Hamiltonian (\ref{VQ}) in the basis (\ref{Qbasis}).
 By making a $\pi/2$ rotation around the $y$ axis, we can write the resulting Hamiltonian in the bosonized form  as
\begin{equation}
{\cal H}_{\rm SBQ}=h_0- \frac{\Delta}{2}\hat\tau_x+  \hat \tau_z \sqrt{\alpha}
\sum_{0\le k \le k_c} g_k \left(b_k^\dag+b_k\right),
\label{SBQ}
\end{equation}
with $g_k=\hbar v_f(\pi k/L)^{1/2}$.
This corresponds to the spin-boson model for an Ohmic spectral density~\cite{Leggett_Chakravarty_rmphys_1987},  a model introduced to study the dissipative dynamics of a TLS. Although exact analytical results are not available, it admits nonperturbative  solutions for some phase space parameters~\cite{Leggett_Chakravarty_rmphys_1987}.  Furthermore, these approximate solutions are in good agreement with numerical analysis based on numerical renormalization group~\cite{Hur_annphys_2008}.

The second case of interest corresponds to half-integer anisotropic spin with uniaxial anisotropy, described by Hamiltonian (\ref{VC}) in the basis of classical states (\ref{Cstates}) for $E=0$: 
\begin{equation}
{\cal H}_{\rm SBC}= h_0
+\hat\tau_z \sqrt{\alpha}\sum_{0\le k\le k_c} g_k(b_k^\dag+b_k).
\label{SBC}
\end{equation}
The spin-boson model is mathematically equivalent to a set of displaced harmonic oscillators~\cite{Leggett_Chakravarty_rmphys_1987}. The sign of the displacement is provided by $\hat\tau_z$, which in the case of half-integer spins is the orientation of the magnetic  moment along the easy axis.

It is worth mentioning that, in general, even a  single spin Kondo-coupled to the electron gas leads to a more complex result, since it may involve $s_{x,y}(0)$ operators that requires a more complex treatment~\cite{Leggett_Chakravarty_rmphys_1987}. In addition, in the case spin  arrays, the coupling constant is no longer momentum independent,  which results in deviations from the Ohmic behavior~\cite{Prokoev_Stamp_rpp_2000,Delgado_Rossier_inprep16}.

\section{Steady-state solution of the Bloch equation for the TLS \label{AppBE}}
In this appendix  derive  the steady state solution of  the Bloch equations 
equations (\ref{B1}) and (\ref{B2}). We assume 
${\cal V}_1^{ba}(t)=V_{ba} \cos\omega t$ and we 
 look for their steady state solution.    For that matter, we assume that the diagonal entries of $\hat\rho(t)$ are time independent whereas the off-diagonal part oscillates in phase with the perturbation, i. e.
 \begin{equation}
 \rho_{aa}(t) \equiv P_a, \,\,\, \rho_{bb}(t) \equiv P_b, \,\,\,\,\,  \rho_{ab}(t)= c e^{i\omega t} 
 \label{ansatz}
 \end{equation}
where $c$ is complex and $P_a$ and $P_b$ are real constants. The ansatz (\ref{ansatz}) is compatible with the rotating wave approximation, where the fast rotating terms $e^{\pm i(\omega_{ab}+\omega)t}$ are removed from the equation of motion of $\hat \rho(t)$. In terms of a classical magnetization field dynamics, this amounts to describe the evolution of the magnetization in a frame rotating at the Larmor frequency.
Then, the steady state version of Eqs. (\ref{B1}-\ref{B2}) reduces to 
    \begin{eqnarray}
       \label{B1st}
  0 &=&   \frac{1}{2T_1}\left( 
 (P_b-P_a) -(P_b^{\rm eq}-P_a^{\rm eq})   \right) + 
  \frac{i}{2}\Omega\left(c - c^*  \right)
  \\
  \label{B2st}
i\omega c &=&  - \frac{c}{T_2} 
+i\frac{\Delta}{\hbar} c
-  \frac{i}{2}\Omega 
  \left( 
  P_b-P_a\right),
    \end{eqnarray} 
where for simplicity we have assumed a real Rabi frequency $\Omega =V_{ba}/\hbar$.    
 We can write these equations  down as: 
 \begin{eqnarray}
 \left(
 \begin{array}{ccc}
1/T_2 -i \delta & 0&   -i\Omega/2 
 \\
 0  &   1/T_2 + i\delta &  i\Omega/2 
 \\
 i\Omega/2 &  -i\Omega/2  & 1/(2T_1)
 \end{array}
 \right)
 \left(
 \begin{array}{c}
 c \\ c^* \\ P_b-P_a
 \end{array}
 \right)=
 \left(
 \begin{array}{c}
 0\\ 0  \\ (P_b^{\rm eq}-P_a^{\rm eq})/(2T_1)
 \end{array}
 \right),
 \end{eqnarray} 
 where we have introduced the detuning $\delta=\Delta/\hbar-\omega$.
If we define $\delta P=P_b-P_a$ (and $\delta P^{\rm eq}= P_b^{\rm eq}-P_a^{\rm eq}$), the steady state occupation imbalance is given by: 
\begin{equation}
\left(\frac{\delta P-\delta P^{\rm eq}}{\delta P^{\rm eq} }\right)=-\frac{  \Omega^2 T_1T_2}{1+ T_2^2 \delta^2 +  T_1T_2 \Omega^2}.
\label{P3ss}
\end{equation}
Notice that, by definition, $-1 \leqslant \delta P \leqslant 1$, while $|c|\leqslant 1/2$. We also notice that the microscopic Bloch equations (\ref{B1}) and (\ref{B2}) slightly differ from the macroscopic dynamical evolution of the magnetization components after making the identification (\ref{defSa}). Basically, the role of the relaxation time $T_1$ of the occupations is substituted by $2\tau_1$, with $\tau_1$ the relaxation time of the $z$-component of the magnetization.


\end{document}